# Towards Automatic Resource Bound Analysis for OCaml

Jan Hoffmann Ankush Das Shu-Chun Weng
Carnegie Mellon University Yale University

**Abstract**

This article presents a resource analysis system for OCaml programs. This system automatically derives worst-case resource bounds for higher-order polymorphic programs with user-defined inductive types. The technique is parametric in the resource and can derive bounds for time, memory allocations and energy usage. The derived bounds are multivariate resource polynomials which are functions of different size parameters that depend on the standard OCaml types. Bound inference is fully automatic and reduced to a linear optimization problem that is passed to an off-the-shelf LP solver. Technically, the analysis system is based on a novel multivariate automatic amortized resource analysis (AARA). It builds on existing work on linear AARA for higher-order programs with user-defined inductive types and on multivariate AARA for first-order programs with built-in lists and binary trees. For the first time, it is possible to automatically derive polynomial bounds for higher-order functions and polynomial bounds that depend on user-defined inductive types. Moreover, the analysis handles programs with side effects and even outperforms the linear bound inference of previous systems. At the same time, it preserves the expressivity and efficiency of existing AARA techniques. The practicality of the analysis system is demonstrated with an implementation and integration with Inria's OCaml compiler. The implementation is used to automatically derive resource bounds for 411 functions and 6018 lines of code derived from OCaml libraries, the CompCert compiler, and implementations of textbook algorithms. In a case study, the system infers bounds on the number of queries that are sent by OCaml programs to DynamoDB, a commercial NoSQL cloud database service.

## 1 Introduction

The quality of software crucially depends on the amount of resources —such as time, memory, and energy—that are required for its execution. Statically understanding and controlling the resource usage of software continues to be a pressing issue in software development. Performance bugs are very common and among the bugs that are most difficult to detect [41, 50] and large software systems are plagued by performance problems. Moreover, many security vulnerabilities exploit the space and time usage of software [43, 21].

Developers would greatly profit from high-level resource-usage information in the specifications of software libraries and other interfaces, and from automatic warnings about potentially high-resource usage during code review. Such information is particularly relevant in contexts of mobile applications and cloud services, where resources are limited or resource usage is a major cost factor.

Recent years have seen fast progress in developing frameworks for statically reasoning about the resource usage of programs. Many advanced techniques for imperative integer programs apply abstract interpretation to generate numerical invariants. The obtained *size-change information* forms the basis for the computation of actual bounds on loop iterations and recursion depths; using counter instrumentation [27], ranking functions [6, 2, 15, 52], recurrence relations [4, 1], and abstract interpretation itself [58, 18]. Automatic resource analysis techniques for functional programs are based on sized types [54], recurrence relations [23], term-rewriting [9], and amortized resource analysis [35, 42, 30, 51].

Despite major steps forward, there are still many obstacles to overcome to make resource analysis technologies available to developers. On the one hand, typed functional programs are particularly well-suited for automatic resource-bound analysis since the use of pattern matching and recursion often results in a relatively regular code structure. Moreover, types provide detailed information about the shape of data structures. On the other hand, existing automatic techniques for higher-order programs can only infer linear bounds [54, 42]. Furthermore, techniques that can derive polynomial bounds are limited to bounds that depend on predefined lists and binary trees [33, 30] or integers [15, 52]. Finally, resource analyses for functional programs have been implemented for custom languages that are not supported by mature tools for compilation and development [35, 54, 42, 30, 51].

The goal of a long term research effort is to overcome these obstacles by developing Resource Aware ML (RAML), a resource-aware version of the functional programming language OCaml. RAML is based on an automatic amortized resource analysis (AARA) that derives multivariate polynomials that are functions of the sizes of the inputs. In this paper, we report on *three main contributions* that are part of this effort.

1. We present the first implementation of an AARA that is integrated with an industrial-strength compiler.

2. We develop the first automatic resource analysis system that infers multivariate polynomial bounds that depend on size parameters of complex user-defined data structures.

3. We present the first AARA that infers polynomial bounds for higher-order functions.

The techniques we develop are not tied to a particular resource but are parametric in the resource of interest. RAML infers tight bounds for many complex example programs such as sorting algorithms with complex comparison functions, Dijkstra's single-source shortest-path algorithm, and the most common higher-order functions such as (sequences) of nested maps, and folds. The technique is naturally compositional, tracks size changes of data across function boundaries, and can deal with amortization effects that arise, for instance, from the use of a functional queue. Local inference rules generate linear constraints and reduce bound inference to off-the-shelf linear program (LP) solving, despite deriving polynomial bounds.

To ensure compatibility with OCaml's syntax, we reuse the parser and type inference engine from Inria's OCaml compiler [47]. We extract a type-annotated syntax tree to perform (resource preserving) code transformations and the actual resource-bound analysis. To precisely model the evaluation of OCaml, we introduce a novel operational semantics that makes the efficient handling of function closures in Inria's compiler explicit. The semantics is complemented by a new type system that refines function types.

To express a wide range of bounds, we introduce a novel class of multivariate resource polynomials that map data of a given type to a non-negative number. The set of multivariate resource polynomials that is available for bound inference depends on the types of input data. It can be parametric in integers, lengths of lists, or number of particular nodes in an inductive data type. As a special case, a resource polynomial can contain conditional additive factors. These novel multivariate resource polynomials are a substantial generalization of the resource polynomials that have been previously defined for lists and binary trees [30]. To deal with realistic OCaml code, we develop a novel multivariate AARA that handles higher-order functions. To this end, we draw inspirations from multivariate AARA for first-order programs [30] and linear AARA for higher-order programs [42]. However, our new solution is more than the combination of existing techniques. For instance, we infer linear bounds for the curried append function for lists, which has not been possible previously [42]. Moreover, we address specifics of Inria's OCaml compiler such as the evaluation order of function arguments to efficiently avoid function-closure creation.

We performed experiments on more than 6018 lines of OCaml code. We still do not support all language features of OCaml and it is thus not straightforward to automatically analyze complete existing applications. However, the automatic analysis performs well on code that only uses supported language features. For instance, we applied RAML to OCaml's standard list library *list.ml*: RAML automatically derives evaluation-step bounds for 47 of the 51 top-level functions. All derived bounds are asymptotically tight.

It is also easy to develop and analyze real OCaml applications if we keep the current capabilities of the system in mind. In Section 9, we present a case study in which we automatically bound the number of queries that an OCaml program issues to Amazon's DynamoDB NoSQL cloud database service. Such bounds are interesting since Amazon charges DynamoDB users based on the number of queries made to a database.

Our experiments are easily reproducible: The source code of RAML, the OCaml code for the experiments, and an easy-to-use interactive web interface are available online [29].

# 2 Overview

Before we describe the technical development, we give a short overview of the challenges and achievements of our work.

**Example Bound Analysis (Running Example).** To demonstrate user interaction with RAML, Figure 1 contains an example bound analysis. The OCaml code in Figure 1 will serve as a running example in this article. The function *abmap* is a polymorphic map function for a user-defined list that contains *Acons* and *Bcons* nodes. It takes two functions *f* and *g* as arguments and applies *f* to data stored in the A-nodes and *g* to data stored in the B-nodes. The function *asort* takes a comparison function and an A-B-list in which the A-nodes contain lists. It then uses *quicksort* (the code of *quicksort* is also automatically analyzed and available online [29]) to sort the lists in the A-nodes. The B-nodes are left unchanged. The function *asort'* is a variation of *asort* that raises an exception if it encounters a B-node in the list.

To derive a worst-case resource bound with RAML, the user needs to pick a maximal degree of the search space of polynomials and a resource metric. In the example analysis in Figure 1 we picked degree 4 and the *steps* metric which counts the number of evaluation steps in the big-step semantics. After 0.23 seconds, RAML reports a bound for each of the top-level functions. The shown output is only an excerpt. In this case, all derived bounds are tight in the sense that there are inputs for every size that exactly result in the reported number of evaluation steps.

In the derived bound, for *abmap* RAML assumes that the resource cost of $f$ and $g$ is 0. So we get a linear bound. In the case of *asort* we derive a bound which is quadratic in the maximal length of the lists that are stored in the A-nodes ($22K + 13K^2$) for every A-node in the

```
type ('a,'b) ablist =  Acons of 'a * ('a,'b) ablist
                     | Bcons of 'b * ('a,'b) ablist
                     | Nil

let rec abmap f g abs = match abs with
  | Acons (a,abs') → Acons(f a, abmap f g abs')
  | Bcons (b,abs') → Bcons(g b, abmap f g abs')
  | Nil → Nil

let asort gt abs = abmap (quicksort gt) (fun x → x) abs

let asort' gt abs = abmap (quicksort gt) (fun _ → raise Inv_arg) abs

let btick = abmap (fun a → a) (fun b → Raml.tick 2.5; b)
```

Excerpt of the RAML output for analyzing evaluation steps ($0.23s$ run time):

```
Simplified bound for abmap:
   3.00 + 12.00*L + 12.00*N
Simplified bound for asort:
   11.00 + 22.00*K*N + 13.00*K^2*N + 13.00*L + 15.00*N
Simplified bound asort':
   13.00 + 22.00*K*N + 13.00*K^2*N + 15.00*N
 where
   L is the number of Bcons-nodes of the 2nd (3rd) component of the argument
   N is the number of Acons-nodes of the 2nd (3rd) component of the argument
   K is the maximal number of ::-nodes in the Acons-nodes of the 2nd component
     of the argument
```

**Figure 1:** The function *abmap* will serve as a running example in this article. When deriving the linear bound for *abmap*, we assume that the higher-order arguments *f* and *g* have no resource consumption. If *abmap* is applied to concrete functions, like in *asort* and *asort'* then the cost of the concrete application is bounded. Only the *Acons* node contribute to the cubic cost in the bound of *asort*. Moreover, the number of *Bcons* nodes do not contribute to the linear factor in *asort'*.

list ($(22K+13K^2)N$) plus an additional linear factor that also depends on the number of B-nodes that are simply traversed ($13L+15N$). For *asort'* this linear factor only depends on the number of A-nodes: RAML automatically deduces that the traversal is aborted in case we encounter a B-node.

The *tick* metric can be used to derive bounds on user defined metrics. An instructive example is the function *btick*. With the tick metric, RAML derives the bound $2.5L$ where $L$ is the number of B-nodes in the argument list. This is a tight bound on the sum of "ticks" that are executed in an evaluation of *btick*. Ticks can also be negative to express that resources become available.

Note that RAML does not make guarantees about the precision of the derived bounds. Since an evaluation-step bound proves termination, bound analysis is an undecidable problem. So there are many functions for which RAML cannot derive a bound either because no (polynomial) bound exists or the analysis is not able to find a bound. In these cases, RAML terminates with a message like *"A bound for abmap could not be derived."*

**Currying and Function Closures.** Currying and function closures pose a challenge to automatic resource analysis systems that has not been addressed in the past. To see why, assume that we want to design a type system to verify resource usage. Now consider for example the curried append function which has the type *append* : $\alpha\,\text{list} \to \alpha\,\text{list} \to \alpha\,\text{list}$ in OCaml. At first glance, we might say that the time complexity of *append* is $O(n)$ if $n$ is the length of the first argument. But a closer inspection of the definition of *append* reveals that this is a gross simplification. In fact, the complexity of the partial function call *app_par = append* $\ell$ is constant. Moreover, the complexity of the function *app_par* is linear—not in the length of the argument but in the length of the list $\ell$ that is captured in the function closure. We are not aware of any existing approach that can automatically derive a worst-case time bound for the curried append function. For example, previous AARA systems would fail without deriving a bound [42, 30].

In general, we have to describe the resource consumption of a curried function $f : A_1 \to \cdots \to A_n \to A$ with $n$ expressions $c_i(a_1,\ldots,a_i)$ such that $c_i$ describes the complexity of the computation that takes place after $f$ is applied to $i$ arguments $a_1,\ldots,a_i$. In Inria's OCaml implementation, the situation is even more complex since the resource usage (time and space) depends on how a function is used at its call sites. If *append* is partially applied to one argument then a function closure is created as expected. However—and this is one of the reasons for OCaml's great performance—if *append* is applied to both of its arguments at the same time then the intermediate closure is not created and the performance of the function is even better than that of the curried version since we do not have to create a pair before the application.

To model the resource usage of curried functions accurately we refine function types to capture how functions are used at their call sites. For example, *append* can have both of the following types

$$\alpha\,\text{list} \to \alpha\,\text{list} \to \alpha\,\text{list} \quad \text{and} \quad [\alpha\,\text{list}, \alpha\,\text{list}] \to \alpha\,\text{list}\,.$$

The first type implies that the function is partially applied and the second type implies that the function is applied to both arguments at the same time. Of course, it is possible that the function has both types (technically we achieve this using let polymorphism). For the second type, our system automatically derives tight time and space bounds that are linear in the first argument. However, our system fails to derive a bound for the first type. The reason is that we made the design decision to not derive bounds that asymptotically depend on data captured in function closures to keep the complexity of the system at a manageable level.

Fortunately, *append* belongs to a large set of OCaml functions in the standard library that are defined in the form *let rec f x y z = e*. If such a function is partially applied, the only computation that happens is the creation of a closure. As a result, *eta expansion* does not change the resource behavior of programs. This means for example that we can safely replace the expression *let app_par = append* $\ell$ *in e* with the expression *let app_par x = append* $\ell$ *x in e* prior to the analysis. Consequently, we can always use the type $[\alpha\,\text{list}, \alpha\,\text{list}] \to \alpha\,\text{list}$ of *append* that we can successfully analyze.

The conditions under which functions can be analyzed might look complex at first but they can be boiled down to a simple principle:

> The worst-case resource usage of a function must be expressible as a function of the sizes of its *arguments*.

**Higher-Order Arguments.** The other main challenge with higher-order resource analysis is functions with higher-order arguments. To a large extent, this problem has been successfully solved for linear resource bounds in previous work [42]. Basically, the higher-order case is reduced to the first-order case if the higher-order arguments are available. It is not necessary to reanalyze such higher-order functions for every call site since we can abstract the resource usage with a constraint system that has holes for the constraints of the function arguments. However, a presentation of the system in such a way mixes type checking with the constraint-based type inference. Therefore, we chose to present the analysis system in a more declarative way in which the bound of a function with higher-order arguments is derived with respect to a given set of resource behaviors of the argument functions.

A concrete advantage of our declarative view is that we can derive a meaningful type for a function like *map* for lists even when the higher-order argument is not available. The function *map* can have the following types.

$$(\alpha \to \beta) \to \alpha\,\text{list} \to \beta\,\text{list} \qquad [\alpha \to \beta, \alpha\,\text{list}] \to \beta\,\text{list}$$

Unlike *append*, the resource usage of *map* does not depend on the size of the first argument. So both types are equivalent in our system except for the cost of creating an intermediate closure. If the higher-order argument is not available then previous systems [42] produce a constraint system that is not meaningful to a user. An innovation in this work is that we are also able to report a meaningful resource bound for *map* if the arguments are not available. To this end, we assume that the argument function does not consume resources. For example, we report in the case of *map* that the number of evaluation steps needed is $11n+3$ and the number of heap cells needed is $4n+2$ where $n$ is the length of the input list. Such bounds are useful for two purposes. First, a developer can see the cost that *map* itself contributes to the total cost of a program. Second, the time bound for *map* proves that *map* is guaranteed to terminate if the higher-order argument terminates for every input.

In contrast, consider the function *rec_scheme* : $(\alpha\,\text{list} \to \alpha\,\text{list}) \to \alpha\,\text{list} \to \beta\,\text{list}$ that is defined as follows.

```
let rec rec_scheme f l =
  match l with | [] → []
               | x::xs → rec_scheme f (f l)
let g = rec_scheme tail
```

Here, RAML is not able to derive an evaluation-step bound for *rec_scheme* since the number of evaluation steps (and even termination) depends on the argument $f$. However, RAML derives the tight evaluation-step bound $12n+7$ for the function $g$.

**Polynomial Bounds and Inductive Types.** Existing AARA systems are either limited to linear bounds [35, 42] or to polynomial bounds that are functions of the sizes of simple predefined lists and binary-tree data structures [30]. In contrast, this work presents the first analysis that can derive polynomial bounds that depend on size parameters of complex user-defined data structures.

The bounds we derive are multivariate resource polynomials that can take into account individual sizes of inner data structures. While it is possible to simplify the resource polynomials in the user output, it is essential to have this more precise information for intermediate results to derive tight whole-program bounds.

In general, the resource bounds are built of functions that count the number of specific tuples that one can form from the nodes in a tree-like data structure. In their simplest form (i.e., without considering the data stored inside the nodes), they have the form

$$\lambda a.|\{\vec{a} \mid a_i \text{ is an } A_{k_i}\text{-node in } a \text{ and if } i < j \text{ then } a_i <^{a}_{pre} a_j\}|$$

where $a$ is an inductive data structure with constructors $A_1,\ldots,A_m$, $\vec{a} = (a_1,\ldots,a_n)$, and $<^a_{pre}$ denotes the pre-order (tree traversal) on the tree $a$. We are able to keep track of changes of these quantities in pattern matches and data construction fully automatically by generating linear constraints. At the same time, they allow us to accurately describe the resource usage of many common functions in the same way it has been done previously for simple types [28]. As an interesting special case, we can also derive conditional bounds that describe the resource usage as a conditional statement. For instance, for an expression such as

```
match x with | True → quicksort y | False → y
```

we derive a bound that is quadratic in the length of $y$ if $x$ is *True* and constant if $x$ is *False*.

**Effects.** Our analysis handles references and arrays by ensuring that resource cost does not asymptotically depend on values that have been stored in mutable cells. While it has been shown that it is possible to extend AARA to handle mutable state [17], we decided not to add the feature in the current system to focus on the presentation of the main contributions. There are still a lot of possible interactions with mutable state, such as storing functions in references.

# 3 Setting the Stage

We describe and formalize the new resource analysis using Core RAML, a subset of the intermediate language that we use to perform the analysis. Expressions in Core RAML are in share-let-normal form, which means that syntactic forms allow only variables instead of arbitrary terms whenever possible without restricting expressivity. We automatically transform user-level OCaml programs to Core RAML without changing their resource behavior before the analysis.

**Syntax.** For the purpose of this article, the syntax of Core RAML expressions is defined by the following grammar. The actual core expressions also contain constants and operators for primitive data types such as integer, float, and boolean; arrays and built-in operations for arrays; conditionals; and *free* versions of syntactic forms. These free versions are semantically identical to the standard versions but do not contribute to the resource cost. This is needed for the resource preserving translation of user-level code to share-let-normal form.

$$\begin{aligned}
e ::=\ & x \mid x\, x_1 \cdots x_n \mid C\, x \mid \lambda x.e \mid \mathsf{ref}\ x \mid !x \mid x_1 := x_2 \mid \mathsf{fail} \mid \mathsf{tick}\,(q) \\
& \mid \mathsf{match}\ x\ \mathsf{with}\ C\, y \to e_1 \mid e_2 \\
& \mid (x_1,\ldots,x_n) \mid \mathsf{match}\ x\ \mathsf{with}\ (x_1,\ldots,x_n) \to e \\
& \mid \mathsf{share}\ x\ \mathsf{as}\ (x_1,x_2)\ \mathsf{in}\ e \mid \mathsf{let}\ x = e_1\ \mathsf{in}\ e_2 \mid \mathsf{let\ rec}\ F\ \mathsf{in}\ e \\
F ::=\ & f = \lambda x.e \mid F_1\ \mathsf{and}\ F_2
\end{aligned}$$

The syntax contains forms for variables, function application, data constructors, lambda abstraction, references, tuples, pattern matching, and (recursive) binding. For simplicity, we only allow recursive definitions of functions. In the function application we allow the application of several arguments at once. This is useful to statically determine the cost of closure creation but also introduces ambiguity. The type system will determine if an expression like $f\ x_1\ x_2$ is parsed as $(f\ x_1\ x_2)$ or $(f\ x_1)\ x_2$. The sharing expressions $\mathsf{share}\ x\ \mathsf{as}\ (x_1,x_2)\ \mathsf{in}\ e$ is not standard and used to explicitly introduce multiple occurrences of a variable. It binds the free variables $x_1$ and $x_2$ in $e$. The expression *fail* is used to model exceptions. The expression *tick(q)* contains a floating point constant $q$. It can be used with the tick metric to specify a constant cost. A negative floating point number $q$ means that resources become available.

We focus on this set of language features since it is sufficient to present the main contributions of our work. We sometimes take the liberty to describe examples in user level syntax and to use features such as built-in data types that are not described in this article.

**Big-Step Operational Cost Semantics.** The resource usage of RAML programs is defined by a big-step operational cost semantics. The semantics has three interesting non-standard features. First, it measures (or defines) the resource consumption of the evaluation of a RAML expression by using a resource metric that defines a constant cost for each evaluation step. If this cost is negative then resources are returned. Second, it models terminating and diverging executions by inductively describing finite subtrees of infinite execution trees. Third, it models OCaml's stack-based mechanism for function application, which avoids creation of intermediate function closures.

The semantics of Core RAML is formulated with respect to a stack (to store arguments for function application), an environment, and a heap. Let *Loc* be an infinite set of *locations* modeling memory addresses. A *heap* is a finite partial mapping $H : \mathit{Loc} \rightharpoonup \mathit{Val}$ that maps locations to values. An *environment* is a finite partial mapping $V : \mathit{Var} \rightharpoonup \mathit{Loc}$ from variable identifiers to locations. An *argument stack* $S ::= \cdot \mid \ell{::}S$ is a finite list of locations. We assume every heap $H$ contains a distinguished location $\mathrm{Null} \in \mathrm{dom}(H)$ such that $H(\mathrm{Null}) = \mathrm{Null}$.

The set of RAML *values Val* is given by

$$v ::= \ell \mid (\ell_1, \dots, \ell_k) \mid (\lambda x.e, V) \mid (C, \ell)$$

A value $v \in \mathit{Val}$ is either a location $\ell \in \mathit{Loc}$, a tuple of locations $(\ell_1, \dots, \ell_k)$, a function closure $(\lambda x.e, V)$, or a node of a data structure $(C, \ell)$ where $C$ is a constructor and $\ell$ is a location. In a function closure $(\lambda x.e, V)$, $V$ is an environment, $e$ is an expression, and $x$ is a variable.

Since we also consider resources like memory that can become available during an evaluation, we have to track the *watermark* of the resource usage, that is, the maximal number of resource units that are simultaneously used during an evaluation. To derive a watermark of a sequence of evaluations from the watermarks of the sub evaluations one has to also take into account the number of resource units that are available after each sub evaluation.

The big-step operational evaluation rules Figure 2 and Figure 3 are formulated with respect to a resource metric $M$. They define an evaluation judgment of the form

$$S, V, H \;{}_M\!\vdash e \Downarrow (\ell, H') \mid (q, q') \,.$$

It expresses the following. If the argument stack $S$, the environment $V$, and the initial heap $H$ are given then the expression $e$ evaluates to the location $\ell$ and the new heap $H'$. The evaluation of $e$ needs $q \in \mathbb{Q}_0^+$ resource units (watermark) and after the evaluation there are $q' \in \mathbb{Q}_0^+$ resource units available. The actual resource consumption is then $\delta = q - q'$. The quantity $\delta$ is negative if resources become available during the execution of $e$.

There are two other behaviors that we have to express in the semantics: failure (i.e., array access outside array bounds) and divergence. To this end, our semantic judgement not only evaluates expressions to values but also to an error $\bot$ and to incomplete computations expressed by $\circ$. The judgement has the general form

$$S, V, H \;{}_M\!\vdash e \Downarrow w \mid (q, q') \quad \text{where} \quad w ::= (\ell, H) \mid \bot \mid \circ \,.$$

Intuitively, this evaluation statement expresses that the watermark of the resource consumption after some number of evaluation steps is $q$ and there are currently $q'$ resource units left. A

$$\frac{S \neq \cdot \qquad H(V(x)) = (\lambda x.e, V') \qquad S, V', H_M \vdash \lambda x.e \Downarrow w \mid (q, q')}{S, V, H_M \vdash x \Downarrow w \mid M^{\mathsf{var}} \cdot (q, q')} \text{ (E:VarApp)}$$

$$\frac{V(x) = \ell}{\cdot, V, H_M \vdash x \Downarrow (\ell, H) \mid M^{\mathsf{var}}} \text{ (E:Var)} \qquad \frac{}{S, V, H_M \vdash e \Downarrow \circ \mid 0} \text{ (E:Abort)}$$

$$\frac{V(x_1) :: \cdots :: V(x_n), V, H_M \vdash x \Downarrow w \mid (q, q') \qquad S = \cdot \lor w \in \{\bot, \circ\}}{S, V, H_M \vdash x\, x_1 \cdots x_n \Downarrow w \mid M_n^{\mathsf{app}} \cdot (q, q')} \text{ (E:App)}$$

$$\frac{\begin{array}{c} S \neq \cdot \qquad H(\ell) = (\lambda x.e, V') \\ V(x_1) :: \cdots :: V(x_n), V, H_M \vdash x \Downarrow \ell \mid (q, q') \qquad S, V', H_M \vdash \lambda x.e \Downarrow w \mid (p, p') \end{array}}{S, V, H_M \vdash x\, x_1 \cdots x_n \Downarrow w \mid M_n^{\mathsf{app}} \cdot (q, q') \cdot (p, p')} \text{ (E:AppApp)}$$

$$\frac{S, V[x \mapsto \ell], H_M \vdash e \Downarrow w \mid (q, q')}{\ell :: S, V, H_M \vdash \lambda x.e \Downarrow w \mid M^{\mathsf{bind}} \cdot (q, q')} \text{ (E:AbsBind)}$$

$$\frac{H' = H, \ell \mapsto (\lambda x.e, V)}{\cdot, V, H_M \vdash \lambda x.e \Downarrow (\ell, H') \mid M^{\mathsf{abs}}} \text{ (E:AbsClos)}$$

$$\frac{\cdot, V, H_M \vdash e_1 \Downarrow w \mid (q, q') \qquad w \in \{\bot, \circ\}}{S, V, H_M \vdash \mathsf{let}\ x = e_1\ \mathsf{in}\ e_2 \Downarrow w \mid M_1^{\mathsf{let}} \cdot (q, q')} \text{ (E:Let1)}$$

$$\frac{\cdot, V, H_M \vdash e_1 \Downarrow (\ell, H') \mid (q, q') \qquad S, V[x \mapsto \ell], H'_M \vdash e_2 \Downarrow w \mid (p, p')}{S, V, H_M \vdash \mathsf{let}\ x = e_1\ \mathsf{in}\ e_2 \Downarrow w \mid M_1^{\mathsf{let}} \cdot (q, q') \cdot M_2^{\mathsf{let}} \cdot (p, p')} \text{ (E:Let2)}$$

$$\frac{\begin{array}{c} F \triangleq f_1 = \lambda x_1.e_1\ \mathsf{and} \cdots \mathsf{and}\ f_n = \lambda x_n.e_n \qquad V' = V[f_1 \mapsto \ell_1, \ldots, f_n \mapsto \ell_n] \\ H' = H, \ell_1 \mapsto (\lambda x_1.e_1, V'), \ldots, \ell_n \mapsto (\lambda x_n.e_n, V') \qquad S, V', H'_M \vdash e_0 \Downarrow w \mid (q, q') \end{array}}{S, V, H_M \vdash \mathsf{let\ rec}\ F\ \mathsf{in}\ e_0 \Downarrow w \mid M^{\mathsf{rec}} \cdot (q, q')} \text{ (E:LetRec)}$$

$$\frac{H' = H, \ell \mapsto (C, V(x))}{\cdot, V, H_M \vdash C\, x \Downarrow (\ell, H') \mid M^{\mathsf{cons}}} \text{ (E:Cons)}$$

$$\frac{H(V(x)) = (C, \ell) \qquad S, V[y \mapsto \ell], H_M \vdash e_1 \Downarrow w \mid (q, q')}{S, V, H_M \vdash \mathsf{match}\ x\ \mathsf{with}\ C\, y \to e_1 \mid e_2 \Downarrow w \mid M_1^{\mathsf{mat}} \cdot (q, q')} \text{ (E:Mat1)}$$

$$\frac{H(V(x)) \neq (C, \ell) \qquad S, V, H_M \vdash e_2 \Downarrow w \mid (q, q')}{S, V, H_M \vdash \mathsf{match}\ x\ \mathsf{with}\ C\, y \to e_1 \mid e_2 \Downarrow w \mid M_2^{\mathsf{mat}} \cdot (q, q')} \text{ (E:Mat2)}$$

**Figure 2:** Rules of the operational big-step semantics (part 1 of 2).

$$\frac{H' = H, \ell \mapsto (V(x_1),\ldots,V(x_n))}{\cdot, V, H\,{}_M\!\vdash (x_1,\ldots,x_n) \Downarrow (\ell, H') \mid M^{\mathsf{tuple}}}\ \text{(E:TUPLE)}$$

$$\frac{H(V(x)) = (\ell_1,\ldots,\ell_n) \qquad S, V[x_1 \mapsto \ell_1,\ldots,x_n \mapsto \ell_n], H\,{}_M\!\vdash \Downarrow w \mid (q,q')}{S, V, H\,{}_M\!\vdash \mathsf{match}\ x\ \mathsf{with}\ (x_1,\ldots,x_n) \to e \Downarrow w \mid M^{\mathsf{matT}} \cdot (q,q')}\ \text{(E:MATT)}$$

$$\frac{V(x) = \ell \qquad S, V[x_1 \mapsto \ell, x_2 \mapsto \ell], H\,{}_M\!\vdash e \Downarrow w \mid (q,q')}{S, V, H\,{}_M\!\vdash \mathsf{share}\ x\ \mathsf{as}\ (x_1,x_2)\ \mathsf{in}\ e \Downarrow w \mid M^{\mathsf{share}} \cdot (q,q')}\ \text{(E:SHARE)}$$

$$\frac{H' = H, \ell \mapsto V(x)}{\cdot, V, H\,{}_M\!\vdash \mathsf{ref}\ x \Downarrow (\ell, H') \mid M^{\mathsf{ref}}}\ \text{(E:REF)} \qquad \frac{\ell = H(V(x))}{S, V, H\,{}_M\!\vdash !x \Downarrow (\ell, H) \mid M^{\mathsf{dref}}}\ \text{(E:DREF)}$$

$$\frac{H' = H[V(x_1) \mapsto V(x_2)]}{\cdot, V, H\,{}_M\!\vdash x_1 := x_2 \Downarrow (\mathrm{Null}, H') \mid M^{\mathsf{assign}}}\ \text{(E:ASSIGN)} \qquad \frac{}{S, V, H\,{}_M\!\vdash \mathsf{fail} \Downarrow \bot \mid M^{\mathsf{fail}}}\ \text{(E:UNDEF)}$$

$$\frac{}{\cdot, V, H\,{}_M\!\vdash \mathsf{tick}(q) \Downarrow (\mathrm{Null}, H) \mid M^{\mathsf{tick}}(q)}\ \text{(E:TICK)}$$

**Figure 3:** Rules of the operational big-step semantics (part 2 of 2).

resource metric $M : K \times \mathbb{N} \to \mathbb{Q}$ defines the resource consumption in each evaluation step of the big-step semantics where $K$ is a set of constants. We write $M^k_n$ for $M(k,n)$ and $M^k$ for $M(k,0)$.

It is handy to view the pairs $(q,q')$ in the evaluation judgments as elements of a monoid $\mathcal{Q} = (\mathbb{Q}^+_0 \times \mathbb{Q}^+_0, \cdot)$. The neutral element is $(0,0)$, which means that resources are neither needed before the evaluation nor returned after the evaluation. The operation $(q,q') \cdot (p,p')$ defines how to account for an evaluation consisting of evaluations whose resource consumptions are defined by $(q,q')$ and $(p,p')$, respectively. We define

$$(q,q') \cdot (p,p') = \begin{cases} (q+p-q',\ p') & \text{if } q' \le p \\ (q,\ p'+q'-p) & \text{if } q' > p \end{cases}$$

If resources are never returned (as with time) then we only have elements of the form $(q,0)$ and $(q,0)\cdot(p,0)$ is just $(q+p,0)$. We identify a rational number $q$ with an element of $\mathcal{Q}$ as follows: $q \ge 0$ denotes $(q,0)$ and $q < 0$ denotes $(0,-q)$. This notation avoids case distinctions in the evaluation rules since the constants $K$ that appear in the rules can be negative. In the semantic rules we use the notation $H' = H, \ell \mapsto v$ to indicate that $\ell \notin \mathrm{dom}(H)$, $\mathrm{dom}(H') = \mathrm{dom}(H) \cup \{\ell\}$, $H'(\ell) = v$, and $H'(x) = H(x)$ for all $x \neq \ell$.

For efficiency reasons, Inria's OCaml compiler evaluates function applications $e\ e_1 \cdots e_n$ from right to left, that is, it starts with evaluating $e_n$. In this way, one can avoid the expensive creation of intermediate function closures. A naive implementation would create $n$ function closures when evaluating the aforementioned expression: one for $e$, one for the application to the first argument, etc. By starting with the last argument, we are able to put the results of the evaluation on an argument stack and access them when we encounter a function abstraction during the evaluation. In this case, we do not create a closure but simply bind the value on the stack to the name in the abstraction.

To model the treatment of function application in the OCaml compiler, we use a stack $S$ on which we store the locations of function arguments. The only rules that push locations to $S$ are E:APP and E:APPAPP. To pop locations from the stack we modify the leaf rules that can return a function closure, namely, the rules E:VAR and E:ABS for variables and lambda abstractions: Whenever we would return a function closure $(\lambda x.e, V)$ we inspect the argument stack $S$. If $S$ contains a location $\ell$ then we pop it from the stack $S$, bind it to the argument $x$, and evaluate the function body $e$ in the new environment $V[x \mapsto \ell]$. This is defined by the rule E:ABSBIND and indirectly by the rule E:VARAPP. Another rule that modifies the argument stack is E:LET2. Here, we evaluate the subexpression $e_1$ with an empty argument stack because the arguments on the stack when evaluating the let expressions are consumed by the result of the evaluation of $e_2$.

The argument stack accurately captures Inria's OCaml compiler's behavior to avoid the creation of intermediate function closures. It also extends naturally to the evaluation of expressions that are not in share-let-normal form. As we will see in Section 6, the argument stack is also necessary to prove the soundness of the multivariate resource bound analysis.

Another important feature of the big-step semantics, is that it can model failing and diverging evaluations by allowing partial derivation judgments that can be used to derive the resource usage after $n$ steps. Technically, this is realized by the rule E:ABORT which can be applied at any point to abort the current evaluation without additional resource cost. The mechanism of aborting an evaluation is most visible in the rules E:LET1 and E:LET2: During the evaluation of a let expression we have two possibilties. The first possibility is that the evaluation of the subexpression $e_1$ is aborted using E:ABORT at some point. We can then apply the rule E:LET1 to pass on the resource usage before the abort. The second possibility is that $e_1$ evaluates to a location $\ell$. We can then apply the E:LET2 to bind $\ell$ to the variable $x$ and evaluate the expression $e_2$.

**Example Evaluation (Running Example).** We use the running example defined in Figure 1 to illustrate how the operational cost semantics works. To this end, we use the metric *steps* which assigns cost 1 to every evaluation step and the metric *tick* which assigns cost 0 to every evaluation step but *Raml.tick(q)*.

Let *abs* ≡ *Acons ([1;2],Bcons (3, Bcons (4, Nil)))* is a A-B-list and let $e_1$ the expression that arises by concatenating appending the expression *asort (>) abs* to the code in Figure 1. Then for every $H$ and $V$ there exists $H'$ and $\ell$ such that $\cdot, V, H \;{}_{\text{tick}}{\vdash}\; e_1 \Downarrow (\ell, H') \mid (0,0)$ and $\cdot, V, H \;{}_{\text{steps}}{\vdash}\; e_1 \Downarrow (\ell, H') \mid (186,0)$. Moreover, $\cdot, V, H \;{}_{\text{steps}}{\vdash}\; e_1 \Downarrow \circ \mid (n,0)$ for every $n < 186$.

Let $e_2$ be the expression that results from appending *btick abs* to the code in Figure 1. Then for every $H$ and $V$ there exists $H'$ and $\ell$ such that $\cdot, V, H \;{}_{\text{tick}}{\vdash}\; e_2 \Downarrow (\ell, H') \mid (5,0)$ and $\cdot, V, H \;{}_{\text{tick}}{\vdash}\; e_2 \Downarrow \circ \mid (n,0)$ for every $n \in \{2.5, 0\}$.

# 4 Stack-Based Type System

In this section, we introduce a type system that is a refinement of OCaml's type system. In this type system, we mirror the resource-aware type system and introduce some particularities that explain features of the resource-aware types. For the purpose of this article, we define simple types as follows.

$$T ::= \text{unit} \mid X \mid T\ \text{ref} \mid T_1 * \cdots * T_n \mid [T_1, \ldots, T_n] \to T \\ \mid \mu X. \langle C_1 : T_1 * X^{n_1}, \ldots, C_k : T_k * X^{n_k} \rangle$$

A (simple) type $T$ is the unit type, an uninterpreted type variable $X \in \mathcal{X}$, a type $T$ ref of references of type $T$, a tuple type $T_1 * \cdots * T_n$, a function type $[T_1,\ldots,T_n] \to T$, or an inductive data type $\mu X.\langle C_1 : T_1 {*} X^{n_1},\ldots,C_k : T_k {*} X^{n_k}\rangle$.

Two parts of this definition are non-standard and deserve further explanation. First, bracket function types $[T_1,\ldots,T_n] \to T$ correspond to the standard function type $T_1 \to \cdots \to T_n \to T$. The meaning of $[T_1,\ldots,T_n] \to T$ is that the function is applied to its first $n$ arguments at the same time. The type $T_1 \to \cdots \to T_n \to T$ indicates that the function is applied to its first $n$ arguments one after another. These two uses of a function can result in a very different resource behavior. For instance, in the latter case we have to create $n-1$ function closures. Also we have $n$ different costs to account for: the evaluation cost after the first argument is present, the cost of the closure when the second argument is present, etc. Of course, it is possible that a function is used in different ways in program. We account for that with let polymorphism (see the following subsection). Also note that $[T_1,\ldots,T_n] \to T$ still describes a higher-order function while $T_1 * \cdots * T_n \to T$ describes a first-order function with $n$ arguments.

Second, inductive types are required to have a particular tree-like form. This makes it possible to track costs that depend on size parameters of values of such types. It is of course possible to allow arbritary inductive types and not to track such cost. Such an extension is straighforward and we do not present it in this article.

We assume that each constructor $C \in \mathcal{C}$ is part of at most one recursive type. Furthermore we assume that each recursive type has at least one constructor. For an inductive type $T = \mu X.\langle C_1 : T_1 {*} X^{n_1},\ldots,C_k : T_k {*} X^{n_k}\rangle$ we sometimes write $T = \langle C_1 : (T_1,n_1),\ldots,C_k : (T_k,n_k)\rangle$. We say that $T_i$ is the node type and $n_i$ is the branching number of the constructor $C_i$. The maximal branching number $n = \max\{n_1,\ldots,n_k\}$ of the constructors is the branching number of $T$.

**Let Polymorphism and Sharing.** Following the design of the resource-aware type system, our stack-based type system is affine. That means that a variable in a context can be used at most once in an expression. However, we enable multiple uses of a variable with the sharing expression $\mathsf{share}\ x\ \mathsf{as}\ (x_1,x_2)\ \mathsf{in}\ e$ that denotes that $x$ can be used twice in $e$ using the (different) names $x_1$ and $x_2$. For input programs we allow multiple uses of a variable $x$ in an expression $e$ in RAML. We then introduce sharing constructs, and replace the occurrences of $x$ in $e$ with the new names before the analysis.

Interestingly, this mechanism is closely related to let polymorphism. To see this relation, first note that our type system is polymorphic but that a value can only be used with a single type in an expression. In practice, that would mean for instance that we have to define a different map function for every list type. A simple and well-known solution to this problem that is often applied in practice is let polymorphism. In principle, let polymorphism replaces variables with their definitions before type checking. For our map function it would mean to type the expression $[\mathrm{map} \mapsto e_{\mathrm{map}}]e$ instead of typing the expression $\mathsf{let}\ \mathrm{map} = e_{\mathrm{map}}\ \mathsf{in}\ e$.

In principle, it would be possible to treat sharing of variables in a similar way as let polymorphism. But if we start from an expression $\mathsf{let}\ x = e_1\ \mathsf{in}\ e_2$ and replace the occurrences of $x$ in the expression $e_2$ with $e_1$ then we also change the resource consumption of the evaluation of $e_2$ because we evaluate $e_1$ multiple times. Interestingly, this problem coincides with the treatment of let polymorphism for expressions with side effects (the so called value restriction).

In RAML, we support let polymorphism for function closures only. Assume we have a function definition $\mathsf{let}\ f = \lambda x.e_f\ \mathsf{in}\ e$ that is used twice in $e$. Then the usual approach to enable the analysis in our system would be to use sharing

$$\mathsf{let}\ f = \lambda x.e_f\ \mathsf{in}\ \mathsf{share}\ f\ \mathsf{as}\ (f_1,f_2)\ \mathsf{in}\ e'\ .$$

To enable let polymorphism, we will however define $f$ twice and ensure that we only pay once for the creation of the closure and the let binding:

$$\mathsf{let}\ f_1 = \lambda x.e_f\ \mathsf{in}\ \mathsf{let}\ f_2 = \lambda x.e_f\ \mathsf{in}\ e'$$

The functions $f_1$ and $f_2$ can now have different types. This method can cause an exponential blow up of the size of the expression. It is nevertheless appealing because it enables us to treat resource polymorphism in the same way as let polymorphism.

**Type Judgements.** Type judgements have the form

$$\Sigma;\Gamma \vdash e : T$$

where $\Sigma = T_1,\ldots,T_n$ is a list of types, $\Gamma : \mathit{Var} \rightharpoonup \mathcal{T}$ is a type context that maps variables to types, $e$ is a core expression, and $T$ is a (simple) type. The intuitive meaning (which is formalized later in this section) is as follows. Given an evaluation environment that matches the type context $\Gamma$ and an argument stack that matches the type stack $\Sigma$ then $e$ evaluates to a value of type $T$ (or does not terminate).

The most interesting feature of the type judgements is the handling of bracket function types $[T_1,\ldots,T_n] \to T$. Even though function types can have multiple forms, a well-typed expression often has a unique type (in a given type context). This type is derived from the way a function is used. For instance, we have $\lambda f.\lambda x.\lambda y.f\,x\,y : ([T_1,T_2] \to T) \to T_1 \to T_2 \to T$ and $\lambda f.\lambda x.\lambda y.(f\,x)\,y : (T_1 \to T_2 \to T) \to T_1 \to T_2 \to T$, and the two function types are both unique.

A type $T$ of an expression $e$ has a unique type derivation that produces a type judgement $\cdot,\Gamma \vdash e : T$ with an empty type stack. We call this *canonical type derivation* for $e$ and a *closed type judgement*. If $T$ is a function type $\Sigma \to T'$ then there is a second type derivation for $e$ that we call an *open type derivation*. It derives the *open type judgement* $\Sigma;\Gamma \vdash e : T'$ where $|\Sigma| > 0$. The following lemma can be proved by induction on the type derivations.

**Lemma 1.** $\cdot;\Gamma \vdash e : \Sigma \to T$ *if and only if* $\Sigma;\Gamma \vdash e : T$.

Open and canonical type judgements are *not* interchangeable. An open type judgement $\Sigma;\Gamma \vdash e : T$ can only appear in a derivation with an open root of the form $\Sigma',\Sigma;\Gamma \vdash e : T$, or in a subtree of a derivation whose root is a closed judgement of the form $\cdot;\Gamma \vdash e : \Sigma'',\Sigma \to T$ where $|\Sigma''| > 0$. In other words, in an open derivation $\Sigma;\Gamma \vdash e : T$, the expression $e$ is a function that has to be applied to $n > |\Sigma|$ arguments at the same time. In a given type context and for a fixed function type, a well-typed expression has as most one open type derivation.

**Type Rules.** Figure 4 presents selected type rules of the type system. As usual $\Gamma_1,\Gamma_2$ denotes the union of the type contexts $\Gamma_1$ and $\Gamma_2$ provided that $\mathrm{dom}(\Gamma_1) \cap \mathrm{dom}(\Gamma_2) = \emptyset$. We thus have the implicit side condition $\mathrm{dom}(\Gamma_1) \cap \mathrm{dom}(\Gamma_2) = \emptyset$ whenever $\Gamma_1,\Gamma_2$ occurs in a typing rule. Especially, writing $\Gamma = x_1{:}T_1,\ldots,x_k{:}T_k$ means that the variables $x_i$ are pairwise distinct.

There is a close correspondence between the evaluation rules and the type rules in the sense that every evaluation rule corresponds to exactly one type rule. (We view the two rules for pattern match and let binding as one rule, respectively.) The type stack is modified by the rules T:VARPUSH, T:APPPUSH, T:ABSPUSH, and T:ABSPOP. For every leaf rule that can return a function type, such as T:VAR, T:APP, and T:APPPUSH, we add a second rule that derives the equivalent open type. The reason becomes clear in the resource-aware type system in Section 6. The rules that directly control the shape of the function types are T:ABSPUSH and T:ABSPOP for lambda abstraction. While the other rules are (deterministically) syntax driven, the rules for

$$\frac{}{\cdot; x:T \vdash x:T}\ (\text{T:Var}) \qquad \frac{}{\Sigma; x:\Sigma \to T \vdash x:T}\ (\text{T:VarPush}) \qquad \frac{\Sigma;\Gamma, x{:}T_1 \vdash e:T_2}{T_1{::}\Sigma;\Gamma \vdash \lambda x.e:T_2}\ (\text{T:AbsPush})$$

$$\frac{}{\Sigma; x:[T_1,\ldots,T_n]\to\Sigma\to T, x_1:T_1,\ldots,x_n:T_n \vdash x\,x_1\cdots x_n:T}\ (\text{T:AppPush})$$

$$\frac{}{\cdot; x:[T_1,\ldots,T_n]\to T, x_1{:}T_1,\ldots,x_n{:}T_n \vdash x\,x_1\cdots x_n:T}\ (\text{T:App}) \qquad \frac{\Sigma;\Gamma \vdash \lambda x.e:T}{\cdot;\Gamma \vdash \lambda x.e:\Sigma \to T}\ (\text{T:AbsPop})$$

$$\frac{T=\mu X.\langle\ldots C:U{*}X^n\ldots\rangle \qquad \Sigma;\Gamma, y:U{*}T^n, \vdash e_1:T' \qquad \Sigma;\Gamma, x:T \vdash e_2:T'}{\Sigma;\Gamma, x:T \vdash \mathsf{match}\ x\ \mathsf{with}\ C\,y \to e_1 \mid e_2 : T'}\ (\text{T:Mat})$$

$$\frac{T=\mu X.\langle\ldots C:U{*}X^n\ldots\rangle}{\cdot; x:U{*}T^n \vdash C\,x:T}\ (\text{T:Cons}) \qquad \frac{T=T_1*\cdots*T_n}{\cdot; x_1:T_1,\ldots,x_n:T_n \vdash (x_1,\ldots,x_n):T}\ (\text{T:Tuple})$$

$$\frac{T=T_1*\cdots*T_n \qquad \Sigma;\Gamma, x_1:T_1,\ldots,x_n:T_n \vdash e:T'}{\Sigma;\Gamma, x:T \vdash \mathsf{match}\ x\ \mathsf{with}\ (x_1,\ldots,x_n) \to e:T'}\ (\text{T:MatT})$$

$$\frac{\Sigma;\Gamma, x_1:T, x_2:T \vdash e:T'}{\Sigma;\Gamma, x:T \vdash \mathsf{share}\ x\ \mathsf{as}\ (x_1,x_2)\ \mathsf{in}\ e:T'}\ (\text{T:Share})$$

$$\frac{\cdot;\Gamma_1 \vdash e_1:T_1 \qquad \Sigma;\Gamma_2, x:T_1 \vdash e_2:T_2}{\Sigma;\Gamma_1,\Gamma_2 \vdash \mathsf{let}\ x=e_1\ \mathsf{in}\ e_2:T_2}\ (\text{T:Let}) \qquad \frac{\Sigma;\Gamma \vdash e:B}{\Sigma;\Gamma, x{:}A \vdash e:B}\ (\text{T:Weak})$$

$$\frac{\begin{array}{c} F \triangleq f_1=\lambda x_1.e_1\ \mathsf{and}\ \cdots\ \mathsf{and}\ f_n=\lambda x_n.e_n \\ \Delta = f_1:T_1,\ldots,f_n:T_n \qquad \forall i: \cdot;\Gamma_i,\Delta \vdash \lambda x_i.e_i:T_i \qquad \Sigma;\Gamma_0,\Delta \vdash e:T \end{array}}{\Sigma;\Gamma_0,\ldots,\Gamma_n \vdash \mathsf{let\ rec}\ F\ \mathsf{in}\ e:T}\ (\text{T:LetRec})$$

$$\frac{}{\cdot; x:T \vdash \mathsf{ref}\ x:T\,\mathrm{ref}}\ (\text{T:Ref}) \qquad \frac{}{\cdot; x:T\,\mathrm{ref} \vdash !x:T}\ (\text{T:DRef})$$

$$\frac{}{\Sigma; x:(\Sigma \to T)\,\mathrm{ref} \vdash !x:T}\ (\text{T:DRefPush}) \qquad \frac{}{\cdot; x_1:T\,\mathrm{ref}, x_2:T \vdash x_1 := x_2:\mathrm{unit}}\ (\text{T:Assign})$$

$$\frac{}{\Sigma;\cdot \vdash \mathsf{fail}:\mathrm{B}}\ (\text{T:Fail}) \qquad \frac{}{\cdot;\cdot \vdash \mathsf{tick}(q):\mathrm{unit}}\ (\text{T:Tick})$$

**Figure 4:** Rules of the stack-based affine type system.

$$\frac{X \in \mathcal{X} \qquad \ell \in \mathrm{dom}(H)}{H \vDash \ell \mapsto \ell : X}\;(\text{V:TVAR}) \qquad \frac{}{H \vDash \mathrm{Null} \mapsto () : \mathrm{unit}}\;(\text{V:UNIT})$$

$$\frac{H(\ell) = \ell' \qquad H \vDash \ell' \mapsto a : T}{H \vDash \ell \mapsto R(a) : T\ \mathrm{ref}}\;(\text{V:REF}) \qquad \frac{H(\ell) = (\lambda x.e, V) \qquad \exists \Gamma : H \vDash V : \Gamma \wedge \cdot;\Gamma \vdash \lambda x.e : \Sigma \to T}{H \vDash \ell \mapsto (\lambda x.e, V) : \Sigma \to T}\;(\text{V:FUN})$$

$$\frac{H(\ell) = (\ell_1, \ldots, \ell_n) \qquad \forall i : H \vDash \ell_i \mapsto a_i : T_i}{H \vDash \ell \mapsto (a_1, \ldots, a_n) : T_1 * \cdots * T_n}\;(\text{V:TUPLE})$$

$$\frac{B = \mu X.\langle \ldots, C : T * X^n, \ldots \rangle \qquad H(\ell) = (C, \ell') \qquad H \vDash \ell' \mapsto (a, b_1, \ldots, b_n) : T * B^n}{H \vDash \ell \mapsto C(a, b_1, \ldots, b_n) : B}\;(\text{V:CONS})$$

**Figure 5:** Coinductively relating heap cells to semantic values.

lambda abstraction introduce a choice that shapes functions types. However, there is often only one possible choice depending on how the abstracted function is used.

As mentioned, the type system is affine and every variable in a context can at most be used once in the typed expression. Multiple uses have to be introduced explicitly using the rule T:SHARE. The only exception is the rule T:LETREC. Here we allow the use of the context $\Delta$ in the body of all defined functions. The reason for this is apparent in the resource aware version: sharing of function types is always possible without any restrictions.

**Well-Formed Environments.** For each simple type $T$ we inductively define a set $[\![T]\!]$ of values of type $T$. Our goal here is not to advance the state of the art in denotational semantics but rather to capture the tree structure of data structures stored on the heap. To this end, we distinguish mainly inductive types (possible inner nodes of the trees) and other types (leaves). For the formulation of type soundness, we also require that function closures are well-formed. We simply interpret polymorphic data with the set of locations *Loc*.

$$\begin{aligned}
[\![X]\!] &= \mathit{Loc} \\
[\![\mathrm{unit}]\!] &= \{()\} \\
[\![T\ \mathrm{ref}]\!] &= \{R(a) \mid a \in [\![T]\!]\} \\
[\![\Sigma \to T]\!] &= \{(\lambda x.e, V) \mid \exists \Gamma : H \vDash V{:}\Gamma \wedge \cdot;\Gamma \vdash \lambda x.e : \Sigma {\to} T\} \\
[\![T_1 * \cdots * T_n]\!] &= [\![T_1]\!] \times \cdots \times [\![T_n]\!] \\
[\![B]\!] &= \mathit{Tr}(B) \text{ if } B = \langle C_1{:}(T_1, n_1), \ldots, C_n{:}(T_k, n_k)\rangle
\end{aligned}$$

Here, $\mathcal{T} = \mathit{Tr}(\langle C_1{:}(T_1, n_1), \ldots, C_n{:}(T_k, n_k)\rangle)$ is the set of trees $\tau$ with node labels $C_1, \ldots, C_k$ which are inductively defined as follows. If $i \in \{1, \ldots, k\}$, $a_i \in [\![T_i]\!]$, and $\tau_j \in \mathcal{T}$ for all $1 \le j \le n_i$ then $C_i(a_i, \tau_1, \ldots, \tau_{n_i}) \in \mathcal{T}$.

If $H$ is a heap, $\ell$ is a location, $A$ is a type, and $a \in [\![A]\!]$ then we write $H \vDash \ell \mapsto a{:}A$ to mean that $\ell$ defines the semantic value $a \in [\![A]\!]$ when pointers are followed in $H$ in the obvious way. The judgment is formally defined in Figure 5. For a heap $H$ there may exist different semantic values $a$ and simple types $A$ such that $H \vDash \ell \mapsto a{:}A$. However, if we fix a simple type $A$ and a heap $H$ then there exists at most one value $a$ such that $H \vDash \ell \mapsto a{:}A$.

**Proposition 1.** *Let $H$ be a heap, $\ell \in Loc$, and let $A$ be a simple type. If $H \vDash \ell \mapsto a{:}A$ and $H \vDash \ell \mapsto a'{:}A$ then $a = a'$.*

We write $H \vDash \ell : A$ to indicate that there exists a, necessarily unique, semantic value $a \in [\![A]\!]$ so that $H \vDash \ell \mapsto a : A$. An environment $V$ and a heap $H$ are *well-formed* with respect to a context $\Gamma$ if $H \vDash V(x) : \Gamma(x)$ holds for every $x \in \mathrm{dom}(\Gamma)$. We then write $H \vDash V : \Gamma$. Similarly, an argument stack $S = \ell_1, \ldots, \ell_n$ is well-formed with respect to a type stack $\Sigma = T_1, \ldots, T_n$ in heap $H$, written $H \vDash S : \Sigma$, if $H \vDash \ell_i : T_i$ for all $1 \leq i \leq n$.

Note that the rules in Figure 5 are interpreted coinductively. The reason is that in the rule V:Fun, the location $\ell$ can be part of the closure environment $V$ if the closure has been created with the rule E:LetRec. The influence of the coinductive definition on the proofs is minimal since all proofs in this article are by induction.

**Type Preservation.** Theorem 1 shows that the evaluation of a well-typed expression in a well-formed environment results in a well-formed environment.

**Theorem 1.** *If* $\Sigma; \Gamma \vdash e : T$, $H \vDash V : \Gamma$, $H \vDash S : \Sigma$, *and* $S, V, H\ {}_M\!\vdash e \Downarrow (\ell, H') \mid (q, q')$ *then* $H' \vDash V : \Gamma$, $H' \vDash S : \Sigma$, *and* $H' \vDash \ell : T$.

Theorem 1 is proved by induction on the evaluation judgement.

# 5 Multivariate Resource Polynomials

In this section we define the set of resource polynomials which is the search space of our automatic resource bound analysis. A resource polynomial $p : [\![T]\!] \to \mathbb{Q}_0^+$ maps a semantic value of some simple type $T$ to a non-negative rational number.

An analysis of typical polynomial computations operating on a list $[a_1, \ldots, a_n]$ shows that they often consist of operations that are executed for every $k$-tuple $(a_{i_1}, \ldots, a_{i_k})$ with $1 \leq i_1 < \cdots < i_k \leq n$. The simplest examples are linear map operations that perform some operation for every $a_i$. Other common examples are sorting algorithms that perform comparisons for every pair $(a_i, a_j)$ with $1 \leq i < j \leq n$ in the worst case.

In this article, we generalize this observation to user-defined tree-like data structures. In lists of different node types with constructors $C_1, C_2$ and $C_3$, a linear computation is for instance often carried out for all $C_1$-nodes, all $C_2$-nodes, or all $C_1$ and $C_3$ nodes. In general, a typical polynomial computation is carried out for all tuples $(a_1, \ldots, a_k)$ such that $a_i$ is a list element with constructor $C_j$ for some $j$ and $a_i$ appears in the list before $a_{i+1}$ for all $i$.

As in previous work, which considered binary trees, we will essentially interpret all tree-like data structures as lists with different nodes by flattening them in pre-order. As a result, our resource polynomials only depend on the number of nodes of a certain kind in the tree but not on structural measures like the height of the tree. To include the height into resource polynomials in a general way, we would need a way to express a maximum (or a choice) in the resource polynomials. We leave this for future research in favor of compositionality and modularity. For compositionality, it is useful that the potential of a data structure is invariant under changes in the structure of the tree.

**Base Polynomials and Indices.** In Figure 6, we define for each simple type $T$ a set $P(T)$ of functions $p : [\![T]\!] \to \mathbb{N}$ that map values of type $T$ to natural numbers. The resource polynomials for type $T$ are then given as non-negative rational linear combinations of these *base polynomials*.

Let $B = \langle C_1 : (T_1, n_1), \ldots, C_m : (T_m, n_m) \rangle$ be an inductive type. Let $\overline{C} = [C_{j_1}, \ldots, C_{j_k}]$ be a list of $B$-constructors and $b \in [\![B]\!]$. We inductively define a set $\tau_B(\overline{C}, b)$ of $k$-tuples as follow: $\tau_B(\overline{C}, b)$ is

$$\frac{}{\lambda a.1 \ \in P(T)} \qquad \frac{\forall i : p_i \in P(T_i)}{\lambda \vec{a}. \prod_{i=1,\ldots,k} p_i(a_i) \ \in P(T_1 * \cdots * T_k)}$$

$$\frac{B = \langle C_1 : (T_1, n_1), \ldots, C_m : (T_m, n_m)\rangle \qquad \overline{C} = [C_{j_1}, \ldots, C_{j_k}] \qquad \forall i : p_i \in P(T_{j_i})}{\lambda b. \sum_{\vec{a} \in \tau_B(\overline{C}, b)} \prod_{i=1,\ldots,k} p_i(a_i) \ \in P(B)}$$

**Figure 6:** Defining the set $P(T)$ of base polynomials for type $T$.

$$\frac{}{\star \ \in \mathcal{I}(T)} \qquad \frac{\forall j : I_j \in \mathcal{I}(T_j)}{(I_1, \ldots, I_k) \ \in \mathcal{I}(T_1 * \cdots * T_k)}$$

$$\frac{B = \langle C_1 : (T_1, n_1), \ldots, C_m : (T_m, n_m)\rangle \qquad \forall i : I_{j_i} \in \mathcal{I}(T_{j_i})}{[\langle I_1, C_{j_1}\rangle, \ldots, \langle I_k, C_{j_k}\rangle] \ \in \mathcal{I}(B)}$$

**Figure 7:** Defining the set $\mathcal{I}(T)$ of indices for type $T$.

the set of $k$-tuples $(a_1, \ldots, a_k)$ such that $C_{j_1}(a_1, \vec{b}_1), \ldots, C_{j_k}(a_k, \vec{b}_k)$ are nodes in the tree $b \in [\![B]\!]$ and $C_{j_1}(a_1, \vec{b}_1) <_{\text{pre}} \cdots <_{\text{pre}} C_{j_k}(a_k, \vec{b}_k)$ for the pre-order $<_{\text{pre}}$ on $b$.

Like in the lambda calculus, we use the notation $\lambda a. e(a)$ for the anonymous function that maps an argument $a$ to the natural number that is defined by the expression $e(a)$. Every set $P(T)$ contains the constant function $\lambda a.1$. In the case of an inductive data type $B$ this constant function arises also for $\overline{C} = []$ (one element sum, empty product).

In Figure 7, we inductively define for each simple type $T$ a set of indices $\mathcal{I}(T)$. For tuple types $T_1 * \cdots * T_k$ we identify the index $\star$ with the index $(\star, \ldots, \star)$. Similarly, we identify the index $\star$ with the index [] for inductive types.

Let $T$ be a base type. For each index $i \in \mathcal{I}(T)$, we define a base polynomial $p_i : [\![T]\!] \to \mathbb{N}$ as follows.

$$\begin{aligned}
p_\star(a) &= 1 \\
p_{(I_1,\ldots,I_k)}(a_1, \ldots, a_k) &= \prod_{j=1,\ldots,k} p_{I_j}(a_j) \\
p_{[\langle I_1,C_1\rangle,\ldots,\langle I_k,C_k\rangle]}(b) &= \sum_{\vec{a} \in \tau_B([C_1,\ldots,C_k],b)} \ \prod_{j=1,\ldots,k} p_{I_j}(a_j)
\end{aligned}$$

**Examples.** To illustrate the definitions, we construct the set of base polynomials for different data types.

- We first consider the inductive type singleton that has only one constructor without arguments.

  $$\text{singleton} = \mu X \langle \mathsf{Nil} : \text{unit}\rangle$$

  Then we have

$$[\![\text{singleton}]\!] = \{\mathsf{Nil}\,(())\} \qquad \text{and} \qquad P(\text{singleton}) = \{\lambda\, a.\, 1, \lambda\, a.\, 0\}\,.$$

To see why, we first examine the set of tuples $\mathcal{T}(\overline{C}) = \tau_{\text{singleton}}(\overline{C}, \mathsf{Nil}\,(()))$ for different list of constructors $\overline{C}$. If $|\overline{C}| > 1$ then $\mathcal{T}(\overline{C}) = \emptyset$ because the tree $\mathsf{Nil}\,(())$ does not contain any tuples of size 2. Thus we have $p_{[\langle I_1,C_1\rangle,\ldots,\langle I_k,C_k\rangle]}(\mathsf{Nil}\,(())) = 0$ in this case (empty sum). The only remaining constructor lists $\overline{C}$ are $[]$ and $[\langle\star, \mathsf{Nil}\rangle]$. As always $p_{[]}(\mathsf{Nil}\,(())) = 1$ (singleton sum). Furthermore $p_{[\langle\star,\mathsf{Nil}\rangle]}(\mathsf{Nil}\,(())) = 1$ because $\tau_{\text{singleton}}([\langle\star, \mathsf{Nil}\rangle], \mathsf{Nil}\,(())) = \{\mathsf{Nil}\,(())\}$ and $P(\text{unit}) = \{\lambda\, a.\, 1\}$.

- Let us now consider the usual sum type

  $$\text{sum}(T_1, T_2) = \mu X\, \langle \mathsf{Left} : T_1, \mathsf{Right} : T_2\rangle\,;.$$

  Then $[\![\text{sum}(T_1, T_2)]\!] = \{\mathsf{Left}\,(a) \mid a \in [\![T_1]\!]\} \cup \{\mathsf{Right}\,(b) \mid b \in [\![T_2]\!]\}$. If we define

  $$\sigma_C(p)(C'(a)) \begin{cases} p(a) & \text{if } C = C' \\ 0 & \text{otherwise} \end{cases}$$

  then $P(\text{sum}(T_1, T_2)) = \{x \mapsto 1, x \mapsto 0\} \cup \{\sigma_{\mathsf{Left}}(p) \mid p \in P(T_1)\} \cup \{\sigma_{\mathsf{Right}}(p) \mid p \in P(T_2)\}$.

- The next example is the list type

  $$\text{list}(T) = \mu X\, \langle \mathsf{Cons} : T * X, \mathsf{Nil} : \text{unit}\rangle\ .$$

  Then $[\![\text{list}(T)]\!] = \{\mathsf{Nil}\,(()), \mathsf{Cons}\,(a_1, \mathsf{Nil}\,(())), \ldots\}$ and we write

  $$[\![\text{list}(T)]\!] = \{[], [a_1], [a_1, a_2], \ldots \mid a_i \in [\![T]\!]\}\,.$$

  We have $\tau_{\text{list}}([\langle\star, \mathsf{Cons}\rangle], [a_1, \ldots, a_n]) = \{a_1, \ldots, a_n\}$ and furthermore

  $$\tau_{\text{list}}([\langle\star, \mathsf{Cons}\rangle, \langle\star, \mathsf{Cons}\rangle], [a_1, \ldots, a_n]) = \{(a_i, a_j) \mid 1 \le i < j \le n\}\,.$$

  More generally, let $\overline{C} = [\langle\star, \mathsf{Cons}\rangle, \ldots, \langle\star, \mathsf{Cons}\rangle]$ or $\overline{C} = [\langle\star, \mathsf{Cons}\rangle, \ldots, \langle\star, \mathsf{Cons}\rangle, \langle\star, \mathsf{Nil}\rangle]$ for lists of length $k$ and $k+1$, respectively. Then $\tau_{\text{list}}(\overline{C}, [a_1, \ldots, a_n]) = \{(a_{i_1}, \ldots, a_{i_k}) \mid 1 \le i_1 < \cdots < i_k \le n\}$. On the other hand, $\tau_{\text{list}}(\overline{D}, [a_1, \ldots, a_n]) = \emptyset$ if $\overline{D} = \langle\star, \mathsf{Nil}\rangle :: \overline{D'}$ for some $\overline{D'} \neq []$. Since $\sum_{\vec{a} \in \tau_{\text{list}}(\overline{C}, [a_1, \ldots, a_n])} 1 = \binom{n}{k}$ and $\lambda\, a.\, 1 \in P(T)$ we have

  $$\{\lambda\, b. \binom{|b|}{n} \mid n \in \mathbb{N}\} \subseteq P(\text{list}(T))\,.$$

- Finally consider a list type with two different $\mathsf{Cons}$-nodes (as in the running example in Figure 1)

  $$\text{list2}(T_1, T_2) = \mu X\, \langle \mathsf{C_1} : T_1 * X, \mathsf{C_2} : T_2 * X, \mathsf{Nil} : \text{unit}\rangle\ .$$

  Then we write (similarly as for list(T))

  $$[\![\text{list2}(T_1, T_2)]\!] = \{[], [a_1], [a_1, a_2], \ldots \mid a_i \in (\{C_1\} \times [\![T_1]\!]) \cup (\{C_2\} \times [\![T_2]\!])\}\,.$$

  Let $b = [b_1, \ldots, b_n]$. We have for example $\tau_{\text{list2}}([\langle\star, \mathsf{C_1}\rangle], b) = \{b_1, \ldots, b_n \mid \forall i \exists a : b_i = (\mathsf{C_1}, a)\}$ and $\tau_{\text{list2}}([\langle\star, \mathsf{C_1}\rangle, \langle\star, \mathsf{C_2}\rangle], [b_1, \ldots, b_n]) = \{(b_i, b_j) \mid \forall i, j \exists a, a' : b_i = (\mathsf{C_1}, a) \wedge b_j = (\mathsf{C_2}, a') \wedge 1 \le i < j \le n\}$.

If $\overline{C} = [\langle\star, C_1\rangle, \ldots, \langle\star, C_1\rangle]$ and $|\overline{C}| = k$ then $\sum_{\vec{a}\in\tau_{\text{list2}}(\overline{C},b)} 1 = \binom{|b|_{C_1}}{k}$ where $|b|_{C_1}$ denotes the number of $C_1$-nodes in the list $b$. Therefore we have

$$\{\lambda\, b. \binom{|b|_{C_1}}{n} \mid n \in \mathbb{N}\} \subseteq P(\text{list2}(T)) \text{ and } \{\lambda\, b. \binom{|b|_{C_2}}{n} \mid n \in \mathbb{N}\} \subseteq P(\text{list2}(T))\,.$$

Now consider the set $\mathcal{D}$ of constructor lists $\overline{D}$ such that $D$ contains exactly $k_1$ elements of the form $\langle\star, C_1\rangle$ and $k_2$ elements of the form $\langle\star, C_2\rangle$. If $S = \bigcup_{\overline{D}\in\mathcal{D}} \tau_{\text{list2}}(\overline{D}, b)$ then $\sum_{\vec{a}\in S} 1 = \binom{|b|_{C_1}}{k_1}\binom{|b|_{C_2}}{k_2}$. This means that such products of binomial coefficients are sums of base polynomials.

- Coinductive types like $\text{stream}(T) = \mu X \langle \mathsf{St} : T * X\rangle$ are not inhabited in our language since we interpret them inductively. A data structure of such a type cannot be created since we allow recursive definitions only for functions.

**Spurious Indices.** The previous examples illustrate that for some inductive data structures, different indices encode the same resource polynomial. For example, for the type list$(T)$ we have $p_{[\langle\star,\mathsf{Nil}\rangle]}(a) = p_{[]}(a) = 1$ for all lists $a$. Additionally, some indices encode a polynomial that is constantly zero. For the type list$(T)$ this is for example the case for $p_{\langle\star,\mathsf{Nil}\rangle::\overline{C}}$ if $|\overline{C}| > 0$. We call such indices *spurious*.

In practice, it is not beneficial to have spurious indices in the index sets since they slow down the analysis without being useful components of bounds. It is straightforward to identify spurious indices from the data type definition. The index $[\langle I_1, C_1\rangle, \ldots, \langle I_k, C_k\rangle]$ is for example spurious if $k > 1$ and the branching number of $C_i$ is 0 for an $i \in \{1, \ldots, k-1\}$.

**Resource Polynomials.** A *resource polynomial* $p : [\![T]\!] \to \mathbb{Q}_0^+$ for a simple type $T$ is a non-negative linear combination of base polynomials, i.e.,

$$p = \sum_{i=1,\ldots,m} q_i \cdot p_i$$

for $m \in \mathbb{N}$, $q_i \in \mathbb{Q}_0^+$ and $p_i \in P(T)$. We write $R(T)$ for the set of resource polynomials for the base type $T$.

**Running Example.** Consider again our running example from Figure 1. For the function *abmap*, we derived the evaluation-step bound $3 + 12L + 12N$. It corresponds to the following resource polynomial. $12p_{(\star,\star,[\langle\star,\mathsf{Acons}\rangle])} + 12p_{(\star,\star,[\langle\star,\mathsf{Bcons}\rangle])} + 3p_{(\star,\star,[])}$.

For the function *asort'*, we derived the evaluation-step bound $13 + 22KN + 13K^2N + 15N$, which corresponds to the resource polynomial

$$26p_{(\star,[\langle[\langle\star,::\rangle,\langle\star,::\rangle],\mathsf{Acons}\rangle])} + 35p_{(\star,[\langle[\langle\star,::\rangle],\mathsf{Acons}\rangle])} + 15p_{(\star,[\langle[],\mathsf{Acons}\rangle])} + 13p_{(\star,[])}\,.$$

**Selecting a Finite Index Set.** Every resource polynomial is defined by a finite number of base polynomials. In an implementation, we also have to fix a finite set of indices to make possible an effective analysis. The selection of the indices to track can be customized for each inductive data type and for every program. However, we currently allow the user only to select a maximal degree of the bounds and then track all indices that correspond to polynomials of the same or a smaller degree.

# 6 Resource-Aware Type System

In this section, we describe the resource-aware type system. Essentially, we annotate the simple type system from Section 4 with resource annotations so that type derivations correspond to proofs of resource bounds.

**Type Annotations.** We use the indices and base polynomials to define type annotations and resource polynomials.

A *type annotation* for a simple type $T$ is defined to be a family

$$Q_T = (q_I)_{I \in \mathcal{I}(T)} \text{ with } q_I \in \mathbb{Q}_0^+$$

We write $\mathcal{Q}(T)$ for the set of type annotations for the type $T$.

An *annotated type* is a pair $(A, Q)$ where $Q$ is a type annotation for the simple type $|A|$ where $A$ and $|A|$ are defined as follows.

$$A ::= \text{unit} \mid X \mid A\,\text{ref} \mid A_1 * \cdots * A_n \mid \langle [A_1,\ldots,A_n] \to B, \Theta \rangle$$
$$\mid \mu X.\langle C_1 : A_1 {*} X^{n_1}, \ldots, C_k : A_k {*} X^{n_k} \rangle$$

We define $|A|$ to be the simple type $T$ that can be obtained from $A$ by removing all type annotations from function types.

A function type $\langle [A_1,\ldots,A_n] \to B, \Theta \rangle$ is annotated with a set

$$\Theta \subseteq \{(Q_A, Q_B) \mid Q_A \in \mathcal{Q}(|A_1 * \cdots * A_n|) \land Q_B \in \mathcal{Q}(|B|)\}.$$

The set $\Theta$ can contain multiple valid resource annotations for arguments and the result of the function.

**Potential of Annotated Types and Contexts.** Let $(A, Q)$ be an annotated type. Let $H$ be a heap and let $v$ be a value with $H \vDash \ell \mapsto a : |A|$. Then the type annotation $Q$ defines the *potential*

$$\Phi_H(v : (A, Q)) = \sum_{I \in \mathcal{I}(T)} q_I \cdot p_I(a)$$

where only finitely many $q_I$'s are non-zero. Usually, we define type annotations $Q$ by only stating the values of the non-zero coefficients $q_I$.

If $a \in [\![ |A| ]\!]$ and $Q \in \mathcal{Q}(|A|)$ is a type annotation then we also write $\Phi(a : (A, Q))$ for $\sum_I q_I \cdot p_I(a)$.

For use in the type system we need to extend the definition of resource polynomials to type contexts and stacks. We treat them like tuple types. Let $\Gamma = x_1{:}A_1, \ldots, x_n{:}A_n$ be a type context and let $\Sigma = B_1, \ldots, B_m$ be a list of types. The index set $\mathcal{I}(\Sigma;\Gamma)$ is defined through

$$\mathcal{I}(\Sigma;\Gamma) = \{(I_1,\ldots,I_m,J_1,\ldots,J_n) \mid I_j \in \mathcal{I}(|B_j|), J_i \in \mathcal{I}(|A_i|\}.$$

A *type annotation* $Q$ for $\Sigma;\Gamma$ is a family

$$Q = (q_I)_{I \in \mathcal{I}(\Sigma;\Gamma)} \text{ with } q_I \in \mathbb{Q}_0^+.$$

We denote a *resource-annotated context* with $\Sigma;\Gamma;Q$. Let $H$ be a heap and $V$ be an environment with $H \vDash V : \Gamma$ where $H \vDash V(x_j) \mapsto a_{x_j} : |\Gamma(x_j)|$. Let furthermore $S = \ell_1, \ldots, \ell_m$ be an argument stack with $H \vDash S : \Sigma$ where $H \vDash \ell_i \mapsto b_i : |B_i|$ for all $i$. The potential of $\Sigma;\Gamma;Q$ with respect to $H$ and $V$ is

$$\Phi_{S,V,H}(\Sigma;\Gamma;Q) = \sum_{\vec{I} \in \mathcal{I}(\Sigma;\Gamma)} q_{\vec{I}} \prod_{j=1}^{m} p_{I_j}(b_j) \prod_{j=m+1}^{m+n} p_{I_j}(a_{x_j})$$

Here, $\vec{I} = (I_1, \cdots, I_{m+n})$. In particular, if $\Sigma = \Gamma = \cdot$ then $\mathcal{I}(\Sigma;\Gamma) = \{()\}$ and $\Phi_{V,H}(\Sigma;\Gamma;q_{()}) = q_{()}$. We sometimes also write $q_\star$ for $q_{()}$.

**Folding of Potential Annotations.** A key notion in the type system is the *folding* for potential annotations that is used to assign potential to typing contexts that result from a pattern match (unfolding) or from the application of a constructor of an inductive data type (folding). Folding of potential annotations is conceptually similar to folding and unfolding of inductive data types in type theory.

Let $B = \mu X.\langle \dots, C : A{*}X^n, \dots\rangle$ be an inductive data type. Let $\Sigma$ be a type stack, $\Gamma, b{:}B$ be a context and let $Q = (q_I)_{I \in \mathcal{I}(\Sigma;\Gamma,y:B)}$ be a context annotation. The *C-unfolding* $\lhd_B^C(Q)$ of $Q$ with respect to $B$ is an annotation $\lhd_B^C(Q) = (q'_I)_{I\in\mathcal{I}(\Sigma;\Gamma')}$ for a context $\Gamma' = \Gamma, x{:}A{*}B^n$ that is defined by

$$q'_{(I,(J,L_1,\dots,L_n))} = \begin{cases} q_{(I,\langle J,C\rangle :: L_1 \cdots L_n)} + q_{(I,L_1\cdots L_n)} & J = 0 \\ q_{(I,\langle J,C\rangle :: L_1\cdots L_n)} & J \neq 0 \end{cases}$$

Here, $L_1 \cdots L_n$ is the concatenation of the lists $L_1, \dots, L_n$.

**Lemma 2.** *Let $B = \mu X.\langle \dots, C : A{*}X^n, \dots\rangle$ be an inductive data type. Let $\Sigma;\Gamma,x{:}B;Q$ be an annotated context, $H \vDash V : \Gamma, x{:}B$, $H \vDash S : \Sigma$, $H(V(x)) = (C,\ell)$, and $V' = V[y \mapsto \ell]$. Then $H \vDash V' : \Gamma, y{:}A{*}B^n$ and $\Phi_{S,V,H}(\Sigma;\Gamma,x{:}B;Q) = \Phi_{S,V',H}(\Sigma;\Gamma,y{:}A{*}B^n;\lhd_B^C(Q))$.*

**Sharing.** Let $\Sigma;\Gamma,x_1{:}A,x_2{:}A;Q$ be an annotated context. The *sharing operation* $\curlyvee Q$ defines an annotation for a context of the form $\Sigma;\Gamma,x{:}A$. It is used when the potential is split between multiple occurrences of a variable. Lemma 3 shows that sharing is a linear operation that does not lead to any loss of potential.

**Lemma 3.** *Let $A$ be a data type. Then there are natural numbers $c_k^{(i,j)}$ for $i,j,k \in \mathcal{I}(|A|)$ such that the following holds. For every context $\Sigma;\Gamma,x_1{:}A,x_2{:}A;Q$ and every $H,V$ with $H \vDash V : \Gamma, x{:}A$ and $H \vDash S : \Sigma$ it holds that $\Phi_{S,V,H}(\Sigma,\Gamma,x{:}A;Q') = \Phi_{S,V',H}(\Sigma;\Gamma,x_1{:}A,x_2{:}A;Q)$ where $V' = V[x_1,x_2 \mapsto V(x)]$ and $q'_{(\ell,k)} = \sum_{i,j\in\mathcal{I}(A)} c_k^{(i,j)} q_{(\ell,i,j)}$.*

The coefficients $c_k^{(i,j)}$ can be computed effectively. We were however not able to derive a closed formula for the coefficients. The proof is similar as in previous work [31]. For a context $\Sigma;\Gamma,x_1{:}A,x_2{:}A;Q$ we define $\curlyvee Q$ to be $Q'$ from Lemma 3.

**Type Judgements.** A resource-aware type judgement has the form

$$\Sigma;\Gamma;Q \,{}_M\!\vdash e : (A,Q')$$

where $\Sigma;\Gamma;Q$ is an annotated context, $M$ is a resource metric, $A$ is an annotated type, and $Q'$ is a type annotation for $|A|$. The intended meaning of this judgment is that if there are more than $\Phi(\Sigma;\Gamma;Q)$ resource units available then this is sufficient to cover the evaluation cost of $e$ under metric $M$. In addition, there are at least $\Phi(v{:}(A,Q'))$ resource units left if $e$ evaluates to a value $v$.

**Notations.** Families that describe type and context annotations are denoted with upper case letters $Q,P,R,\dots$ with optional superscripts. We use the convention that the elements of the families are the corresponding lower case letters with corresponding superscripts, i.e., $Q = (q_I)_{I\in\mathcal{I}}$ and $Q' = (q'_I)_{I\in\mathcal{I}}$.

If $Q,P$ and $R$ are annotations with the same index set $\mathcal{I}$ then we extend operations on $\mathbb{Q}$ pointwise to $Q,P$ and $R$. For example, we write $Q \le P + R$ if $q_I \le p_I + r_I$ for every $I \in \mathcal{I}$. For $K \in \mathbb{Q}$ we write $Q = Q' + K$ to state that $q_\star = q'_\star + K \ge 0$ and $q_I = q'_I$ for $I \neq \star \in \mathcal{I}$. Let $\Gamma = \Gamma_1,\Gamma_2$ be a context, let $I = (I_1,\dots,I_k) \in \mathcal{I}(\Gamma_1)$ and $J = (J_1,\dots,J_\ell) \in \mathcal{I}(\Gamma_2)$ . We write $(I,J)$ for the index $(I_1,\ldots,I_k,J_1,\ldots,J_\ell) \in \mathcal{I}(\Gamma)$. Let $Q$ be an annotation for a context $\Sigma;\Gamma_1,\Gamma_2$. For $J \in \mathcal{I}(\Gamma_2)$ we define the *projection* $\pi^{\Gamma_1}_{(J,J')}(Q)$ of $Q$ to $\Gamma_1$ to be the annotation $Q'$ for $\cdot;\Gamma_1$ with $q'_I = q_{(J,I,J')}$. In the same way, we define the annotations $\pi^{\Sigma}_{J}(Q)$ for $\Sigma;\cdot$ and $\pi^{\Sigma;\Gamma_1}_{J}(Q)$ for $\Sigma;\Gamma_1$.

**Cost Free Types.** We write $\Sigma;\Gamma;Q \,{}_{\mathsf{cf}}\!\vdash e : (A,Q')$ to refer to cost-free type judgments where cf is the cost-free metric with $\mathsf{cf}(K) = 0$ for constants $K$. We use it to assign potential to an extended context in the let rule. More info is available in previous work [32].

**Subtyping.** As usual, subtyping is defined inductively so that types have to be structurally identical. The most interesting rule is the one for function types:

$$\frac{\Theta' \subseteq \Theta \qquad \forall i : A'_i <: A_i \qquad B <: B'}{\langle [A_1,\ldots,A_n] \to B,\Theta\rangle <: \langle [A'_1,\ldots,A'_n] \to B',\Theta'\rangle}\ (\text{S:Fun})$$

A function type is a subtype of another function type if it allows more resource behaviors ($\Theta' \subseteq \Theta$). Result types are treated covariant and arguments are treated contravariant.

Unsurprisingly, our type system does not have principle types. This is to allow the typing of examples such as *rec_scheme* from Section 2. In a principle type, we would have to assume the weakest type for the arguments, that is, function types that are annotated with empty sets of type annotations. This would mean that we cannot use functions in the arguments. However, it is possible to derive a principle type $\langle \Sigma \to B,\Theta\rangle$ for fixed argument types $\Sigma$. Here, we would derive all possible annotations $(Q,Q') \in \Theta$ in the function annotation and all possible annotations $(Q,Q')$ that appear in function annotations of the result type.

If we take the more algorithmic view of previous work [42] then we can express a principle type for a function with a set of constraints that has holes for the constraint sets of the higher-order arguments. It is however unclear what such a type means for a user and we prefer a more declarative view that clearly separates type checking and type inference. An open problem with constraint based principle types is polymorphism.

**Type Rules.** Figure 8 and Figure 9 contain the type rules for annotated types. We integrated the new concepts so that the rules look similar to the rules in previous papers [42, 31, 34].

The rule A:Var can only be applied if the type stack $\Sigma$ is empty. It then simply accounts for the cost $M^{\mathsf{var}}$ and passes the potential that is assigned to the variable by the type context to the result type. If the type stack is not empty then the rule A:VarPush has to be applied. In this case, the variable x must have a function type. We then look up a possible type annotation for the arguments and the result $(P,P') \in \Theta$ in the type context, account for the cost of variable look-up ($M^{\mathsf{var}}$) and behave as specified by $(P,P')$. We do not account for the cost of the "function application" because is cost is handled in the rules A:App and A:AppPush.

The rules A:App and A:AppPush correspond to the simple type rules T:App and T:AppPush. In A:App we assume that the type stack is empty. We account for the cost $M^{\mathsf{app}}_n$ of applying a function to $n$ arguments and look up valid potential annotations $(P,P')$ for the function body in the function annotation $\Theta$. We then require that we have the potential specified by $P$ available and return potential as specified by $P'$. In the rule A:AppPush we account for two applications: We first account for the function application as in the rule A:App. We then assume that the return type is a function type and apply the arguments that are stored on the type stack $\Sigma$ as we do in the rule A:VarPush.

The rules A:AbsPush and A:AbsPop for lambda abstraction correspond the rules T:AbsPush and T:AbsPop. As in the simple type system we can use them to non-deterministically pop

$$\frac{Q = Q' + M^{\mathsf{var}}}{\cdot; x{:}B; Q \;{}_M\!\vdash x : (B, Q')}\ (\text{A:Var}) \qquad \frac{(P,P') \in \Theta \qquad \pi^{\Sigma;\cdot}_{\star}(Q) = P + M^{\mathsf{var}} \qquad P' = Q'}{\Sigma; x{:}\langle \Sigma \to B, \Theta\rangle; Q \;{}_M\!\vdash x : (B, Q')}\ (\text{A:VarPush})$$

$$\frac{\Gamma = x_1{:}A_1,\dots,x_n{:}A_n \qquad (P,P') \in \Theta \qquad \pi^{\Gamma}_{\star}(Q) = P + M^{\mathsf{app}}_n \qquad Q' = P'}{\cdot; x{:}\langle [A_1,\dots,A_n] \to B, \Theta\rangle, \Gamma; Q \;{}_M\!\vdash x\, x_1 \cdots x_n : (B, Q')}\ (\text{A:App})$$

$$\frac{\begin{array}{c}\Gamma = x_1{:}A_1,\dots,x_n{:}A_n\\ (P,P') \in \Theta \qquad (R,R') \in \Theta' \qquad \pi^{\Gamma}_{\star}(Q) = P + M^{\mathsf{app}}_n \qquad \pi^{\Sigma}_{\star}(Q) - q_\star + p'_\star = R \qquad R' = Q'\end{array}}{\Sigma; x{:}\langle [A_1,\dots,A_n] \to \langle \Sigma \to B, \Theta'\rangle, \Theta\rangle, \Gamma; Q \;{}_M\!\vdash x\, x_1 \cdots x_n : (B, Q')}\ (\text{A:AppPush})$$

$$\frac{\Sigma;\Gamma, x{:}A; P \;{}_M\!\vdash e : (B, Q') \qquad Q = R + M^{\mathsf{bind}} \qquad \forall\, I, \vec{J} : r_{(I,\vec{J})} = p_{(\vec{J},I)}}{A{::}\Sigma; \Gamma; Q \;{}_M\!\vdash \lambda x.e : (B, Q')}\ (\text{A:AbsPush})$$

$$\frac{Q = Q' + M^{\mathsf{abs}} \qquad \forall (P,P') \in \Theta : \ \Sigma;\Gamma; R \;{}_M\!\vdash \lambda x.e : (B, P') \wedge r_{(\vec{I},\vec{J})} = \begin{cases} p_{\vec{I}} & \text{if } \vec{J} = \vec{\star} \\ 0 & \text{otherwise}\end{cases}}{\cdot;\Gamma; Q \;{}_M\!\vdash \lambda x.e : (\langle \Sigma \to B, \Theta\rangle, Q')}\ (\text{A:AbsPop})$$

$$\frac{\begin{array}{c} B = \mu X.\langle \dots C : A * X^n \dots\rangle \\ \Sigma;\Gamma, y{:}A*B^n; P \;{}_M\!\vdash e_1 : (A', P') \qquad \Sigma;\Gamma, x{:}B; R \;{}_M\!\vdash e_2 : (A', R') \\ \lhd^{C}_{B}(Q) = P + M^{\mathsf{mat}}_1 \qquad P' = Q' \qquad Q = R + M^{\mathsf{mat}}_2 \qquad R' = Q' \end{array}}{\Sigma;\Gamma, x{:}B; Q \;{}_M\!\vdash \mathsf{match}\ x\ \mathsf{with}\ C\,y \to e_1 \mid e_2 : (A', Q')}\ (\text{A:Mat})$$

$$\frac{B = \mu X.\langle \dots C : A * X^n \dots\rangle \qquad Q = \lhd^{C}_{B}(Q') + M^{\mathsf{cons}}}{\cdot; x{:}A*B^n; Q \;{}_M\!\vdash C\,x : (B, Q')}\ (\text{A:Cons})$$

$$\frac{\Sigma;\Gamma, x_1{:}A, x_2{:}A; P \;{}_M\!\vdash e : (B, Q') \qquad Q = M^{\mathsf{share}} + \curlyvee(P)}{\Sigma;\Gamma, x{:}A; Q \;{}_M\!\vdash \mathsf{share}\ x\ \mathsf{as}\ (x_1, x_2)\ \mathsf{in}\ e : (B, Q')}\ (\text{A:Share})$$

$$\frac{\begin{array}{c} \Sigma;\Gamma_2,\Gamma_1; P \;{}_M\!\vdash e_1 \rightsquigarrow \Sigma;\Gamma_2, x{:}A; P' \\ \Sigma;\Gamma_2, x{:}A; R \;{}_M\!\vdash e_2 : (B, Q') \qquad Q = P + M^{\mathsf{let}}_1 \qquad P' = R + M^{\mathsf{let}}_2 \end{array}}{\Sigma;\Gamma_2,\Gamma_1; Q \;{}_M\!\vdash \mathsf{let}\ x = e_1\ \mathsf{in}\ e_2 : (B, Q')}\ (\text{A:Let})$$

$$\frac{\begin{array}{c} F \triangleq f_1 = \lambda x_1.e_1\ \mathsf{and} \cdots \mathsf{and}\ f_n = \lambda x_n.e_n \\ \Delta = f_1{:}A_1,\dots,f_n{:}A_n \qquad \forall i : \cdot;\Gamma_i,\Delta; P_i \;{}_M\!\vdash \lambda x_i.e_i : (A_i, P'_i) \\ \pi^{\Sigma;\Gamma_0}_{\star}(Q) = \pi^{\Sigma;\Gamma_0}_{\star}(P) + M^{\mathsf{rec}} + n\cdot M^{\mathsf{abs}} \qquad \Sigma;\Gamma_0,\Delta; P \;{}_M\!\vdash e : (B, Q') \end{array}}{\Sigma;\Gamma_0,\dots,\Gamma_n; Q \;{}_M\!\vdash \mathsf{let\ rec}\ F\ \mathsf{in}\ e : (B, Q')}\ (\text{A:LetRec})$$

◇ ◇ ◇

$$\frac{\begin{array}{c} \forall j \in \mathcal{I}(\Sigma;\Delta): \\ j = \vec{\star} \implies \cdot;\Gamma; \pi^{\Gamma}_{j}(Q) \;{}_M\!\vdash e : (A, \pi^{x:A}_{j}(Q')) \qquad j \neq \vec{\star} \implies \cdot;\Gamma; \pi^{\Gamma}_{j}(Q) \;{}_{\mathsf{cf}}\!\vdash e : (A, \pi^{x:A}_{j}(Q')) \end{array}}{\Sigma;\Delta,\Gamma; Q \;{}_M\!\vdash e \rightsquigarrow \Sigma;\Delta, x{:}A; Q'}\ (\text{B:Bind})$$

**Figure 8:** Type rules for annotated types (part 1 of 2).

$$\frac{B = A_1 * \cdots * A_n \qquad Q = Q' + M^{\mathsf{tuple}}}{\cdot; x_1{:}A_1,\ldots,x_n{:}A_n; Q \;{}_M\!\vdash (x_1,\ldots,x_n) : (B, Q')} \text{(A:TUPLE)}$$

$$\frac{A = A_1 * \cdots * A_n \qquad \Sigma;\Gamma, x_1{:}A_1,\ldots,x_n{:}A_n; P \;{}_M\!\vdash e : (B,Q') \qquad Q = P + M^{\mathsf{matT}}}{\Sigma;\Gamma, x{:}A; Q \;{}_M\!\vdash \mathsf{match}\ x\ \mathsf{with}\ (x_1,\ldots,x_n) \to e : (B,Q')} \text{(A:MATT)}$$

$$\frac{q_\star = q'_\star + M^{\mathsf{fail}}}{\Sigma;\cdot;Q \vdash \mathsf{fail} : \text{(B,Q)'}} \text{(A:FAIL)} \qquad \frac{q_0 = q'_\star + M^{\mathsf{tick}}(q)}{\cdot;\cdot;Q \vdash \mathsf{tick}(q) : (\text{unit}, Q')} \text{(A:TICK)}$$

$$\frac{q_\star = q'_\star + M^{\mathsf{ref}}}{\cdot; x : A; Q \;{}_M\!\vdash \mathsf{ref}\ x : (A \text{ref}, Q')} \text{(A:REF)} \qquad \frac{q'_I = \begin{cases} q_\star - M^{\mathsf{dref}} & \text{if } I = \star \\ 0 & \text{otherwise} \end{cases}}{\cdot; x : A \text{ref}; Q \;{}_M\!\vdash !x : (A, Q')} \text{(A:DREF)}$$

$$\frac{(P,P') \in \Theta \qquad \pi_\star^{\Sigma;\cdot}(Q) = P + M^{\mathsf{dref}} \qquad P' = Q'}{\Sigma; x : \langle \Sigma \to A, \Theta \rangle \text{ ref}; Q \;{}_M\!\vdash !x : (A, Q')} \text{(A:DREFPUSH)}$$

$$\frac{q_\star = q'_\star + M^{\mathsf{assign}}}{\cdot; x_1 : A \text{ref}, x_2 : A; Q \;{}_M\!\vdash x_1 := x_2 : (\text{unit}, Q')} \text{(A:ASSIGN)}$$

$$\frac{\Sigma;\Gamma;P \vdash e : (B,P') \qquad Q \geq P + c \qquad Q' \leq P' + c}{\Sigma;\Gamma;Q \vdash e : (B,Q')} \text{(A:WEAK-A)}$$

$$\frac{\Sigma;\Gamma;\pi_\star^\Gamma(Q) \;{}_M\!\vdash e : (B,Q')}{\Sigma;\Gamma, x{:}A; Q \;{}_M\!\vdash e : (B,Q')} \text{(A:WEAK-C)} \qquad \frac{\Sigma;\Gamma;Q \;{}_M\!\vdash e : (B',Q') \qquad B' <: B}{\Sigma;\Gamma;Q \;{}_M\!\vdash e : (B,Q')} \text{(A:SUBTYPE-R)}$$

$$\frac{\Sigma;\Gamma, x{:}A'; Q \;{}_M\!\vdash e : (B,Q') \qquad A <: A'}{\Sigma;\Gamma, x{:}A; Q \;{}_M\!\vdash e : (B,Q')} \text{(A:SUBTYPE-C)}$$

**Figure 9:** Type rules for annotated types (part 2 of 2).

the type stack $\Sigma$. When we do so in the rule A:AbsPop, we create the function annotation $\Theta$ by essentially deriving $\Sigma;\Gamma;P\,{}_M\!\vdash \lambda x.e : (B,P')$ for every $(P,P') \in \Theta$. However, we throw away all potential that depends on the context $\Gamma$ and only use the potential that is assigned the arguments $\Sigma$ (annotation $R$).

The rule A:Cons assigns potential to a new node of an inductive data structure. The additive shift $\lhd_B^C(Q')$ transforms the annotation $Q'$ to an annotation $Q$ for the context $\cdot; x{:}A{*}B^n$. Lemma 2 shows that potential is neither gained nor lost by this operation. The potential $Q$ of the context has to pay for both the potential $Q'$ of the resulting list and the resource cost $M^{\mathsf{cons}}$ of the construction of the new node.

The rule A:Mat shows how to treat pattern matching. The initial potential defined by the annotation $Q$ of the context $\Sigma;\Gamma,x{:}B$ has to be sufficient to pay the costs of the evaluation of $e_1$ or $e_2$ (depending on whether the matched succeeds) and the potential defined by the annotation $Q'$ of the result type. To type the expression $e_2$ we basically just use the annotation $Q$ (after paying for the constant match cost). To type the expression $e_1$ we rely on the additive shift $\lhd_B^C(Q)$ that results in an annotation for the context $\Sigma;\Gamma,y{:}A{*}B^n$. Again there is no loss of potential (see Lemma 2). The equalities relate the potential before and after the evaluation of $e_1$ or $e_2$, to the potential before and after the evaluation of the match operation by incorporating the respective resource cost for the matching.

The rule A:Share uses the sharing operation $\curlyvee P$ to related the potentials defined by $\Sigma;\Gamma,x{:}A;Q$ and $\Sigma;\Gamma,x_1{:}A,x_2{:}A;P$. As with matching, there is no loss of potential (see Lemma 3).

In the rule A:Let the result of the evaluation of an expression $e_1$ is bound to a variable $x$. The problem that arises is that the resulting annotated context $\Sigma;\Gamma_2,x{:}A;R$ features potential functions whose domain consists of data that is referenced by $x$ as well as data that is referenced in the type context $\Gamma_2$. This potential has to be related to data that is referenced by $\Gamma_2$ and the free variables in $e_1$ (i.e., the variables in the type context $\Gamma_1$).

To express the relations between mixed potentials before and after the evaluation of $e_1$, we introduce a new auxiliary binding judgement of the from

$$\Sigma;\Delta,\Gamma;Q\,{}_M\!\vdash e \rightsquigarrow \Sigma;\Delta,x{:}A;Q'$$

in the rule B:Bind. The intuitive meaning of the judgement is the following. Assume that $e$ is evaluated in the context $\Delta,\Gamma$, $\mathrm{FV}(e) \in \mathrm{dom}(\Gamma)$, and that $e$ evaluates to a value that is bound to the variable $x$. Then the initial potential $\Phi(\Sigma;\Delta,\Gamma;Q)$ is larger than the cost of evaluating $e$ in the metric $M$ plus the potential of the resulting context $\Phi(\Sigma,\Delta,x{:}A;Q')$. Lemma 4 formalizes this intuition.

**Lemma 4.** *Let $H \vDash V{:}\Delta,\Gamma$, $H \vDash S : \Sigma$, and $\Sigma;\Delta,\Gamma;Q\,{}_M\!\vdash e \rightsquigarrow \Sigma;\Delta,x{:}A;Q'$.*

1. *If $S,V,H\,{}_M\!\vdash e \Downarrow (\ell,H') \mid (p,p')$ then $\Phi_{S,V,H}(\Delta,\Gamma;Q) \geq p + \Phi_{S,V',H'}(\Delta,x{:}A;Q')$ where $V' = V[x \mapsto \ell]$.*
2. *If $S,V,H\,{}_M\!\vdash e \Downarrow \rho \mid (p,p')$ then $p \leq \Phi_{S,V,H}(\Gamma;Q)$.*

Formally, Lemma 4 is a consequence of the soundness of the type system (Theorem 2). In the inductive proof of Theorem 2, we use a weaker version of Lemma 4 in which the soundness of the type judgements in Lemma 4 is an additional precondition.

The rule A:LetRec is similar the rule T:LetRec for standard type systems. The cost of the creation of the $n$ closures is accounted for by $n{\cdot}M^{\mathsf{abs}}$. It is not difficult to relate this cost to the number of captured variables in the closure but we refrain from doing so in favor of simplicity. The initial potential, defined by $\Sigma;\Gamma_0,\ldots,\Gamma_n;Q$, only flows into the potential $\Sigma;\Gamma_0,\Delta;P$ that is used to pay for the cost of evaluating the expression $e$. The potential annotation $P_i$ and $P_i'$ that are used in the typing of the recursive functions are unconstrained. This is not a bug but uses the fact that $P_i$ can only be used to pay for the cost of the closure creation in the rule A:ABSPOP.

In the rule A:FAIL, we only require that the constant potential $M^{\mathsf{fail}}$ is available. In contrast to the other rules we do not relate the initial potential $Q$ with the resulting potential $Q'$. Intuitively, this is sound because the program is aborted when evaluating the expression *fail*. A consequence of the rule T:FAIL is that we can type the expression $\mathsf{let}\ x = \mathsf{fail}\ \mathsf{in}\ e$ with constant initial potential $M^{\mathsf{fail}}$ regardless of the resource cost of the expression $e$.

In the rule A:TICK we simply require that $M^{\mathsf{tick}}(q)$ constant potential is available.

In the rule A:REF, we only require the availability of the constant potential $M^{\mathsf{ref}}$. We discard the remaining potential that is assigned to $x$ by $Q$. Since references do not carry potential in our system, $q'_\star$ is the only coefficient in $Q'$. In the rule A:ASSIGN we simply pay for the cost of the operation ($M^{\mathsf{assign}}$) and discard the potential that is assigned to the arguments. Since the return value is *()*, $q'_\star$ is the only coefficient in $Q'$. In the rule T:ADREF, we again discard the potential of the arguments and also require that the non-linear coefficients of the annotation of the result are zero. Again, this is because references do not carry potential.

The structural rules A:WEAK-A, A:WEAK-C, A:SUBTYPE-R, and A:SUBTYPE-C apply to every expression. In the implementation, they are integrated into the syntax directed rules to enable automatic type inference. As expected, they are used at the exact same places at which you would use corresponding rules in a standard type system; for instance, when combining branches (weakening and subtyping) of match expressions or when constructing inductive data structures (subtyping). The rule A:WEAK-A relies on the fact that an annotated type remains sound if we add more potential to the context and remove potential from the result. Similarly, the rule A:WEAK-C states that we can add variables with arbritary to the type context. The rules A:SUBTYPE-R and A:SUBTYPE-C are similar to the standard rules of subtyping.

**Soundness.** Our goal is to prove the following soundness statement for type judgements. Intuitively, it says that the initial potential is an upper bound on the watermark resource usage, no matter how long we execute the program.

If $\Sigma;\Gamma;Q\ {}_M\!\vdash e : (A,Q')$ and $S,V,H\ {}_M\!\vdash e \Downarrow w \mid (p,p')$ then $p \le \Phi_{S,V,H}(\Sigma;\Gamma;Q)$.

To prove this statement by induction, we need to prove a stronger statement that takes into account the return value and the annotated type $(A,Q')$ of $e$. Moreover, the previous statement is only true if the values in $S$, $V$ and $H$ respect the types required by $\Sigma$ and $\Gamma$. Therefore, we adapt our definition of well-formed environments to annotated types. We simply replace the rule V:FUN in Figure 5 with the following rule. Of course, $H \vDash V : \Gamma$ refers to the newly defined judgment.

$$\frac{H(\ell) = (\lambda x.e, V) \qquad \exists \Gamma, Q, Q' : H \vDash V : \Gamma \wedge \qquad \cdot;\Gamma;Q\ {}_M\!\vdash \lambda x.e : (\langle \Sigma \to B, \Theta\rangle, Q')}{H \vDash \ell \mapsto (\lambda x.e, V) : \langle \Sigma \to B, \Theta\rangle}\ (\text{V:FUN})$$

In addition to the aforementioned soundness, Theorem 2 states a stronger property for terminating evaluations. If an expression $e$ evaluates to a value $v$ in a well-formed environment then the difference between initial and final potential is an upper bound on the resource usage of the evaluation.

**Theorem 2** (Soundness). *Let* $H \vDash V : \Gamma$, $H \vDash S : \Sigma$, *and* $\Sigma;\Gamma;Q\ {}_M\!\vdash e : (B,Q')$.

1. *If* $S,V,H\ {}_M\!\vdash e \Downarrow (\ell,H') \mid (p,p')$ *then* $p \le \Phi_{S,V,H}(\Sigma;\Gamma;Q)$ *and* $p - p' \le \Phi_{S,V,H}(\Sigma;\Gamma;Q) - \Phi_{H'}(\ell{:}(B,Q'))$ *and* $H \vDash \ell : B$.

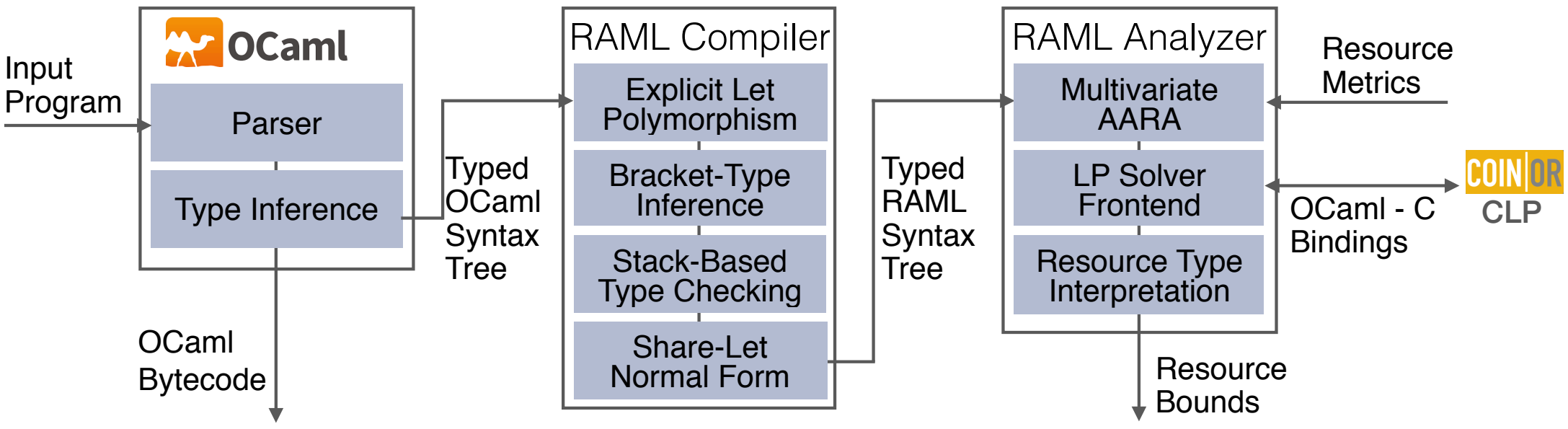

**Figure 10:** Implementation of RAML.

2. *If* $S,V,H\ {}_M\!\vdash e \Downarrow \circ \mid (p,p')$ *then* $p \leq \Phi_{S,V,H}(\Sigma;\Gamma;Q)$.

Theorem 2 is proved by a nested induction on the derivation of the evaluation judgment and the type judgment $\Sigma;\Gamma;Q \vdash e{:}(B,Q')$. The inner induction on the type judgment is needed because of the structural rules. There is one proof for all possible instantiations of the resource constants. An sole induction on the type judgement fails because the size of the type derivation can increase in the case of the function application in which we retrieve a type derivation for the function body from the well-formed judgement as defined by the (updated) rule V:FUN.

The structure of the proof matches the structure of the previous soundness proofs for type systems based on AARA [35, 42, 31, 34]. The induction case of many rules is similar to the induction cases of the corresponding rules for multivariate AARA for first-order programs [31] and linear AARA for higher-order programs [42]. For one thing, additional complexity is introduced by the new resource polynomials for user-defined data types. We designed the system so that this additional complexity is dealt with locally in the rules A:MAT, A:CONS, and A:SHARE. The soundness of these rules follows directly from an application of Lemma 2 and Lemma 3, respectively. As in previous work [35] the well-formed judgement that captures type derivations enables us to treat function abstraction and application in a very similar fashion as in the first-order case [31]. The coinductive definition of the well-formedness judgement does not cause any difficulties. A major novel aspect in the proof is the typed argument stack $S:\Sigma$ that also carries potential. Surprisingly, this typed stack is simply treated like a typed environment $V:\Gamma$ in the proof. It is already incorporated in the shift and share operations (Lemma 2 and Lemma 3).

We deal with the mutable heap by requiring that array elements do not influence the potential of an array. As a result, we can prove the following lemma, which is used in the proof of Theorem 2.

**Lemma 5.** *If* $H \vDash V{:}\Gamma$, $H \vDash S : \Sigma$, $\Sigma;\Gamma;Q\ {}_M\!\vdash e : (B,Q')$ *and* $S,V,H\ {}_M\!\vdash e \Downarrow (\ell,H') \mid (p,p')$ *then* $\Phi_{S,V,H}(\Gamma;Q) = \Phi_{S,V,H'}(\Gamma;Q)$.

# 7 Implementation and Bound Inference

Figure 10 shows an overview of the implementation of RAML. It consists of about 12000 lines of OCaml code, excluding the parts that we reused from Inria's OCaml implementation. The development took around 8 person months. We found it very helpful to develop the implementation and the theory in parallel, and many theoretical ideas have been inspired by implementation challenges.

We reuse the parser and type inference algorithm from OCaml 4.01 to derive a typed OCaml syntax tree from the source program. We then analyze the function applications to introduce bracket function types. To this end, we copy a lambda abstraction for every call site. We still have to implement a unification algorithm since functions, such as *let g = f x*, that are defined by partial application may be used at different call sites. Moreover, we have to deal with functions that are stored in references.

In the next step, we convert the typed OCaml syntax tree into a typed RAML syntax tree. Furthermore, we transform the program into share-let-normal form without changing the resource behavior. For this purpose, each syntactic form has a *free* flag that specifies whether it contributes to the cost of the original program. For example, all share forms that are introduced are *free*. We also insert eta expansions whenever they do not influence resource usage.

After this compilation phase, we perform the actual multivariate AARA on the program in share-let-normal form. Resource metrics can be easily specified by a user. We include a metric for heap cells, evaluation steps, and *ticks*. Ticks allows the user to flexibly specify the resource cost of programs by inserting tick commands *Raml.tick(q)* where $q$ is a (possibly negative) floating-point number.

In principle, the actual bound inference works similarly as in previous AARA systems [35, 30]: First, we fix a maximal degree of the bounds and annotate all types in the derivation of the simple types with variables that correspond to type annotations for resource polynomials of that degree. Second, we generate a set of linear inequalities, which express the relationships between the added annotation variables as specified by the type rules. Third, we solve the inequalities with Coin-Or's fantastic LP solver CLP. A solution of the linear program corresponds to a type derivation in which the variables in the type annotations are instantiated according to the solution. The objective function contains the coefficients of the resource annotation of the program inputs to minimize the initial potential. Modern LP solvers provide support for iterative solving that allows us to express that minimization of higher-degree annotations should take priority.

The type system we use in the implementation significantly differs from the declarative version we describe in this article. For one thing, we have to use algorithmic versions of the type rules in the inference in which the non-syntax-directed rules are integrated into the syntax-directed ones [31]. For another thing, we annotate function types not with a set of type annotations but with a function that returns an annotation for the result type if presented with an annotation of the return type. The annotations here are symbolic and the actual numbers are yet to be determined by the LP solver. Function annotations have the side effect of sending constraints to the LP solver. It would be possible to keep a constraint set for the respective function in memory and to send a copy with fresh variables to the LP solver at every call. However, it is more efficient to lazily trigger the constraint generation from the function body at every call site when the function is provided with a return annotation.

To make the resource analysis more expressive, we also allow resource-polymorphic recursion. This means that we need a type annotation in the recursive call that differs from the annotations in the argument and result types of the function. To infer such types we successively infer type annotations of higher degree. Details can be found in previous work [32].

For the most part, our constraints have the form of a so-called network (or network-flow) problem [53]. LP solvers can handle network problems very efficiently and in practice CLP solves the constraints RAML generates in linear time. Because our problem sizes are large, we can save memory and time by reducing the number of constraints that are generated during typing. A representative example of an optimization is that we try to reuse constraint names instead of producing constraints like $p = q$.

RAML provides two ways of analyzing a program. In *main mode* RAML derives a bound for evaluation cost of the main expression of the program, that is, the last expression in the top-level list of let bindings. In *module mode*, RAML derives a bound for every top-level let binding that has a function type.

Apart from the analysis itself, we also implemented the conversion of the derived resource polynomials into easily-understood polynomial bounds and a pretty printer for RAML types and expressions. Additionally, we implemented an efficient RAML interpreter that we use for debugging and to determine the quality of the bounds.

# 8 Experimental Evaluation

The development of RAML was driven by an ongoing experimental evaluation with OCaml code. Our goal has been to ensure that the derived bounds are precise, that different programming styles are supported, that the analysis is efficient, and that existing code can be analyzed. In an experimental evaluation, we applied our automatic resource bound analysis to 411 functions and 6018 lines of code. The source code of RAML as well as all OCaml files used in the experiments are available online [29]. The website also provides an easy-to-use interactive web interface that can be used to experiment with RAML.

**Analyzed Code and Limitations.** The experiments have been performed with four sets of source code, extracted OCaml code from Coq specifications in CompCert [46], an OCaml tutorial [48], code from the OCaml standard library, and handwritten code. For the handwritten part, we mostly implemented classical textbook algorithms and use cases inspired from real-word applications. The textbook algorithms include algorithms for matrices, graph algorithms, search algorithms, and classic examples from amortized analysis such as functional queues and binary counters. The use cases include energy management in an autonomous mobile device and calling Amazon's Dynamo DB from OCaml (see Section 9).

OCaml is a complex programming language and RAML does not yet support all language features of OCaml. This includes modules, object-oriented features, record types, built-in equality, strings, nested patterns, and calls to native C functions. Therefore it is currently hard to apply RAML directly to existing OCaml code. However, the support of many of these additional features is not a theoretical limitation of the analysis. but rather caused by a lack of resources on the implementation side. If RAML can be applied to existing code then the results are very satisfactory. For instance, we applied RAML to OCaml's standard list library *list.ml*: RAML automatically derives evaluation-step bounds for 47 of the 51 top-level functions. All derived bounds are asymptotically tight. The 4 functions that cannot be bounded by RAML all use a variation of merge sort whose termination (and thus resource usage) depends on an arithmetic shift which is currently unsupported. The file *list.ml* consists of 428 lines of code and the analysis takes 3.2 s on a MacBook Pro.

Also note that our technique extends all previous works on AARA for strict, sequential evaluation. Thus we can handle all examples that have been previously evaluated. The quality of the derived bounds is identical to the previous ones and the efficiency of the analysis is improved.

RAML fails if the resource usage can only be bounded by a measure that depends on a semantic property of the program or a measure that depends on the difference of the sizes of two data structures. Loose bounds are often the result of inter-procedural dependencies. For instance, the worst-case behaviors of two functions $f$ and $g$ might be triggered by different

```
let comp f x g = fun z → f x (g z)

let rec walk f xs =
  match xs with | [] → (fun z →  z)
    | x::ys → match x with | Left _ →
         fun y → comp (walk f) ys (fun z → x::z) y
       | Right l →
         let x' = Right (quicksort f l) in
         fun y → comp (walk f) ys (fun z → x'::z) y

let rev_sort f l = walk f l []
```

RAML output for *rev_sort* (after 0.68*s* run time):

```
10 + 23*K*M + 32*L*M + 20*L*M*Y + 13*L*M*Y^2
where
  M is the num. of ::-nodes of the 2nd comp. of the argument
  L is the fraction of Right-nodes in the ::-nodes of the 2nd component of the arg.
  Y is the maximal number of ::-nodes in the Right-nodes in the ::-nodes of the 2nd
       component of the argument
  K is the fraction of Left-nodes in the ::-nodes of the  2nd component of the arg.
```

**Figure 11:** Modified challenge example from Avanzini et al. [9] and shortened output of the automatic bound analysis performed by RAML for the function *rev_sort*. The derived bound is a tight bound on the number of evaluation steps in the big-step semantics if we do not take into account the cost of the higher-order argument *f*.

inputs. However, the analysis would add up the worst-case behaviors of both functions in a program such as *f(a);g(a)*. Another reason for loose bounds is that a tight bound cannot be represented by a multivariate resource polynomial.

**Example Experiment.** To give an impression of the experiments we performed, Figure 11 contains the output of an analysis of a challenging function in RAML. The code is an adoption of an example that has been recently presented by Avanzini et al. [9] as a function that can not be handled by existing tools. To illustrate the challenges of resource analysis for higher-order programs, Avanzini et al. implemented a (somewhat contrived) reverse function *rev* for lists using higher-order functions. RAML automatically derives a tight linear bound on the number of evaluation steps used by *rev*.

To show more features of our analysis, we modified Avanzini et al.'s *rev* in Figure 11 by adding an additional argument *f* and a pattern match to the definition of the function *walk*. The resulting type of walk is

$$\begin{aligned}(\alpha \to \alpha \to \text{bool}) \;&\to [(\beta * \alpha\,\text{list})\text{ either list}; (\beta * \alpha\,\text{list})\text{ either list}] \\ &\to (\beta * \alpha\,\text{list})\text{ either list}\end{aligned}$$

Like before the modification, *walk* is essentially the *append_reverse* function for lists. However, we assume that the input lists contain nodes of the form *Left a* or *Right b* so that *b* is a list. During the reverse process of the first list in the argument, we sort each list that is contained in a *Right*-node using the standard implementation of quick sort (not given here). RAML derives the tight evaluation-step bound that is shown in Figure 11. Since the comparison function for

| | Metric | #Funs | LOC | Time | #Const | #Lin | #Quad | #Cubic | #Poly | #Fail | Asym. Tight |
|---|---|---|---|---|---|---|---|---|---|---|---|
| | *steps* | 243 | 3218 | 72.10s | 16 | 130 | 60 | 28 | 239 | 4 | 225 |
| | *heap* | 243 | 3218 | 70.36s | 41 | 112 | 60 | 22 | 239 | 4 | 225 |
| | *tick* | 174 | 2144 | 64.68s | 19 | 79 | 53 | 19 | 174 | 0 | 160 |
| *CompCert* | steps | 164 | 2740 | 1300.91s | 32 | 99 | 7 | 0 | 138 | 26 | 137 |

**Table 1:** Overview of experimental results.

*quicksort* (argument *f*) is not available, RAML assumes that it does not consume any resources during the analysis. If *rev_sort* is applied to a concrete argument *f* then the analysis is repeated to derive a bound for this instance.

**CompCert Evaluation.** We also performed an evaluation with the OCaml code that is created by Coq's code extraction mechanism during the compilation of the verified CompCert compilers [46]. We sorted the files topologically from their dependency requirements, and analyzed 13 files from the top. [1] We could not process the files further down the dependency graph because they heavily relied on modules which we do not currently support. Using the evaluation-step metric, we analyzed 164 functions, 2740 LOC in 1300 seconds.

Figure 12 shows example functions from the CompCert code base. As an artifact from the Coq code extraction, CompCert uses two implementations of the reverse function for lists. The function *rev* is a naive quadratic implementation that uses *append* and the function *rev'* is an efficient tail-recursive linear implementation. RAML automatically derives precise evaluation step bounds for both functions. As a result, a Coq user who is inspecting the derived bounds for the extracted OCaml code is likely to spot performance problems resulting from the use of *rev*.

**Summary of Results.** Table 1 contains a compilation of the experimental results. The first 3 rows show the results for OCaml libraries, handwritten code, and the OCaml tutorial [48]. The last column shows the results for CompCert [46]. The column *LOC* contains the total number of lines of OCaml code that has been analyzed with the respective metric. Similarly, the column *Time* contains the total time of all analyses with this metric. The column *#Poly* contains the number of functions for which RAML automatically derived a bound. The columns *#Const*, *#Lin*, *#Quad*, and *#Cubic* show the number of derived bounds that are constant, linear, quadratic, and cubic. Finally, columns *#Fail* and *Asym.Tight* contain the number of examples for which RAML is unable to derive a bound and the number of bounds that are asymptotically tight, respectively. We also experimented with example inputs to determine the precision of the constant factors in the bounds. In general, the bounds are very precise and often match the actual worse-case behavior. However, we did not yet perform a systematic evaluation with automatically generated example inputs.

The reported numbers result from the analysis of 140 non-trivial functions that are (with a few exceptions) recursive and higher order. Appendix A contains a short description of every function that is part of the evaluation, along with its type, the run time of the analysis, and the derived bounds. The functions have been automatically analyzed using the *steps* metric that counts the number of evaluation steps and the *heap* metric that counts the number of allocated heap cells. Moreover, we have used the *tick* metric to add custom cost measures to some of the functions. These measures vary from program to program and include number of function calls,

[1] A list of analyzed files and functions is included in the TR.

```
let rec app l m =
  match l with
    | [] → m
    | a :: l1 → a :: (app l1 m)

let rec rev = function →
  | [] → []
  | x :: l’ → app (rev l’) (x :: [])

let rec rev_append l l’ =
  match l with
    | [] → l’
    | a :: l0 → rev_append l0 (a :: l’)

let rev’ l = rev_append l []
```

RAML output for *rev* (0.1*s* run time; steps metric):

```
3 + 9.5*M + 4.5*M^2
```

RAML output for *rev’* (0.05*s* run time; steps metric):

```
7 + 9*M
```

**Figure 12:** Two implementations of *rev* from CompCert [46]. Both the derived bounds are tight, one is linear and the other is quadratic.

energy consumption, and amount of data sent to the cloud. Details can be found in the source code [29].

There are two main reasons for the difference between the runtime of the analysis per function for the CompCert code (7.9*s*) and the other evaluated code (0.29*s*). First, the CompCert code contains more complex data structures and we thus track more coefficients. Second, there is a larger percentage of functions for which we cannot derive a bound (15.8% vs. 1.6%). As a result, RAML looks for bounds of higher degree before giving up. Both leads to a larger number of constraints to solve for the LP solver. Finally, there are a few outlier functions that cause an unusually long analysis time. This is possibly due to a performance bug.

In general, the analysis is very efficient. RAML is slowing down if the analyzed program contains many variables or functions with many arguments. Another source of complexity is the maximal degree of polynomials in the search space. Depending on the complexity of the program, the analysis becomes unusable when searching for bounds with maximal degree 7 – 9. The efficiency could be improved by combining amortized resource analysis with data-flow analyses and heuristics that predict the parts of the input that cause higher-degree resource usage.

## 9 Case Study: Bounds for DynamoDB Queries

Having integrated the analysis with Inria’s OCaml compiler enables us to analyze and compile real programs. An interesting use case of our resource bound analysis is to infer worst-case bounds on DynamoDB queries. DynamoDB is a commercial NoSQL cloud database service, which is part of Amazon Web Services (AWS). Amazon charges DynamoDB users on a combination of number of queries, transmitted fields, and throughput. Since DynamoDB is a NoSQL service, it is often only possible to retrieve the whole table—which can be expensive for large data sets—or single entries that are identified by a key value. The DynamoDB API is available through the Opam package *aws*. We make the API available to the analysis by using *tick* functions that specify resource usage. Since the query cost for different tables can be different, we provide one function per action and table.

```
let db_query student_id course_id =
  Raml.tick(1.0); Awslib.get_item ...
```

In the following, we describe the analysis of a specific OCaml application that uses a database that contains a large table that stores grades of students for different courses. Our first function computes the average grade of a student for a given list of courses.

```
let avge_grade student_id course_ids =
  let f acc cid =
    let (length,sum) = acc in
    let grade = match db_query student_id cid with
      | Some q → q
      | None → raise (Not_found (student_id,cid))
    in
    (length +. 1.0, sum +. grade)
  in
  let (length,sum) = foldl f (0.0,0.0) course_ids in
  sum /. length
```

In $0.03s$ RAML computes the tight bound $1 \cdot m$ where $m$ is the length of the argument *course_ids*. We omit the standard definitions of functions like *foldl* and *map*. However, they are not built-in into our systems but the bounds are derived form first principles.

Next, we sort a given list of students based on the average grades in a given list of classes using quick sort. As a first approximation we use a comparison function that is based on *average_grade*.

```
let geq sid1 sid2 cour_ids =
  avge_grade sid1 cour_ids >= avge_grade sid2 cour_ids
```

This results in $O(n^2 m)$ database queries where $n$ is the number of students and $m$ is the number of courses. The reason is that there are $O(n^2)$ comparisons during a run of quick sort. Since the resource usage of quick sort depends on the number of courses, we have to make the list of courses an explicit argument and cannot store it in the closure of the comparison function.

```
let rec partition gt acc l =
  match l with
    | [] → let (cs,bs,_) = acc in (cs,bs)
    | x::xs → let (cs,bs,aux) = acc in
      let acc' = if gt x aux then (cs,x::bs,aux)
                 else (x::cs,bs,aux)
      in partition gt acc' xs

let rec qsort gt aux l = match l with | [] → []
  | x::xs →
    let ys,zs = partition (gt x) ([],[],aux) xs in
    append (qsort gt aux ys) (x::(qsort gt aux zs))

let sort_students s_ids c_ids = qsort geq c_ids s_ids
```

In $0.31s$ RAML computes the tight bound $n^2m - nm$ for *sort_students* where $n$ is the length of the argument *s_ids* and $m$ is the length of the argument *c_ids*. The negative factor arises from the translation of the resource polynomials to the standard basis.

Given the alarming cubic bound, we reimplement our sorting function using memoization. To this end we create a table that looks up and stores for each student and course the grade in the DynamoDB. We then replace the function *db_query* with the function *lookup*.

```
let lookup sid cid table =
  let cid_map = find (fun id → id = sid) table in
  find (fun id → id = cid) cid_map
```

For the resulting sorting function, RAML computes the tight bound $nm$ in $0.87s$.

# 10 Related Work

Our work builds on past research on automatic amortized resource analysis (AARA). AARA has been introduced by Hofmann and Jost for a strict first-order functional language with built-in data types [35]. The technique has been applied to higher-order functional programs and user defined types [42], to derive stack-space bounds [16], to programs with lazy evaluation [51, 56], to object-oriented programs [36, 39], and to low-level code by integrating it with separation logic [8]. All the aforementioned amortized-analysis–based systems are limited to linear bounds. Hoffmann et al. [33, 30, 31] presented a multivariate AARA for a first-order language with built-in lists and binary trees. Hofmann and Moser [38] have proposed a generalization of this system in the context of (first-order) term rewrite systems. However, it is unclear how to automate this system. In this article, we introduce the first AARA that is able to automatically derive (multivariate) polynomial bounds that depend on user-defined inductive data structures. Our system is the only one that can derive polynomial bounds for higher-order functions. Even for linear bounds, our analysis is more expressive than existing systems for strict languages [42]. For instance, we can for the first time derive an evaluation-step bound for the curried append function for lists. Moreover, we integrated AARA for the first time with an existing industrial-strength compiler.

Type systems for inferring and verifying resource bounds have been extensively studied. Vasconcelos et al. [55, 54] described an automatic analysis system that is based on sized-types [40] and derives linear bounds for higher-order functional programs. Here we derive polynomial bounds.

Dal Lago et al. [44, 45] introduced linear dependent types to obtain a complete analysis system for the time complexity of the call-by-name and call-by-value lambda calculus. Crary and Weirich [20] presented a type system for specifying and certifying resource consumption. Danielsson [22] developed a library, based on dependent types and manual cost annotations, that can be used for complexity analyses of functional programs. The advantage of our technique is that it is fully automatic.

Classically, cost analyses are often based on deriving and solving recurrence relations. This approach was pioneered by Wegbreit [57] and is actively studied for imperative languages [3, 7, 25, 5]. These works are not concerned with higher-order functions and bounds do not depend on user-defined data structures.

Benzinger [11] has applied Wegbreit's method in an automatic complexity analysis for Nuprl terms. However, complexity information for higher-order functions has to be provided explicitly. Grobauer [26] reported a mechanism to automatically derive cost recurrences from DML programs using dependent types. Danner et al. [24, 23] propose an interesting technique to derive higher-order recurrence relations from higher-order functional programs. Solving the recurrences is not discussed in these works and in contrast to our work they are not able to automatically infer closed-form bounds.

Abstract interpretation based approaches to resource analysis [27, 12, 58, 52, 18] focus on first-order integer programs with loops. Cicek et al. [19] study a type system for incremental complexity.

In an active area of research, techniques from term rewriting are applied to complexity analysis [10, 49, 15]; sometimes in combination with amortized analysis [37]. These techniques are usually restricted to first-order programs and time complexity. Recently, Avanzini et al. [9] proposed a complexity preserving defunctionaliztion to deal with higher-order programs. While the transformation is asymptotically complexity preserving, it is unclear whether this technique can derive bounds with precise constant factors.

Finally, there exists research that studies cost models to formally analyze parallel programs. Blelloch and Greiner [13] pioneered the cost measures work and depth. There are more advanced cost models that take into account caches and IO (see, e.g., Blelloch and Harper [14]), However, these works do not provide machine support for deriving static cost bounds.

## 11 Conclusion

We have presented important first steps towards a practical automatic resource bound analysis system for OCaml. Our three main contributions are (1) a novel automatic resource analysis system that infers multivariate polynomial bounds that depend on size parameters of user-defined data structures, and (2) the first AARA that infers polynomial bounds for higher-order functions, and (3) the integration of automatic amortized resource analysis with the OCaml compiler.

As the title of this article indicates, there are many open problems left on the way to a resource analysis system for OCaml that can be used in every-day development. In the future, we plan to improve the bound analysis for programs with side-effects and exceptions. We will also work on mechanisms that allow user interaction for manually deriving bounds if the automation fails. Furthermore, we will work on taking into account garbage collection and the runtime system when deriving time and space bounds. Finally, we will investigate techniques to link the high-level bounds with hardware and the low-level code that is produced by the compiler. These open questions are certainly challenging but we now have the tools to further push the boundaries of practical quantitative software verification.

# A Experimental Results

## A.1 Analyzed Functions

| Name | Type | LOC | Description |
|---|---|---|---|
| **File: WorkingWithLists.raml (99 Problems in OCaml)** | | | |
| last | *a list -> ’a option* | 8 | returns the last element of list |
| lastTwo | *a list -> (’a * ’a) option* | 10 | returns last two elements of list |
| at | *int -> ’a list -> ’a option* | 7 | outputs element at a particular location |
| natAt | *nat -> ’a list -> ’a option* | 5 | at implemented using natural numbers defined inductively |
| length | *a list -> int* | 5 | returns length of list |
| rev | *a list -> ’a list* | 6 | reverses the list |
| eqList | *a list -> ’a list -> bool* | 13 | checks equality of two lists |
| isPalindrome | *a list -> bool* | 2 | checks if list is a palindrome |
| flatten | *a node list -> ’a list* | 8 | flattens a tree into a list |
| compress | *a list -> ’a list* | 10 | deletes successive duplicates |
| pack | *a list -> ’a list list* | 16 | packs successive duplicates into an inner list |
| encode | *a list -> (int * ’a) list* | 15 | run-length encoding of list |
| decode | *(int * ’a) list -> ’a list* | 17 | decodes a run-length encoding of list |
| duplicate | *a list -> ’a list* | 4 | duplicates each element of list |
| replicate | *a list -> int -> ’a list* | 15 | replicates each element of list n times |
| drop | *a list -> int -> ’a list* | 7 | drops every n-th element |
| split | *a list -> int -> ’a list * ’a list* | 10 | splits list into two lists |
| slice | *a list -> int -> int -> ’a list* | 15 | extracts a slice from list |
| concat | *a list -> ’a list -> ’a list* | 4 | concatenates two lists |
| rotate | *a list -> int -> ’a list* | 8 | rotates a list by n positions |
| removeAt | *a list -> int -> ’a list* | 4 | removes a list at n-th position |
| insertAt | *a list -> int -> ’a -> ’a list* | 4 | inserts an element at n-th position |
| constructList | *int -> int -> ’a list* | 7 | constructs a list from 1st to 2nd element |
| random | *int -> int* | 1 | generates a random number |
| min | *int -> int -> int* | 4 | returns min of two integers |
| randSelect | *a list -> int -> ’a list* | 21 | generates a random permutation of list |
| lottoSelect | *int -> int -> int list* | 1 | composes randSelect with constructList |
| snd | *a * ’b -> ’b* | 4 | returns second element of a product |
| fst | *a * ’b -> ’a* | 4 | returns first element of a product |
| map | *(’a -> ’b) -> ’a list -> ’b list* | 4 | applies f to every element of list |
| insert | *(’a -> ’a -> int) -> ’a -> ’a list -> ’a list* | 4 | sort helper |
| sort | *(’a -> ’a -> int) -> ’a list -> ’a list* | 4 | sorts list according to compare function |
| compare | *int -> int -> int* | 5 | compares two integers |
| lengthSort | *a list list -> ’a list list* | 4 | sorts list of lists according to size of list |
| **File: LogicAndCodes.raml** | | | |
| eval2 | *int -> bool -> int -> bool -> boolExpr* | 7 | table2 helper |
| table2 | *int -> int -> boolExpr -> (bool * bool * bool) list* | 12 | constructs truth table of expression |
| assoc | *int -> (int * ’a) list -> ’a* | 8 | returns element of list corresponding to key |
| eval | *(int * bool) list -> boolExpr -> bool* | 13 | evaluates a boolean expression |
| tableMake | *(int * bool) list -> int list -> boolExpr -> ((int * bool) list * bool) list* | 20 | |
| **File: echelon_form.raml** | | | |
| size | *a list -> int* | 4 | returns size of list |
| getElem | *a list -> int -> ’a* | 6 | returns i-th element of list |
| get2Elems | *a list -> ’b list -> int -> ’a * ’b* | 9 | returns i-th element of 2 lists |
| subtract_row_helper | *float list -> float list -> float -> float list* | 8 | echelon helper |
| subtract_row | *float list -> float list -> int -> float list* | 4 | echelon helper |
| subtract_helper | *float list list -> float list -> int -> float list list* | 4 | echelon helper |
| concat | *a list -> 'a list -> 'a list* | 4 | concatenates two lists |
| tail | *a list -> int -> 'a list* | 6 | returns the list excluding the first i elements |
| hd_helper | *a list -> int -> 'a list -> 'a list* | 6 | echelon helper |
| reverse_helper | *a list -> 'a list -> 'a list* | 4 | echelon helper |
| reverse | *a list -> 'a list* | 6 | reverses a list |
| head | *a list -> int -> 'a list* | 14 | returns the first i elements of list |
| split_helper | *a list -> int -> int -> 'a list * 'a list* | 6 | echelon helper |
| split | *a list -> int -> 'a list * 'a list* | 8 | splits the list at i-th position and returns the two lists |
| subtract | *float list list -> int -> float list list* | 3 | subtract a row |
| echelon_helper | *float list list -> int -> 'a list -> float list list* | 3 | echelon helper |
| echelon_form | *float list list -> float list list* | 102 | takes a matrix with m rows and n columns and reduces it to an upper triangular matrix |
| **File: matrix.raml** | | | |
| check_lists | *a list list -> int -> bool* | 7 | matrix helper |
| check_mat | *int -> int -> 'a list list -> bool* | 4 | matrix helper |
| check_matrix | *int * int * 'a list list -> bool* | 3 | matrix helper |
| construct_matrix | *int -> int -> 'a list list -> int * int * 'a list list* | 5 | matrix helper |
| getElemMatrix | *a * 'b * 'c list list -> int -> int -> 'c* | 4 | returns (i,j)-th element of matrix mat |
| op | *int -> int -> int -> int* | 4 | matrix helper |
| rec_list | *int list -> int list -> int -> int list* | 7 | matrix helper |
| rec_mat | *int list list -> int list list -> int -> int list list* | 7 | matrix helper |
| check_sanity | *int * int * 'a list list -> int * int * 'b list list -> bool* | 5 | matrix helper |
| plus | *int * int * int list list -> int * int * int list list -> int * int * int list list* | 49 | adds two matrices m1 and m2 |
| minus | *int * int * int list list -> int * int * int list list -> int * int * int list list* | 49 | subtracts two matrices m1 and m2 |
| append | *a list -> 'a -> 'a list* | 4 | appends x to the end of list l |
| append_col | *a list list -> 'a list -> 'a list list* | 11 | appends column col to matrix m |
| transpose_helper | *a list list -> 'a list list -> 'a list list* | 4 | matrix helper |
| transpose | *a list list -> 'a list list* | 17 | takes transpose of matrix m |
| prod | *int list -> int list -> int* | 7 | matrix helper |
| prod_mat | *int list -> int list list -> int list list* | 4 | matrix helper |
| mult_slow | *int list list -> int list list -> int list list* | 33 | multiplies matrices m1 and m2 (naive implementation) |
| lineMult | *int -> int list -> int list -> int list* | 7 | matrix helper |
| computeLine | *int list -> int list list -> int list -> int list list* | 7 | matrix helper |
| mat_mult_jan | *int list list -> int list list -> int list list* | 4 | matrix helper |
| check_mult_sanity | *int * int * 'a list list -> int * int * 'b list list -> bool* | 5 | matrix helper |
| mult | *int * int * int list list -> int * int * int list list -> int * int * int list list* | 50 | multiplies matrices m1 and m2 after performing dimensional sanity checks |
| delete | *a list -> int -> 'a list* | 7 | deletes the i-th element of list |
| submat | *a list list -> int -> int -> 'a list list* | 14 | deletes the i-th row and j-th column of matrix m |
| **File: power_radio.raml** | | | |
| sendmsg msg | *int list -> unit* | 13 | sends a list of integers |
| main1_events | *event list -> unit* | 15 | sends sensor data as soon as it is produced |
| main2_events | *event list -> unit* | 15 | stores sensor data in buffer and sends only at specific events |
| main3_events | *event list -> unit* | 15 | same as main2 with debugging mode |
| main4_events | *event list -> unit* | 15 | same as main3 with function application |
| main5_events | *event list -> unit* | 36 | same as main4 but data is sent only after specific time intervals |
| **File: avanzini.raml** | | | |
| partition | *('a -> bool) -> 'a list -> 'a list * 'a list* | 9 | partitions list l into two depending on whether the elements satisfy function f |
| quicksort | *('a -> 'a -> bool) -> 'a list -> 'a list* | 18 | performs quick sort on list l using comp as comparator |
| rev_sort | *a -> 'a -> bool -> ('b * 'a list) either list -> ('b * 'a list) either list* | 17 | |
| **File: append_all.raml** | | | |
| append_all | *a list list -> 'a list* | 8 | concatenates inner lists |
| append_all2 | *a list list list -> 'a list* | 12 | concatenates innermost lists |
| append_all3 | *a list list list list -to 'a list* | 16 | concatenates innermost lists |
| **File: bfs.raml** | | | |
| dfs | *btree -> int -> btree option* | 15 | depth-first search for binary trees |
| bfs | *btree -> int -> btree option* | 34 | breadth-first search for binary trees |
| **File: rev_pairs.raml** | | | |
| pairs | *a list -> ('a * 'a) list* | 12 | generate ordered pairs from a list |
| **File: binary_counter.raml** | | | |
| add_one | *bin list -> bin list* | 15 | increment a binary counter |
| add_many | *bin list -> Rnat.t -> bin list* | 24 | n increments to a binary counter |
| add_list | *bin list -> 'a list -> bin list* | 28 | n increments to a binary counter |
| **File: array_fun.raml** | | | |
| nat_iterate | *(Rnat.t -> unit) -> Rnat.t -> unit* | 3 | calls a function f (i) for 0<=i<n |
| nat_fold | *('a -> Rnat.t -> 'a) -> 'a -> Rnat.t -> 'a* | 5 | fold for natural numbers |
| apply_all | *('a -> 'a) array -> 'a -> 'a* | 12 | successively apply all functions stored in an array |
| **File: calculator.raml** | | | |
| add | *nat -> nat -> nat* | 4 | recursively add two natural numbers |
| sub | *nat -> nat -> nat* | 8 | recursively subtract two nats |
| mult | *nat -> nat -> nat* | 9 | recursively multiply two nats |
| eval_simpl | *expr -> nat* | 26 | a simple evaluater for arithmetic expressions |
| eval | *expr -> nat* | 35 | a evaluater for arithmetic expressions without aux. funs |
| **File: mergesort.raml** | | | |
| split | *a list -> 'a list * 'a list* | 8 | splits a list in the middle |
| merge | *('a -> 'a -> bool) -> 'a list -> 'a list -> 'a list* | 11 | merges two sorted lists |
| mergesort | *('a -> 'a -> bool) -> Rnat.t -> 'a list -> 'a list* | 32 | merge sort |
| mergesort_list | *Rnat.t -> int list list -> int list list* | 45 | merge sort for lists of lists |
| **File: quicksort.raml** | | | |
| partition | *('a -> bool) -> 'a list -> 'a list * 'a list* | 10 | partition for quick sort |
| quicksort | *('a -> 'a -> bool) -> 'a list -> 'a list* | 19 | quick sort |
| quicksort_pairs | *(int * int) list -> (int * int) list* | 28 | quick sort for integer pairs |
| quicksort_list | *int list list -> int list list* | 32 | quick sort for lists of lists |
| **File: square_mult.raml** | | | |
| square_mult | *int -> binary list -> int* | 30 | exponentiation via squaring |
| **File: subsequence.raml** | | | |
| lcs | *int list -> int list -> int* | 45 | longest common subsequence with dynamic programming |
| **File: running.raml** | | | |
| abmap | *('a -> 'b) -> ('c -> 'd) -> ('a * 'c) ablist -> ('b * 'd) ablist* | 5 | map for AB-lists |
| asort | *('a -> 'a -> bool) -> ('a list * 'b) ablist -> ('a list * 'b) ablist* | 29 | sort inner lists in A-nodes |
| asort’ | (’a -> ’a -> bool) -> (’a list * ’b) ablist -> (’a list * ’c) ablist | 29 | sort inner lists in A-nodes; raise exception on B-node |
| btick | (int list * int) ablist -> (int list * int) ablist | 6 | use map to tick 2.5 at every B-node |
| abfoldr | (’a -> ’b -> ’b) -> (’c -> ’b -> ’b) -> ’b -> (’a * ’c) ablist -> ’b | 10 | fold for AB-lists |
| cons_all | (int * int) ablist -> (’a * int) ablist | 18 | two nested folds |
| **File: ocaml_sort.raml** | | | |
| merge | (’a -> ’a -> bool) -> ’a list -> ’a list -> ’a list | 9 | merge two lists |
| list | (’a -> ’a -> bool) -> ’a list -> ’a list | 31 | interesting variant of merge sort |
| **File: ocaml_list.raml** | | | |
| length | a list -> int | 5 | length of a list |
| cons | a -> ’a list -> ’a list | 1 | list cons |
| hd | a list -> ’a | 3 | head of a list |
| tl | a list -> ’a list | 3 | tail of a list |
| nth | a list -> int -> ’a | 7 | nth element of a list |
| append | a list -> ’a list -> ’a list | 4 | list append |
| rev_append | a list -> ’a list -> ’a list | 4 | |
| rev | a list -> ’a list | 6 | reverses a list |
| flatten | a list list -> ’a list | 7 | flattens a list |
| concat | a list list -> ’a list | 8 | flattens a list |
| map | (’a -> ’b) -> ’a list -> ’b list | 3 | list map |
| mapi | (int -> ’a -> ’b) -> ’a list -> ’b list | 5 | list map with index |
| rev_map | (’a -> ’b) -> ’a list -> ’b list | 6 | reverse and map |
| iter | (’a -> ’b) -> ’a list -> unit | 3 | iterate over a list |
| iteri | (int -> ’a -> ’b) -> ’a list -> unit | 5 | iterate with index |
| fold_left | (’a -> ’b -> ’a) -> ’a -> ’b list -> ’a | 4 | list fold |
| fold_right | (’a -> ’b -> ’b) -> ’a list -> ’b -> ’b | 4 | list fold |
| map2 | (’a -> ’b -> ’c) -> ’a list -> ’b list -> ’c list | 12 | list map with two lists |
| rev_map2 | (’a -> ’b -> ’c) -> ’a list -> ’b list -> ’c list | 15 | reverse and map with two lists |
| iter2 | (’a -> ’b -> ’c) -> ’a list -> ’b list -> unit | 12 | iterate over two lists |
| fold_left2 | (’a -> ’b -> ’c -> ’a) -> ’a -> ’b list -> ’c list -> ’a | 12 | left fold with two lists |
| fold_right2 | (’a -> ’b -> ’c -> ’c) -> ’a list -> ’b list -> ’c -> ’c | 12 | right fold with two lists |
| for_all | (’a -> bool) -> ’a list -> bool | 3 | check condition for all list elems |
| exists | (’a -> bool) -> ’a list -> bool | 3 | check condition for one list elem |
| for_all2 | (’a -> ’b -> bool) -> ’a list -> ’b list -> bool | 12 | as for_all but for two lists |
| exists2 | (’a -> ’b -> bool) -> ’a list -> ’b list -> bool | 12 | as exists but for two lists |
| mem | a -> ’b list -> bool | 3 | checks if elem is in list |
| memq | a -> ’b list -> bool | 3 | same as mem but uses equality |
| assoc | a -> (’b * ’c) list -> ’c | 5 | lookup for key-value pairs |
| assq | a -> (’b * ’c) list -> ’c | 5 | as assoc but uses equality |
| mem_assoc | a -> (’b * ’c) list -> bool | 5 | like mem but for pairs |
| mem_assq | a -> (’b * ’c) list -> bool | 5 | like memq but for pairs |
| remove_assoc | a -> (’b * ’c) list -> (’b * ’c) list | 5 | filter varient using compare |
| remove_assq | a -> (’b * ’c) list -> (’b * ’c) list | 5 | filter varient using equality |
| find | (’a -> bool) -> ’a list -> ’a | 3 | list find |
| find_all | (’a -> bool) -> ’a list -> ’a list | 7 | list find that returns all matches |
| filter | (’a -> bool) -> ’a list -> ’a list | 8 | list filter |
| partition | (’a -> bool) -> ’a list -> ’a list * ’a list | 5 | list partition |
| split | (’a * ’b) list -> ’a list * ’b list | 5 | split a list of pairs |
| combine | a list -> ’b list -> (’a * ’b) list | 12 | zip two lists |
| merge | (’a -> ’a -> int) -> ’a list -> ’a list -> ’a list | 10 | merge for merge sort |
| chop | int -> ’a list -> ’a list | 6 | take the first an elements |
| stable_sort | (’a -> ’a -> int) -> ’a list -> ’a list | 106 | merge sort |
| sort | (’a -> ’a -> int) -> ’a list -> ’a list | 107 | merge sort |
| fast_sort | (’a -> ’a -> int) -> ’a list -> ’a list | 107 | merge sort |
| **File: aws.raml** | | | |
| average_grade | *int -> int list -> float* | 17 | avarage by looking up all grades in DynanmoDB |
| greater_eq | *int -> int -> int list -> bool * int list* | 19 | compare students by looking up grades at DynamoDB |
| sort_students | *int list -> int list -> int list* | 52 | sort students based on avarage grade using greater_eq |
| make_table | *int list -> int list -> (int * (int * float) list) list* | 22 | look up grades in DynamoDB and memoize all the grades in a tables |
| find | *('a -> bool) -> ('a * 'b) list -> 'b* | 9 | find a value by looking up the key |
| lookup | *int -> int -> (int * (int * 'a) list) list -> 'a* | 12 | look up a grade in a table |
| average_grade' | *int -> int list -> (int * (int * float) list) list -> float * (int * (int * float) list) list* | 24 | avarage grade using a look-up table |
| greater_eq' | *int list -> int -> int -> (int * (int * float) list) list] -> bool * (int * (int * float) list) list* | 28 | greater_eq using a look-up table |
| sort_students_efficient | *int list -> int list -> int list* | 76 | sorting using a look-up table |
| **File: PROTOTYPE** | | | |
| **File: appendAll.raml** | | | |
| appendAll | *a list list -> 'a list* | 9 | collapses all elements of a 2D matrix into a list |
| appendAll2 | *a list list list -> 'a list* | 14 | collapses all elements of a 3D matrix into a list |
| appendAll3 | *a list list list list -> 'a list* | 18 | collapses all elements of a 4D matrix into a list |
| **File: duplicates.raml** | | | |
| eq | *int list -> int list -> bool* | 15 | checks if two lists are equal |
| remove | *int list -> int list list -> int list list* | 5 | duplicates helper |
| nub | *int list list -> int list list* | 27 | removes duplicate lists from a list of lists |
| **File: dyade.raml** | | | |
| multList | *int -> int list -> int list* | 4 | multiplies all elements of a list with a constant |
| dyade | *int list -> int list -> int list list* | 8 | multiplies all elements of two lists to form a 2D matrix |
| **File: eratosthenes.raml** | | | |
| filter | *int -> int list -> int list* | 8 | deletes all elements in list divisible by the first argument |
| eratos | *int list -> int list* | 3 | runs the sieve of Eratosthenes algorithm on the list |
| **File: bitvectors.raml** | | | |
| bitToInt' | *int list -> int -> int* | 4 | bitvector helper |
| bitToInt | *int list -> int* | 6 | converts bit vector to integer |
| sum | *int -> int -> int -> int * int* | 7 | bitvector helper |
| add' | *int list -> int list -> int -> int list* | 9 | bitvector helper |
| add | *int list -> int list -> int list* | 26 | adds two bitvectors |
| diff | *int -> int -> int -> int * int* | 2 | bitvector helper |
| sub' | *int list -> int list -> int -> int list * int* | 11 | |
| sub | *int list -> int list -> int list* | 17 | subtracts two bitvectors |
| mult | *int list -> int list -> int list* | 8 | multiplies two bitvectors |
| compare | *int list -> int list -> int* | 14 | bitvector helper |
| leq | *int list -> int list -> bool* | 16 | compares two bitvectors |
| **File: flatten.raml** | | | |
| flatten | *tree -> int list* | 8 | collapses tree into a list |
| insert | *int -> int list -> int list* | 7 | inserts element in a sorted list |
| insertionsort | *int list -> int list* | 12 | performs insertion sort on list |
| flattensort | *tree -> int list* | 20 | performs insertion sort after flattening the tree |
| **File: listsort.raml** | | | |
| isortlist | *int list list -> int list list* | 21 | performs an insertion sort on list of lists, where lists are compared lexicographically |
| **File: longestCommonSubsequence.raml** | | | |
| firstline | *a list -> int list* | 4 | returns first line of zeros |
| right | *int list -> int* | 4 | lcs helper |
| max | *int -> int -> int* | 2 | computes max of two integers |
| newline | *int -> int list -> int list* | 15 | computes new line recursively |
| lcstable | *int list -> int list -> int list list* | 8 | computes length of table |
| lcs | *int list -> int list -> int* | 47 | computes longest common subsequence of two lists |
| **File: mergesort.raml** | | | |
| msplit | *a list -> 'a list * 'a list* | 9 | splits list into two |
| merge | *int list -> int list -> int list* | 10 | merges two sorted lists |
| mergesortBuggy | *int list -> int list* | 15 | buggy version of mergesort |
| mergesort | *int list -> int list* | 19 | correct version of mergesort |
| **File: minsort.raml** | | | |
| findMin | *int list -> int list* | 10 | helper for selection sort |
| minSort | *int list -> int list* | 14 | performs selection sort on list |
| **File: queue.raml** | | | |
| empty | *a -> 'b list * 'c list* | 2 | returns an empty list |
| enqueue | *a -> 'a list * 'b -> 'a list * 'b* | 3 | enqueues element into list |
| enqueues | *a list -> 'a list * 'b -> 'a list * 'b* | 4 | enqueues a list of trees into a queue of trees |
| copyover | *a list * 'a list -> 'b list * 'a list* | 5 | dequeue helper |
| dequeue | *a list * 'a list -> ('a list * 'a list) * 'a list* | 10 | dequeues element from queue |
| children | *a * 'b * 'b list * 'b list -> ('a * 'b) * ('b * 'b * 'b list * 'b list) list* | 17 | constructs a node of tree |
| breadth | *('a * 'a * 'a list * 'a list) list * ('a * 'a * 'a list * 'a list) list -> ('a * 'a) list* | 7 | bfs helper |
| startBreadth | *a list -> ('a * 'a) list* | 4 | performs breadth first search on tree |
| depth | *a * 'a * 'a list * 'a list -> ('a * 'a) list* | 13 | dfs helper |
| startDepth | *a list -> ('a * 'a) list* | 4 | performs depth first search on tree |
| **File: quicksort_mutual.raml** | | | |
| part | *int -> int list -> int list -> int list -> int list* | 6 | partitions the list for performing quick sort |
| quicksortMutual | *int list -> int list* | 10 | performs a mutually recursive implementation of quicksort as presented by Hongwei Xi |
| **File: rationalPotential.raml** | | | |
| zip3 | *a list -> 'b list -> 'c list -> ('a * 'b * 'c) list* | 10 | zips 3 lists together |
| group3 | *a list -> ('a * 'a * 'a) list* | 10 | groups list into list of triples |
| **File: sizechange.raml** | | | |
| r1 | *a list -> 'a list -> 'a list* | 4 | reverse helper |
| rev | *a list -> 'a list* | 2 | reverses a list |
| f | *a list -> 'a list -> 'a list* | 6 | mutual recursion with g |
| g | *a list -> 'a list -> 'a list -> 'a list* | 4 | mutual recursion with f |
| f2 | *a list -> 'a list -> 'a list* | 7 | function in the size change paper |
| last | *a list -> 'a list* | 8 | re-implementation of f2 |
| f2' | *a list -> 'a list -> 'a list* | 7 | f2 reimplemented |
| g3 | *a list -> 'a list -> 'a list* | 4 | helper |
| f3 | *a list -> 'a list -> 'a list* | 4 | late starting descending parameters |
| **File: splitandsort.raml** | | | |
| insert | *a * int -> ('a list * int) list -> ('a list * int) list* | 12 | split helper |
| split | *('a * int) list -> ('a list * int) list* | 4 | splits values according to keys |
| splitqs | *int * int list -> int list * int list* | 11 | quicksort helper |
| quicksort | *int list -> int list* | 7 | performs quicksort on list |
| sortAll | *(int list * 'a) list -> (int list * 'a) list* | 6 | sorts all value lists with minSort |
| splitAndSort | *(int * int) list -> (int list * int) list* | 3 | splits list according to keys, then sorts the inner lists |
| **File: subtrees.raml** | | | |
| subtrees | *tree -> tree list* | 16 | generates a list of all subtrees of a tree |
| **File: tuples.raml** | | | |
| attach | *a -> 'b list -> ('a * 'b) list* | 4 | attaches the first argument to every element of the list |
| pairs | *a list -> ('a * 'a) list* | 4 | generates all distinct pairs in list |
| pairsAux | *a list -> ('a * 'a) list -> ('a * 'a) list* | 4 | helper for pairsSlow |
| pairsSlow | *a list -> ('a * 'a) list* | 4 | slow implementation of pairs |
| triples | *a list -> ('a * 'a * 'a) list* | 4 | generates all distinct triples in list |
| quadruples | *a list -> ('a * 'a * 'a * 'a) list* | 4 | generates all distinct quadruples in list |
| **File: array_dijkstra.raml** | | | |
| makeGraph | *Rnat.t -> (Rnat.t * Rnat.t * int) list -> int array array* | 35 | creates an array based weighted graph from a list |
| dijkstra | *int array array -> Rnat.t -> int array* | 64 | Dijkstra's shortest-path algorithm |
| **File: CompCert** | | | |
| **File: String0.ml** | | | |
| string_dec | *int list -> int list -> bool* | 10 | |
| prefix | *int list -> int list -> bool* | 7 | |
| **File: Tuples.ml** | | | |
| uncurry | *a list -> 'b -> 'c -> 'd* | 4 | |
| **File: Specif.ml** | | | |
| projT1 | *('a * 'b) sigT -> 'a* | 3 | |
| projT2 | *('a * 'b) sigT -> 'b* | 3 | |
| value | *a -> 'a option* | 3 | |
| **File: EquivDec.ml** | | | |
| equiv_dec | *a -> 'a* | 3 | |
| **File: Datatypes.ml** | | | |
| implb | *bool -> bool -> bool* | 2 | |
| xorb | *bool -> bool -> bool* | 2 | |
| negb | *bool -> bool* | 2 | |
| fst | *a * 'b -> 'a* | 2 | |
| snd | *a * 'b -> 'b* | 2 | |
| length | *a list -> nat* | 4 | |
| app | *a list -> 'a list -> 'a list* | 4 | |
| coq_CompOpp | *comparison -> comparison* | 3 | |
| coq_CompareSpec2Type | *comparison -> coq_CompareSpecT* | 3 | |
| coq_CompSpec2Type | *a -> 'b -> comparison -> coq_CompareSpecT* | 2 | |
| **File: Bool.ml** | | | |
| bool_dec | *bool -> bool -> bool* | 2 | |
| eqb | *bool -> bool -> bool* | 2 | |
| iff_reflect | *bool -> reflect* | 2 | |
| **File: Ring.ml** | | | |
| bool_eq | *bool -> bool -> bool* | 2 | |
| **File: Peano.ml** | | | |
| plus | *nat -> nat -> nat* | 4 | |
| max | *nat -> nat -> nat* | 7 | |
| min | *nat -> nat -> nat* | 7 | |
| nat_iter | *nat -> ('a -> 'a) -> 'a -> 'a* | 4 | |
| **File: List0.ml** | | | |
| hd | *a -> 'a list -> 'a* | 3 | |
| tl | *a list -> 'a* | 3 | |
| in_dec | *('a -> 'b -> bool) -> 'b -> 'a list -> bool* | 3 | |
| nth_error | *a list -> nat -> 'a option* | 7 | |
| remove | *('a -> 'b -> bool) -> 'a -> 'b list -> 'b list* | 4 | |
| rev | *a list -> 'a list* | 3 | |
| rev_append | *a list -> 'a list -> 'a list* | 4 | |
| rev' | *a list -> 'a list* | 6 | |

| | | |
|---|---|---|
| list_eq_dec | *(('a -> 'b) -> bool) -> 'a list -> 'b list -> bool* | 10 |
| map | *('a -> 'b) -> 'a list -> 'b list* | 3 |
| fold_left | *(('a -> 'b) -> 'a) -> 'b list -> 'a -> 'a* | 4 |
| fold_right | *(('a -> 'b) -> 'b) -> 'b -> 'a list -> 'b* | 3 |
| existsb | *('a -> bool) -> 'a list -> bool* | 3 |
| forallb | *('a -> bool) -> 'a list -> bool* | 3 |
| filter | *('a list -> bool) -> 'a list -> 'a list* | 2 |
| **File: EqNat.ml** | | |
| beq_nat | *nat -> nat -> bool* | 10 |
| **File: Compare_dec.ml** | | |
| le_lt_dec | *nat -> nat -> bool* | 7 |
| le_gt_dec | *nat -> nat -> bool* | 7 |
| nat_compare | *nat -> nat -> comparison* | 10 |
| **File: BinPosDef.ml** | | |
| succ | *positive -> positive* | 5 |
| add | *positive -> positive -> positive* | 17 |
| add_carry | *positive -> positive -> positive* | 17 |
| pred_double | *positive -> positive* | 5 |
| pred | *positive -> positive* | 10 |
| pred_N | *positive -> coq_N* | 10 |
| mask_rect | *a -> (positive -> 'a) -> 'a -> mask -> 'a* | 5 |
| mask_rec | *a -> (positive -> 'a) -> 'a -> mask -> 'a* | 5 |
| succ_double_mask | *mask -> mask* | 5 |
| double_mask | *mask -> mask* | 5 |
| double_pred_mask | *positive -> mask* | 10 |
| pred_mask | *mask -> mask* | 19 |
| sub_mask | *positive -> positive -> mask* | 32 |
| sub_mask_carry | *positive -> positive -> mask* | 38 |
| sub | *positive -> positive -> positive* | 43 |
| mul | *positive -> positive -> positive* | 22 |
| iter | *positive -> ('a -> 'a) -> 'a -> 'a* | 5 |
| pow | *positive -> positive -> positive* | 29 |
| square | *positive -> positive* | 22 |
| div2 | *positive -> positive* | 5 |
| div2_up | *positive -> positive* | 10 |
| size_nat | *positive -> nat* | 5 |
| size | *positive -> positive* | 10 |
| compare_cont | *positive -> positive -> comparison -> comparison* | 17 |
| compare | *positive -> positive -> comparison* | 19 |
| min | *positive -> positive -> positive* | 24 |
| max | *positive -> positive -> positive* | 24 |
| eqb | *positive -> positive -> bool* | 17 |
| leb | *positive -> positive -> bool* | 24 |
| ltb | *positive -> positive -> bool* | 24 |
| sqrtrem_step | *(positive -> 'a) -> ('a -> positive) -> positive * mask -> positive * mask* | 67 |
| sqrtrem | *positive -> positive * mask* | 85 |
| sqrt | *positive -> positive* | 87 |
| gcdn | *nat -> positive -> positive -> positive* | 39 |
| gcd | *positive -> positive -> positive* | 50 |
| ggcdn | *nat -> positive -> positive -> positive * (positive * positive)* | 66 |
| ggcd | *positive -> positive -> positive * (positive * positive)* | 68 |
| coq_Nsucc_double | *coq_N -> coq_N* | 4 |
| coq_Ndouble | *coq_N -> coq_N* | 4 |
| coq_lor | *positive -> positive -> positive* | 17 |
| coq_land | *positive -> positive -> coq_N* | 25 |
| ldiff | *positive -> positive -> coq_N* | 25 |
| coq_lxor | *positive -> positive -> coq_N* | 25 |
| shiftl_nat | *positive -> nat -> positive* | 6 |
| shiftr_nat | *positive -> nat -> positive* | 7 |
| shiftl | *positive -> coq_N -> positive* | 9 |
| shiftr | *positive -> coq_N -> positive* | 9 |
| testbit_nat | *positive -> nat -> bool* | 14 |
| testbit | *positive -> coq_N -> bool* | 14 |
| iter_op | *((’a -> ’a) -> ’a) -> positive -> ’a -> ’a* | 5 |
| to_nat | *positive -> nat* | 14 |
| of_nat | *nat -> positive* | 7 |
| of_succ_nat | *nat -> positive* | 4 |
| digits2_pos | *positive -> positive* | 10 |
| coq_Zdigits2 | *coq_Z -> coq_Z* | 15 |
| **File: BinNat.ml** | | |
| succ_double_n | *coq_N -> coq_N* | 4 |
| double_n | *coq_N -> coq_N* | 4 |
| succ_n | *coq_N -> coq_N* | 4 |
| pred_n | *coq_N -> coq_N* | 4 |
| succ_pos_n | *coq_N -> positive* | 4 |
| add_n | *coq_N -> coq_N -> coq_N* | 7 |
| sub_n | *coq_N -> coq_N -> coq_N* | 11 |
| mul_n | *coq_N -> coq_N -> coq_N* | 29 |
| compare_n | *coq_N -> coq_N -> comparison* | 10 |
| eqb_n | *coq_N -> coq_N -> bool* | 10 |
| leb_n | *coq_N -> coq_N -> bool* | 5 |
| ltb_n | *coq_N -> coq_N -> bool* | 5 |
| min_n | *coq_N -> coq_N -> coq_N* | 5 |
| max_n | *coq_N -> coq_N -> coq_N* | 5 |
| div2_n | *coq_N -> coq_N* | 8 |
| even_n | *coq_N -> bool* | 8 |
| odd_n | *coq_N -> bool* | 10 |
| pow_n | *coq_N -> coq_N -> coq_N* | 7 |
| square_n | *coq_N -> coq_N* | 26 |
| log2_n | *coq_N -> coq_N* | 18 |
| size_n | *coq_N -> coq_N* | 14 |
| size_nat_n | *coq_N -> nat* | 9 |
| pos_div_eucl | *positive -> coq_N -> coq_N * coq_N* | 37 |
| div_eucl | *coq_N -> coq_N -> coq_N * coq_N* | 44 |
| div | *coq_N -> coq_N -> coq_N* | 46 |
| modulo | *coq_N -> coq_N -> coq_N* | 46 |
| gcd_n | *coq_N -> coq_N -> coq_N* | 57 |
| ggcd_n | *coq_N -> coq_N -> coq_N -> coq_N * coq_N* | 75 |
| coq_lor_n | *coq_N -> coq_N -> coq_N* | 24 |
| coq_land_n | *coq_N -> coq_N -> coq_N* | 32 |
| ldiff_n | *coq_N -> coq_N -> coq_N* | 32 |
| coq_lxor_n | *coq_N -> coq_N -> coq_N* | 32 |
| shiftl_nat | *coq_N -> nat -> coq_N* | 10 |
| shiftr_nat | *coq_N -> nat -> coq_N* | 14 |
| shiftl_n | *coq_N -> coq_N -> coq_N* | 13 |
| shiftr_n | *coq_N -> coq_N -> coq_N* | 13 |
| testbit_nat_n | *coq_N -> nat -> bool* | 18 |
| testbit_n | *coq_N -> coq_N -> bool* | 18 |
| to_nat_n | *coq_N -> nat* | 18 |
| of_nat_n | *nat -> coq_N* | 8 |
| iter_n | *coq_N -> (’a -> ’a) -> ’a -> ’a* | 9 |
| discr | *coq_N -> positive option* | 4 |
| binary_rect | *a -> (coq_N -> ’a -> ’a) -> (coq_N -> ’a -> ’a) -> ’a* | 12 |
| binary_rec | *a -> (coq_N -> 'a -> 'a) -> (coq_N -> 'a -> 'a) -> 'a* | 14 |
| leb_spec0 | *coq_N -> coq_N -> reflect* | 9 |
| ltb_spec0 | *coq_N -> coq_N -> reflect* | 9 |
| log2_up | *coq_N -> coq_N* | 41 |
| lcm | *coq_N -> coq_N -> coq_N* | 134 |
| eqb_spec | *coq_N -> coq_N -> reflect* | 16 |
| b2n | *bool -> coq_N* | 2 |
| setbit | *coq_N -> coq_N -> coq_N* | 39 |
| clearbit | *coq_N -> coq_N -> coq_N* | 47 |
| ones | *coq_N -> coq_N* | 19 |
| lnot | *coq_N -> coq_N -> coq_N* | 53 |
| max_case_strong | *coq_N -> coq_N -> ((coq_N -> coq_N -> unit -> 'a) -> 'b) -> (unit -> 'a) -> (unit -> 'a) -> 'b* | 23 |
| max_case | *coq_N -> coq_N -> ((coq_N -> coq_N -> unit -> 'a) -> 'b) -> 'a -> 'a -> 'b* | 25 |
| max_dec | *coq_N -> coq_N -> bool* | 27 |
| min_case_strong | *coq_N -> coq_N -> ((coq_N -> coq_N -> unit -> 'a) -> 'b) -> (unit -> 'a) -> (unit -> 'a) -> 'b* | 23 |
| min_case | *coq_N -> coq_N -> ((coq_N -> coq_N -> unit -> 'a) -> 'b) -> 'a -> 'a -> 'b* | 25 |
| min_dec | *coq_N -> coq_N -> bool* | 27 |
| max_case_strong_pd | *coq_N -> coq_N -> (unit -> 'a) -> (unit -> 'a) -> 'a* | 25 |
| max_case | *coq_N -> coq_N -> ((coq_N -> coq_N -> unit -> 'a) -> 'b) -> 'a -> 'a -> 'b* | 27 |
| max_dec_pd | *coq_N -> coq_N -> bool* | 29 |
| min_case_strong_pd | *coq_N -> coq_N -> (unit -> 'a) -> (unit -> 'a) -> 'a* | 25 |
| min_case | *coq_N -> coq_N -> ((coq_N -> coq_N -> unit -> 'a) -> 'b) -> 'a -> 'a -> 'b* | 27 |
| min_dec_pd | *coq_N -> coq_N -> bool* | 29 |

## A.2 Evaluation-Step Bounds

| Name | Step Bound | Analysis Time | #Constraints |
|---|---|---|---|
| **File: WorkingWithLists.raml (99 Problems in OCaml)** | | | |
| last | $3+9M$ | 0.01 | 41 |
| lastTwo | $3+11M$ | 0.01 | 58 |
| at | $3+18M$ | 0.01 | 49 |
| natAt | $3+6K+4N$ | 0.02 | 48 |
| length | $9+12M$ | 0.01 | 37 |
| rev | $9+11M$ | 0.02 | 42 |
| eqList | $5+L+17M$ | 0.02 | 62 |
| isPalindrome | $20+29M$ | 0.02 | 115 |
| flatten | – | – | $fail$ |
| compress | $3+18M$ | 0.03 | 64 |
| pack | $21+34M$ | 0.05 | 154 |
| encode | $21+37M$ | 0.07 | 151 |
| decode | – | – | $fail$ |
| duplicate | $3+13M$ | 0.06 | 33 |
| replicate | – | – | $fail$ |
| drop | $9+19M$ | 0.08 | 61 |
| split | $23+31M$ | 0.12 | 158 |
| slice | $35+19M$ | 0.13 | 119 |
| concat | $3+9M$ | 0.12 | 33 |
| rotate | $74+43M$ | 0.19 | 290 |
| removeAt | $3+19M$ | 0.18 | 47 |
| insertAt | $3+22M$ | 0.21 | 105 |
| constructList | – | – | $fail$ |
| random | 5 | 0.21 | 8 |
| min | 9 | 0.21 | 15 |
| randSelect | $61+56.5M+16.5M^2$ | 0.46 | 607 |
| lottoSelect | – | – | $fail$ |
| snd | 5 | 0.46 | 6 |
| fst | 5 | 0.45 | 6 |
| map | $3+13M$ | 0.46 | 73 |
| insert | $6+21M$ | 0.52 | 116 |
| sort | $3+7.5M+10.5M^2$ | 0.51 | 203 |
| compare | 15 | 0.45 | 26 |
| lengthSort | $23+12LM+36M+27M^2$ | 0.67 | 1000 |
| **File: LogicAndCodes.raml** | | | |
| eval2 | $17+38K+38L+14M$ | 0.29 | 1044 |
| table2 | $117+152K+152L+56M$ | 1.62 | 5899 |
| assoc | $2+16M$ | 0.06 | 66 |
| eval | $10+11L+16M+16MX+16MY+20X+20Y$ | 0.33 | 1864 |
| tableMake | – | – | $fail$ |
| **File: echelon_form.raml** | | | |
| size | $3+11n$ | 0.01 | 29 |
| getElem | $2+18n$ | 0.01 | 41 |
| get2Elems | $2+19L+2M$ | 0.02 | 59 |
| subtract_row_helper | $3+20L+2M$ | 0.02 | 58 |
| subtract_row | $12+39L+4M$ | 0.03 | 147 |
| subtract_helper | $3+19LM+28M+24MY$ | 0.16 | 850 |
| concat | $3+11n_1$ | 0.03 | 35 |
| tail | $8+18n$ | 0.03 | 44 |
| hd_helper | $8+22M$ | 0.03 | 53 |
| reverse_helper | $3+11M$ | 0.03 | 35 |
| reverse | $7+11n$ | 0.03 | 41 |
| head | $21+33n$ | 0.05 | 103 |
| split_helper | $18+11L+34M$ | 0.06 | 102 |
| split | $24+34n$ | 0.06 | 110 |
| subtract | $45+43LM+63M$ | 0.34 | 1632 |
| echelon_helper | $3+43LMY+63MY+59Y$ | 1.75 | 8563 |
| echelon_form | $8+43m^2n+59m+63m^2$ | 1.81 | 8838 |
| **File: matrix.raml** | | | |
| check_lists | $3+11LM+19M$ | 0.03 | 161 |
| check_mat | $18+11LM+30M$ | 0.05 | 262 |
| check_matrix | $25+11LM+30M$ | 0.05 | 271 |
| construct_matrix | $28+11LM+30M$ | 0.08 | 349 |
| getElemMatrix | $9+18mn+18m$ | 0.05 | 182 |
| op | 13 | 0.03 | 27 |
| rec_list | $3+29L+2M$ | 0.05 | 78 |
| rec_mat | $3+31LM+2M+19Y$ | 0.15 | 526 |
| check_sanity | $79+11LM+30M+11RY+30Y$ | 0.15 | 681 |
| plus | $130+53m_1n_1+32m_1+11m_2n_2+79m_2$ | 0.41 | 1916 |
| minus | $130+53m_1n_1+32m_1+11m_2n_2+79m_2$ | 0.52 | 1916 |
| append | $6+11n$ | 0.26 | 34 |
| append_col | $3+11mn+8m+14l$ | 0.34 | 396 |
| transpose_helper | $3+16.5LM+5.5LM^2+14M+11MRY$ | 0.55 | 1814 |
| transpose | $7+16.5mn+5.5m^2n+14m$ | 0.55 | 1820 |
| prod | $3+16L+2M$ | 0.31 | 51 |
| prod_mat | $3+17L+2LM+16LY$ | 0.37 | 380 |
| mult_slow | $14+28m_1+47.5m_1m_2n_2+5.5m_1m_2^2n_2+14m_1m_2+47.5m_2n_2+5.5m_2^2n_2+14m_2$ | 6.81 | 42274 |
| lineMult | $3+6L+16M$ | 0.37 | 78 |
| computeLine | $3+22LY+18M$ | 0.48 | 476 |
| mat_mult_jan | $3+2LM+18M+22MRY+16MY$ | 1.11 | 3901 |
| check_mult_sanity | $71+11LM+30M+11RY+30Y$ | 0.6 | 665 |
| mult | $121+13m_1n_1+78m_1+27m_1m_2n_2+16m_1m_2+11m_2n_2+30m_2$ | 1.97 | 7534 |
| delete | $2+21n$ | 1.71 | 48 |
| submat | $9+21mn+19m$ | 1.85 | 412 |
| **File: power_radio.raml** | | | |
| sendmsg msg | $12+10n$ | 0.01 | 38 |
| main1_events | $10+9K'+23.00L'+10L'Y+9R'+9Z'$ | 0.09 | 338 |
| main2_events | $11+10K'+16L'+19L'Y+10R'+26Z'$ | 0.2 | 1230 |
| main3_events | $11+13K'+25L'+23L'Y+5L'MY+29R'+29Z'$ | 1.31 | 7351 |
| main4_events | $11+13K'+25L'+28L'Y+17R'+29Z'$ | 0.42 | 2397 |
| main5_events | $18+13K'+44.6L'+19L'Y+13R'+13Z'$ | 0.44 | 2339 |
| **File: avanzini.raml** | | | |
| partition | $5+17n$ | 0.01 | 57 |
| quicksort | $3+20n+13n^2$ | 0.1 | 587 |
| rev_sort | $10+23KM+32L'+20L'Y+13L'Y^2$ | 0.71 | 17704 |
| **File: append_all.raml** | | | |
| append_all | $7+9LM+15M$ | 0.09 | 1011 |
| append_all2 | $7+15LM+18LMY+21M$ | 0.16 | 4869 |
| append_all3 | $7+21LM+27LMRY+15LMY+21M$ | 0.39 | 13435 |
| **File: bfs.raml** | | | |
| dfs | $18+26M$ | 0.04 | 488 |
| bfs | $24+92M$ | 0.07 | 1723 |
| **File: rev_pairs.raml** | | | |
| pairs | $3+7M+10M^2$ | 0.03 | 317 |
| **File: binary_counter.raml** | | | |
| add_one | $8+12L'$ | 0.03 | 155 |
| add_many | $3+12L'+30X$ | 0.03 | 313 |
| add_list | $5+12L'+34Y$ | 0.03 | 341 |
| **File: array_fun.raml** | | | |
| nat_iterate | $3+10N$ | 0.03 | 55 |
| nat_fold | $3+12N$ | 0.02 | 76 |
| apply_all | $15+21N$ | 0.03 | 135 |
| **File: calculator.raml** | | | |
| add | $3+8M$ | 0.05 | 117 |
| sub | $5+2K+8M$ | 0.03 | 156 |
| mult | $3+8M+4M^2$ | 0.06 | 427 |
| eval_simpl | $8+8KLN+5KN+20L+22X$ | 0.21 | 9548 |
| eval | $3+20KLN+14KN+14L+14X$ | 0.22 | 10723 |
| **File: mergesort.raml** | | | |
| split | $5+8.5M$ | 0.04 | 159 |
| merge | $3+17L+17M$ | 0.04 | 407 |
| mergesort | $4-6K+32.5KN+6K^2$ | 0.17 | 4654 |
| mergesort_list | $4-7.5K+33.5KN-7KX+7.5K^2+7K^2X$ | 0.39 | 15055 |
| **File: quicksort.raml** | | | |
| partition | $5+19M$ | 0.04 | 197 |
| quicksort | $3+19M+14M^2$ | 0.08 | 1747 |
| quicksort_pairs | $3+13M+23M^2$ | 0.09 | 2131 |
| quicksort_list | $3-7.5LM+7.5LM^2+19.5M+16.5M^2$ | 0.27 | 8712 |
| **File: square_mult.raml** | | | |
| square_mult | $9+21K'+30L'$ | 0.08 | 243 |
| **File: subsequence.raml** | | | |
| lcs | $21+8L+50LM+27M$ | 0.04 | 900 |
| **File: running.raml** | | | |
| abmap | $3+12M+12N$ | 0.03 | 477 |
| asort | $11+22KN+13K^2N+13L+15N$ | 0.14 | 5656 |
| asort’ | $14+22KN+13K^2N+7L+15N$ | 0.16 | 5655 |
| btick | $3+16L+14N$ | 0.05 | 904 |
| abfoldr | $3+13M+13N$ | 0.05 | 267 |
| cons_all | $13+17M+21MN+26N$ | 0.08 | 1844 |
| **File: ocaml_sort.raml** | | | |
| merge | $3+17L+17M$ | 0.05 | 404 |
| list | $43+30.5M+8.5M^2$ | 0.11 | 3066 |
| **File: ocaml_list.raml** | | | |
| length | $7+10M$ | 0.02 | 59 |
| cons | $4$ | 0.01 | 9 |
| hd | $3$ | 0.02 | 14 |
| tl | $3$ | 0.02 | 16 |
| nth | $14+16M$ | 0.02 | 93 |
| append | $3+9M$ | 0.03 | 77 |
| rev_append | $3+9M$ | 0.02 | 78 |
| rev | $7+9M$ | 0.02 | 84 |
| flatten | $3+9LM+11M$ | 0.03 | 201 |
| concat | $3+9LM+11M$ | 0.04 | 201 |
| map | $3+13M$ | 0.02 | 73 |
| mapi | $8+19M$ | 0.02 | 92 |
| rev_map | $9+11M$ | 0.02 | 103 |
| iter | $3+10M$ | 0.02 | 55 |
| iteri | $8+16M$ | 0.02 | 74 |
| fold_left | $3+10M$ | 0.02 | 59 |
| fold_right | $3+10M$ | 0.03 | 55 |
| map2 | $5+17L$ | 0.03 | 142 |
| rev_map2 | $12+15L$ | 0.03 | 196 |
| iter2 | $5+14M$ | 0.03 | 120 |
| fold_left2 | $5+14L$ | 0.03 | 128 |
| fold_right2 | $5+14M$ | 0.03 | 120 |
| for_all | $3+12M$ | 0.03 | 61 |
| exists | $3+12M$ | 0.03 | 61 |
| for_all2 | $5+16M$ | 0.03 | 126 |
| exists2 | $5+16L$ | 0.04 | 205 |
| mem | $3+18M$ | 0.04 | 67 |
| memq | $3+22M$ | 0.04 | 75 |
| assoc | $2+18M$ | 0.03 | 74 |
| assq | $2+22M$ | 0.03 | 84 |
| mem_assoc | $3+20M$ | 0.04 | 69 |
| mem_assq | $3+24M$ | 0.04 | 77 |
| remove_assoc | $3+21M$ | 0.04 | 92 |
| remove_assq | $3+25M$ | 0.04 | 102 |
| find | $2+10M$ | 0.04 | 66 |
| find_all | $18+22M$ | 0.04 | 312 |
| filter | $18+22M$ | 0.05 | 312 |
| partition | $30+23M$ | 0.06 | 580 |
| split | $5+17M$ | 0.06 | 83 |
| combine | $5+13M$ | 0.06 | 122 |
| merge | $3+21L+21M$ | 0.08 | 253 |
| chop | $8+16M$ | 0.06 | 83 |
| stable_sort | $fail$ | 0.32 | 6606 |
| sort | $fail$ | 0.45 | 6606 |
| fast_sort | $fail$ | 0.68 | 6606 |
| **File: aws.raml** | | | |
| average_grade | $18+34M$ | 0.04 | 185 |
| greater_eq | $49+68M$ | 0.05 | 721 |
| sort_students | $12-34LM+34LM^2+15M$ | 0.32 | 8933 |
| make_table | $9+25LM+21M$ | 0.08 | 949 |
| find | $2+12M$ | 0.05 | 88 |
| lookup | $8+17LM+17M$ | 0.08 | 269 |
| average_grade' | $21+17LM+17LMY+33M$ | 0.09 | 1094 |
| greater_eq' | $61+34LM+34LMY+66M$ | 0.18 | 2572 |
| sort_students_efficient | $fail$ | 0.64 | 19377 |
| **File: PROTOTYPE** | | | |
| **File: appendAll.raml** | | | |
| appendAll | $3+9n_1n_2+13n_1$ | 0.05 | 203 |
| appendAll2 | $3+13n_1n_2+18n_1n_2n_3+18n_1$ | 0.2 | 1565 |
| appendAll3 | $3+18n_1n_2+27n_1n_2n_3n_4+13n_1n_2n_3+18n_1$ | 1.57 | 12785 |
| **File: duplicates.raml** | | | |
| eq | $5+17n_1+n_2$ | 0.01 | 62 |
| remove | $3+21L+18LY$ | 0.1 | 625 |
| nub | $3-9Nn+9N^2n+5.5N+10.5N^2$ | 0.51 | 3919 |
| **File: dyade.raml** | | | |
| multList | $3+15n$ | 0.01 | 36 |
| dyade | $3+15n_1n_2+17n_1$ | 0.05 | 246 |
| **File: eratosthenes.raml** | | | |
| filter | $3+23n$ | 0.01 | 51 |
| eratos | $3+4.5n+11.5n^2$ | 0.04 | 180 |
| **File: bitvectors.raml** | | | |
| bitToInt' | $3+20M$ | 0.02 | 42 |
| bitToInt | $7+20n$ | 0.02 | 48 |
| sum | 33 | 0.01 | 59 |
| add' | $3+51L+2M$ | 0.03 | 112 |
| add | $8+2n_1+51n_2$ | 0.03 | 119 |
| diff | 31 | 0.03 | 47 |
| sub' | $5+59L+2M$ | 0.06 | 127 |
| sub | $14+61n_1$ | 0.06 | 138 |
| mult | $3+53n_1n_2+30n_1$ | 0.12 | 477 |
| compare | $3+28L+2M$ | 0.07 | 80 |
| leq | $13+2n_1+28n_2$ | 0.07 | 94 |
| **File: flatten.raml** | | | |
| flatten | $3+5.5nl+5.5n^2l+24n$ | 0.42 | 3388 |
| insert | $6+17n$ | 0.02 | 55 |
| insertionsort | $3+7.5n+8.5n^2$ | 0.05 | 181 |
| flattensort | $11+13nl+5.5n^2l+8.5n^2l^2+24n$ | 2.34 | 19153 |
| **File: listsort.raml** | | | |
| isortlist | $3-9nl+9n^2l+6.5n+9.5n^2$ | 0.41 | 2676 |
| **File: longestCommonSubsequence.raml** | | | |
| firstline | $3+10M$ | 0.01 | 29 |
| right | 5 | 0.01 | 14 |
| max | 7 | 0.01 | 13 |
| newline | $3+2L+50M$ | 0.03 | 128 |
| lcstable | $11+10L+52LM+25M$ | 0.15 | 865 |
| lcs | $23+10L+52LM+25M$ | 0.16 | 901 |
| **File: mergesort.raml** | | | |
| msplit | $5+9.5M$ | 0.03 | 112 |
| merge | $3+22L+22M$ | 0.06 | 239 |
| mergesortBuggy | – | – | $fail$ |
| mergesort | $3-38.67M+47.33M^2$ | 0.14 | 869 |
| **File: minsort.raml** | | | |
| findMin | $3+21M$ | 0.03 | 125 |
| minSort | $8+25.5M+10.5M^2$ | 0.04 | 219 |
| **File: queue.raml** | | | |
| empty | 3 | 0.01 | 4 |
| enqueue | 10 | 0.03 | 19 |
| enqueues | $3+21M$ | 0.09 | 203 |
| copyover | $7+14M$ | 0.08 | 226 |
| dequeue | $14+39M$ | 0.18 | 678 |
| children | 27 | 0.26 | 333 |
| breadth | $18+118BR+116BRT-LM+116LMY+L^2M+88M+118MY+49R-RT+RT^2$ | 3.99 | 10540 |
| startBreadth | $19+57M+59M^2$ | 4.34 | 10818 |
| depth | $27+42L+53LM$ | 4.33 | 18711 |
| startDepth | $3+15.5M+26.5M^2$ | 4.29 | 18747 |
| **File: quicksort_mutual.raml** | | | |
| part | $24+16L+30LY+15L^2+27M+30MY+15M^2+46Y+15Y^2$ | 0.15 | 896 |
| quicksortMutual | $3+16M+15M^2$ | 0.15 | 896 |
| **File: rationalPotential.raml** | | | |
| zip3 | $3+17L+2M$ | 0.02 | 68 |
| group3 | $3+5.67M$ | 0.02 | 52 |
| **File: sizechange.raml** | | | |
| r1 | $3+11M$ | 0.04 | 124 |
| rev | $7+11M$ | 0.04 | 130 |
| f | $3+11LM+17.33M+3.67M^3$ | 0.12 | 574 |
| g | $15+11L+11LM+17.33M+11MY+3.67M^3$ | 0.12 | 574 |
| f2 | – | – | $fail$ |
| last | $3+14M$ | 0.05 | 108 |
| f2’ | $12+14L+14M$ | 0.09 | 331 |
| g3 | $3+11M$ | 0.04 | 124 |
| f3 | $11+22L+11M$ | 0.08 | 291 |
| **File: splitandsort.raml** | | | |
| insert | $13+23M$ | 0.13 | 437 |
| split | $3+11.5M+11.5M^2$ | 0.18 | 623 |
| splitqs | $7+24M$ | 0.06 | 184 |
| quicksort | $3+16.5M+17.5M^2$ | 0.26 | 1607 |
| sortAll | $3+16.5LM+17.5L^2M+19M$ | 0.51 | 3165 |
| splitAndSort | $11+47M+29M^2$ | 0.69 | 3793 |
| **File: subtrees.raml** | | | |
| subtrees | $3+23.5M+5.5M^2$ | 0.12 | 680 |
| **File: tuples.raml** | | | |
| attach | $3+13M$ | 0.06 | 95 |
| pairs | $3+7M+12M^2$ | 0.27 | 1428 |
| pairsAux | $3+8M+12M^2$ | 0.23 | 1041 |
| pairsSlow | $3+16.17M+M^2+1.83M^3$ | 0.29 | 1428 |
| triples | $3+30.5M-14.5M^2+8M^3$ | 0.95 | 5804 |
| quadruples | $3+0.33M+35.5M^2-14.83M^3+3M^4$ | 2.52 | 15613 |
| **File: array_dijkstra.raml** | | | |
| makeGraph | $35+36K+21N+1N^2$ | 0.03 | 519 |
| dijkstra | $46+33M+111M^2$ | 0.11 | 2808 |
| **File: CompCert** | | | |
| **File: String0.ml** | | | |
| string_dec | $5+14L$ | 0.06 | 205 |
| prefix | $3+14M$ | 0.05 | 170 |
| **File: Tuples.ml** | | | |
| uncurry | – | 0.02 | 106 |
| **File: Specif.ml** | | | |
| projT1 | 3 | 0.01 | 6 |
| projT2 | 3 | 0.01 | 6 |
| value | 2 | 0.01 | 4 |
| **File: EquivDec.ml** | | | |
| equiv_dec | 1 | 0.01 | 2 |
| **File: Datatypes.ml** | | | |
| implb | 3 | 0.01 | 7 |
| xorb | 5 | 0.01 | 12 |
| negb | 7 | 0.01 | 13 |
| fst | 3 | 0.01 | 4 |
| snd | 3 | 0.01 | 4 |
| length | $3+7M$ | 0.04 | 72 |
| app | $3+9M$ | 0.05 | 118 |
| coq_CompOpp | 3 | 0.01 | 25 |
| coq_CompareSpec2Type | 3 | 0.01 | 25 |
| coq_CompSpec2Type | 6 | 0.01 | 30 |
| **File: Bool.ml** | | | |
| bool_dec | 5 | 0.01 | 17 |
| eqb | 5 | 0.01 | 12 |
| iff_reflect | 7 | 0.01 | 15 |
| **File: Ring.ml** | | | |
| bool_eq | 12 | 0.01 | 23 |
| **File: Peano.ml** | | | |
| plus | $3+8M$ | 0.04 | 117 |
| max | $3+8L+2M$ | 0.06 | 180 |
| min | $3+10M$ | 0.05 | 147 |
| nat_iter | $3+9M$ | 0.06 | 64 |
| **File: List0.ml** | | | |
| hd | 3 | 0.03 | 18 |
| tl | 3 | 0.03 | 21 |
| in_dec | $3+14M$ | 0.04 | 91 |
| nth_error | $5+8M$ | 0.08 | 326 |
| remove | $3+15M$ | 0.05 | 179 |
| rev | $3+9.5M+4.5M^2$ | 0.1 | 423 |
| rev_append | $3+9M$ | 0.05 | 122 |
| rev' | $7+9M$ | 0.05 | 128 |
| list_eq_dec | $5+14L$ | 0.08 | 223 |
| map | $3+11M$ | 0.05 | 96 |
| fold_left | $3+10M$ | 0.05 | 77 |
| fold_right | $3+10M$ | 0.07 | 65 |
| existsb | $3+12M$ | 0.06 | 71 |
| forallb | $3+12M$ | 0.06 | 71 |
| filter | $3+13M$ | 0.07 | 176 |
| **File: EqNat.ml** | | | |
| beq_nat | $5+8L$ | 0.08 | 162 |
| **File: Compare_dec.ml** | | | |
| le_lt_dec | $3+6L+2M$ | 0.07 | 127 |
| le_gt_dec | $7+8M$ | 0.07 | 133 |
| nat_compare | $5+8M$ | 0.08 | 183 |
| **File: BinPosDef.ml** | | | |
| succ | $5+7N$ | 0.04 | 124 |
| add | $7+7L+5M+7N+5X$ | 0.64 | 5438 |
| add_carry | $9+7L+5M+7N+5X$ | 0.67 | 5438 |
| pred_double | $7+7M$ | 0.05 | 124 |
| pred | $5+7M$ | 0.06 | 174 |
| pred_N | $6+7M$ | 0.06 | 187 |
| mask_rect | 5 | 0.04 | 41 |
| mask_rec | 5 | 0.04 | 41 |
| succ_double_mask | 6 | 0.04 | 53 |
| double_mask | 6 | 0.04 | 48 |
| double_pred_mask | $8+7M$ | 0.06 | 197 |
| pred_mask | $13+7L$ | 0.09 | 446 |
| sub_mask | $5+9L+7M+7N+9X$ | 0.74 | 6368 |
| sub_mask_carry | $3+8L+8M+8N+8X$ | 0.73 | 6368 |
| sub | $11+9L+7M+7N+9X$ | 0.72 | 6405 |
| mul | $3+7L+17LN+8M+25N+10NX$ | 1.03 | 7722 |
| iter | – | 0.59 | 2849 |
| pow | – | 32.47 | 139223 |
| square | $3+9M+10MN+24.5N+8.5N^2$ | 0.98 | 7592 |
| div2 | 3 | 0.17 | 44 |
| div2_up | $3+7N$ | 0.19 | 169 |
| size_nat | $5+7M+7N$ | 0.2 | 156 |
| size | $3+19M+19N$ | 0.23 | 542 |
| compare_cont | $5+9L+9X$ | 0.38 | 1248 |
| compare | $10+9L+9X$ | 0.33 | 1258 |
| min | $16+9L+9X$ | 0.45 | 2071 |
| max | $16+9L+9X$ | 0.48 | 2071 |
| eqb | $5+8L+8X$ | 0.27 | 454 |
| leb | $16+9L+9X$ | 0.41 | 1283 |
| ltb | $16+9L+9X$ | 0.43 | 1283 |
| sqrtrem_step | – | | |
| sqrtrem | $5+66M+16MN+4M^2+59N+4N^2$ | 17.54 | 135517 |
| sqrt | $13+66M+16MN+4M^2+59N+4N^2$ | 18.25 | 135529 |
| gcdn | – | 19.67 | 97408 |
| gcd | – | 29.64 | 100787 |
| ggcdn | – | 29.04 | 179760 |
| ggcd | – | 29.79 | 183139 |
| coq_Nsucc_double | 6 | 4.81 | 42 |
| coq_Ndouble | 6 | 4.86 | 37 |
| coq_lor | $5+8L+2M+2N+8X$ | 5.12 | 1032 |
| coq_land | $6+16L+16X$ | 4.97 | 1355 |
| ldiff | $5+13L+3M+3N+13X$ | 5.08 | 1435 |
| coq_lxor | $5+13L+3M+3N+13X$ | 5.09 | 1509 |
| shiftl_nat | $8+12X$ | 4.85 | 136 |
| shiftr_nat | $8+12X$ | 5.08 | 186 |
| shiftl | – | 6.42 | 5453 |
| shiftr | – | 6.58 | 7323 |
| testbit_nat | $5+8X$ | 5.15 | 291 |
| testbit | $5+L+23M+23N+7Z$ | 5.3 | 1219 |
| iter_op | $3+10M+13N$ | 5.2 | 187 |
| to_nat | – | 5.86 | 2652 |
| of_nat | $3+21M$ | 5.44 | 244 |
| of_succ_nat | $3+19M$ | 5.27 | 221 |
| digits2_pos | $3+19M+19N$ | 0.47 | 542 |
| coq_Zdigits2 | $2+19K+19L+7+1+19R+7+19Z$ | 0.54 | 1146 |
| **File: BinNat.ml** | | | |
| succ_double_n | 6 | 1.72 | 42 |
| double_n | 6 | 1.75 | 37 |
| succ_n | $11+7K$ | 1.79 | 162 |
| pred_n | $11+7L$ | 1.77 | 214 |
| succ_pos_n | $10+7K$ | 1.75 | 150 |
| add_n | $18+7K+5L+7R+5S$ | 2.48 | 5562 |
| sub_n | $17+8K+8L+8R+8S$ | 2.65 | 6514 |
| mul_n | $2+17KR+10KS+25K+8L+2+7R+8$ | 14.04 | 89507 |
| compare_n | $18+9K+9L$ | 2.36 | 1373 |
| eqb_n | $13+8R+8S$ | 2.4 | 580 |
| leb_n | $27+9K+9L$ | 2.47 | 1398 |
| ltb_n | $27+9K+9L$ | 2.47 | 1398 |
| min_n | $27+9K+9L$ | 2.71 | 2352 |
| max_n | $27+9K+9L$ | 2.71 | 2352 |
| div2_n | 6 | 2.39 | 79 |
| even_n | 5 | 2.29 | 53 |
| odd_n | 17 | 2.45 | 74 |
| pow_n | – | 7.98 | 25024 |
| square_n | – | 3.4 | 7622 |
| log2_n | $5+19K+19L$ | 2.55 | 1165 |
| size_n | $9+19K+19L$ | 2.49 | 575 |
| size_nat_n | $10+7K+7L$ | 2.52 | 182 |
| pos_div_eucl | – | 38.2 | 183270 |
| div_eucl | $2+50KL-12.5K+99K+12.5K^2-12.5L+99L+12.5L^2+2+15$ | 38.13 | 183578 |
| div | – | 34.64 | 183586 |
| modulo | – | 40.36 | 183588 |
| gcd_n | – | 29.84 | 101067 |
| ggcd_n | – | 41.02 | 183702 |
| coq_lor_n | $14+8K+8L+2R+2S$ | 7.51 | 1156 |
| coq_land_n | $14+16R+16S$ | 7.53 | 1442 |
| ldiff_n | $13+3K+3L+13R+13S$ | 7.75 | 1560 |
| coq_lxor_n | $13+3K+3L+13R+13S$ | 7.63 | 1627 |
| shiftl_nat | $12+15Z$ | 7.59 | 164 |
| shiftr_nat | $12+15Z$ | 7.58 | 225 |
| shiftl_n | – | 16.82 | 5551 |
| shiftr_n | – | 16.99 | 8953 |
| testbit_nat_n | $10+9Z$ | 7.79 | 387 |
| testbit_n | $11+23K+23L+7S$ | 7.88 | 1261 |
| to_nat_n | – | 16.25 | 2694 |
| of_nat_n | $3+19M$ | 7.91 | 246 |
| iter_n | – | 16.49 | 2893 |
| discr | 4 | 7.77 | 32 |
| binary_rect | $17+11K+11L$ | 7.99 | 593 |
| binary_rec | $17+11K+11L$ | 8.22 | 593 |
| leb_spec0 | $33+9K+9L$ | 8.35 | 1416 |
| ltb_spec0 | $33+9K+9L$ | 8.45 | 1416 |
| log2_up | $51+19K+26L$ | 8.87 | 3417 |
| lcm | – | 169.2 | 403044 |
| eqb_spec | $22+8R+8S$ | 11.12 | 598 |
| b2n | 8 | 11.02 | 20 |
| setbit | – | 20.78 | 10128 |
| clearbit | – | 22.03 | 10672 |
| ones | – | 23.56 | 5992 |
| lnot | – | 22.39 | 11681 |
| max_case_strong | $70+18K+18L$ | 15.23 | 12294 |
| max_case | $78+18K+18L$ | 15.24 | 12264 |
| max_dec | $86+18K+18L$ | 16.73 | 12270 |
| min_case_strong | $70+18K+18L$ | 17.72 | 12294 |
| min_case | $78+18K+18L$ | 18.31 | 12264 |
| min_dec | $86+9K+9L+9R+9S$ | 20.11 | 12270 |
| max_case_strong_pd | $78+9K+9L+9R+9S$ | 21.86 | 12300 |
| max_case | $14+18K+18L+2+62$ | 32.11 | 52970 |
| max_dec_pd | $86+18K+18L$ | 23.53 | 12270 |
| min_case_strong_pd | $78+9K+9L+9R+9S$ | 25.48 | 12300 |
| min_case | $14 + 18K + 18L + 2 + 62$ | 35.89 | 52970 |
| min_dec_pd | $86 + 18K + 18L$ | 27.49 | 12270 |

## A.3 Heap-Allocation Bounds

| Name | Heap Bound | Analysis Time | #Constraints |
|---|---|---|---|
| **File: WorkingWithLists.raml (99 Problems in OCaml)** | | | |
| last | $2$ | 0.01 | 31 |
| lastTwo | $4$ | 0.02 | 45 |
| at | $2+2M$ | 0.01 | 32 |
| natAt | $2$ | 0.02 | 36 |
| length | $4+M$ | 0.01 | 22 |
| rev | $5+4M$ | 0.01 | 29 |
| eqList | $1$ | 0.02 | 42 |
| isPalindrome | $6+4M$ | 0.02 | 76 |
| flatten | – | – | $fail$ |
| compress | $2+4M$ | 0.03 | 44 |
| pack | $14+14M$ | 0.05 | 109 |
| encode | $13+12M$ | 0.06 | 98 |
| decode | – | – | $fail$ |
| duplicate | $2+8M$ | 0.05 | 24 |
| replicate | – | – | $fail$ |
| drop | $7+5M$ | 0.08 | 39 |
| split | $16+10M$ | 0.11 | 105 |
| slice | $12+6M$ | 0.13 | 72 |
| concat | $4M$ | 0.12 | 25 |
| rotate | $23+11M$ | 0.21 | 176 |
| removeAt | $2+6M$ | 0.18 | 31 |
| insertAt | $3+6M$ | 0.21 | 84 |
| constructList | – | – | $fail$ |
| random | $1$ | 0.2 | 4 |
| min | $0$ | 0.2 | 5 |
| randSelect | $21+9M+5M^2$ | 0.48 | 500 |
| lottoSelect | – | – | $fail$ |
| snd | $0$ | 0.44 | 1 |
| fst | $0$ | 0.42 | 1 |
| map | $2+4M$ | 0.46 | 62 |
| insert | $6+5M$ | 0.46 | 94 |
| sort | $2+3.5M+2.5M^2$ | 0.51 | 169 |
| compare | $1$ | 0.46 | 12 |
| lengthSort | $12+LM+17M+3M^2$ | 0.66 | 876 |
| **File: LogicAndCodes.raml** | | | |
| eval2 | $0$ | 0.29 | 976 |
| table2 | $46$ | 1.54 | 5607 |
| assoc | $0$ | 0.05 | 49 |
| eval | $0$ | 0.31 | 1804 |
| tableMake | – | – | $fail$ |
| **File: echelon_form.raml** | | | |
| size | $1+n$ | 0.01 | 19 |
| getElem | $2n$ | 0.01 | 24 |
| get2Elems | $1+2L$ | 0.02 | 38 |
| subtract_row_helper | $2+4M$ | 0.02 | 38 |
| subtract_row | $3+2L+4M$ | 0.03 | 89 |
| subtract_helper | $2+2LM+7M+4MY$ | 0.16 | 778 |
| concat | $4n_1$ | 0.03 | 25 |
| tail | $1+2n$ | 0.03 | 27 |
| hd_helper | $1+6M$ | 0.03 | 34 |
| reverse_helper | $4M$ | 0.03 | 25 |
| reverse | $2+4n$ | 0.03 | 28 |
| head | $5+10n$ | 0.04 | 65 |
| split_helper | $4+4L+9M$ | 0.06 | 65 |
| split | $7+9n$ | 0.06 | 69 |
| subtract | $9+6LM+15M$ | 0.31 | 1469 |
| echelon_helper | $6LMY+15MY+10Y$ | 1.74 | 8386 |
| echelon_form | $1+6m^2n+10m+15m^2$ | 1.88 | 8657 |
| **File: matrix.raml** | | | |
| check_lists | $1+LM+M$ | 0.03 | 135 |
| check_mat | $2+LM+2M$ | 0.05 | 214 |
| check_matrix | $2+LM+2M$ | 0.05 | 216 |
| construct_matrix | $5+LM+2M$ | 0.07 | 292 |
| getElemMatrix | $2mn+2m$ | 0.05 | 137 |
| op | 1 | 0.03 | 10 |
| rec_list | $2+5L$ | 0.04 | 45 |
| rec_mat | $2+5LM+6Y$ | 0.15 | 477 |
| check_sanity | $4+LM+2M+RY+2Y$ | 0.15 | 542 |
| plus | $12+7m_1n_1+2m_1+m_2n_2+10m_2$ | 0.4 | 1652 |
| minus | $12+7m_1n_1+2m_1+m_2n_2+10m_2$ | 0.51 | 1652 |
| append | $6+4n$ | 0.26 | 24 |
| append_col | $4mn+12l$ | 0.36 | 367 |
| transpose_helper | $10LM+2LM^2+4MRY$ | 0.56 | 1773 |
| transpose | $2+10mn+2m^2n$ | 0.57 | 1776 |
| prod | 1 | 0.31 | 33 |
| prod_mat | $2+5L$ | 0.35 | 63 |
| mult_slow | $4+8m_1+15m_1m_2n_2+2m_1m_2^2n_2+15m_2n_2+2m_2^2n_2$ | 6.78 | 42184 |
| lineMult | $2+6M$ | 0.39 | 49 |
| computeLine | $6LY+2M$ | 0.5 | 431 |
| mat_mult_jan | $2+2LM+6M+6MRY$ | 1.17 | 3844 |
| check_mult_sanity | $4+LM+2M+RY+2Y$ | 0.64 | 534 |
| mult | $11+3m_1n_1+10m_1+7m_1m_2n_2+m_2n_2+2m_2$ | 2.16 | 7270 |
| delete | $6n$ | 1.88 | 30 |
| submat | $2+6mn+2m$ | 1.81 | 365 |
| **File: power_radio.raml** | | | |
| sendmsg msg | 6 | 0.01 | 22 |
| main1_events | $5+6M$ | 0.01 | 73 |
| main2_events | $8+4L'Y+8Z'$ | 0.2 | 1184 |
| main3_events | $11+2K'+2L'+8L'Y+8R'+10Z'$ | 0.42 | 2508 |
| main4_events | $11+2K'+2L'+8L'Y+2R'+10Z'$ | 0.43 | 2331 |
| main5_events | $12+2K'+5.2L'+4L'Y+2R'+2Z'$ | 0.4 | 2226 |
| **File: avanzini.raml** | | | |
| partition | $6+6n$ | 0.01 | 40 |
| quicksort | $2+7n+5n^2$ | 0.09 | 542 |
| rev_sort | $7+12KM+16L'+7L'Y+5L'Y^2$ | 0.7 | 17616 |
| **File: append_all.raml** | | | |
| append_all | $3+4LM$ | 0.07 | 988 |
| append_all2 | $2+8LMY+3M$ | 0.16 | 4823 |
| append_all3 | $2+3LM+12LMRY+2M$ | 0.36 | 13366 |
| **File: bfs.raml** | | | |
| dfs | $12+8M$ | 0.04 | 466 |
| bfs | $12+24M$ | 0.08 | 1669 |
| **File: rev_pairs.raml** | | | |
| pairs | $2-3M+5M^2$ | 0.03 | 290 |
| **File: binary_counter.raml** | | | |
| add_one | $8+6L'$ | 0.04 | 145 |
| add_many | $6L'+14X$ | 0.04 | 292 |
| add_list | $1+6L'+17Y$ | 0.03 | 316 |
| **File: array_fun.raml** | | | |
| nat_iterate | 1 | 0.02 | 45 |
| nat_fold | 0 | 0.02 | 36 |
| apply_all | 2 | 0.02 | 102 |
| **File: calculator.raml** | | | |
| add | $4M$ | 0.04 | 110 |
| sub | 2 | 0.04 | 145 |
| mult | $2-2M+2M^2$ | 0.04 | 411 |
| eval_simpl | $5+4KLN+2X$ | 0.19 | 9500 |
| eval | $11KLN+4.5KN$ | 0.22 | 10688 |
| **File: mergesort.raml** | | | |
| split | $6+5M$ | 0.04 | 146 |
| merge | $4L+4M$ | 0.05 | 383 |
| mergesort | $2+13KN$ | 0.17 | 4589 |
| mergesort_list | $2-0.5K+17KN+0.5K^2$ | 0.37 | 14973 |
| **File: quicksort.raml** | | | |
| partition | $6+6*M$ | 0.04 | 178 |
| quicksort | $2+7M+5M^2$ | 0.07 | 1700 |
| quicksort_pairs | $2+17M+5M^2$ | 0.09 | 2063 |
| quicksort_list | $2-LM+LM^2+17.5M+5.5M^2$ | 0.25 | 8649 |
| **File: square_mult.raml** | | | |
| square_mult | $7+K'$ | 0.08 | 201 |
| **File: subsequence.raml** | | | |
| lcs | $9+5L+9LM+12M$ | 0.04 | 815 |
| **File: running.raml** | | | |
| abmap | $2+4M+4N$ | 0.04 | 459 |
| asort | $8+15KN+5K^2N+4L+6N$ | 0.13 | 5588 |
| asort' | $8+15KN+5K^2N+6N$ | 0.13 | 5588 |
| btick | $2+7L+7N$ | 0.05 | 884 |
| abfoldr | 0 | 0.03 | 242 |
| cons_all | $5+4M+4MN+5N$ | 0.08 | 1772 |
| **File: ocaml_sort.raml** | | | |
| merge | $4L+4M$ | 0.04 | 381 |
| list | $32+17M+2M^2$ | 0.11 | 2997 |
| **File: ocaml_list.raml** | | | |
| length | $1+M$ | 0.02 | 46 |
| cons | 4 | 0.02 | 7 |
| hd | 0 | 0.01 | 11 |
| tl | 0 | 0.01 | 13 |
| nth | $4+2M$ | 0.02 | 69 |
| append | $4M$ | 0.02 | 69 |
| rev_append | $4M$ | 0.02 | 70 |
| rev | $2+4M$ | 0.02 | 73 |
| flatten | $2+4LM$ | 0.02 | 185 |
| concat | $2+4LM$ | 0.03 | 185 |
| map | $2+4M$ | 0.02 | 62 |
| mapi | $3+5M$ | 0.02 | 72 |
| rev_map | $6+4M$ | 0.02 | 90 |
| iter | 1 | 0.02 | 45 |
| iteri | $2+1M$ | 0.02 | 55 |
| fold_left | 0 | 0.03 | 48 |
| fold_right | 0 | 0.02 | 44 |
| map2 | $2+4L$ | 0.03 | 125 |
| rev_map2 | $6+4M$ | 0.03 | 176 |
| iter2 | 1 | 0.02 | 104 |
| fold_left2 | 0 | 0.03 | 111 |
| fold_right2 | 0 | 0.03 | 103 |
| for_all | 1 | 0.03 | 49 |
| exists | 1 | 0.03 | 49 |
| for_all2 | 1 | 0.03 | 108 |
| exists2 | 1 | 0.03 | 169 |
| mem | $1+2M$ | 0.04 | 51 |
| memq | $1+2M$ | 0.04 | 55 |
| assoc | $2M$ | 0.04 | 57 |
| assq | $2M$ | 0.04 | 63 |
| mem_assoc | $1+2M$ | 0.04 | 51 |
| mem_assq | $1+2M$ | 0.04 | 55 |
| remove_assoc | $2+6M$ | 0.04 | 74 |
| remove_assq | $2+6M$ | 0.04 | 80 |
| find | 0 | 0.05 | 55 |
| find_all | $9+8M$ | 0.05 | 280 |
| filter | $9+8M$ | 0.05 | 280 |
| partition | $16+8M$ | 0.07 | 531 |
| split | $6+10M$ | 0.06 | 71 |
| combine | $2+6M$ | 0.06 | 110 |
| merge | $5L+5M$ | 0.07 | 227 |
| chop | $1+2M$ | 0.06 | 68 |
| stable_sort | $fail$ | 0.33 | 6204 |
| sort | $fail$ | 0.44 | 6204 |
| fast_sort | $fail$ | 0.66 | 6204 |
| **File: aws.raml** | | | |
| average_grade | $7+6M$ | 0.04 | 143 |
| greater_eq | $16+12M$ | 0.06 | 625 |
| sort_students | $4-6LM+6LM^2+11.5M+13.5M^2$ | 0.36 | 8765 |
| make_table | $7+9LM+11M$ | 0.08 | 911 |
| find | 0 | 0.06 | 75 |
| lookup | 4 | 0.07 | 226 |
| average_grade' | $10+8M$ | 0.11 | 1009 |
| greater_eq' | $22+16M$ | 0.22 | 2384 |
| sort_students_efficient | $fail$ | 0.64 | 19074 |
| **File: PROTOTYPE** | | | |
| **File: appendAll.raml** | | | |
| appendAll | $2+4n_1n_2$ | 0.04 | 185 |
| appendAll2 | $2+8n_1n_2n_3+2n_1$ | 0.21 | 1527 |
| appendAll3 | $2+2n_1n_2+12n_1n_2n_3n_4+2n_1$ | 1.57 | 12727 |
| **File: duplicates.raml** | | | |
| eq | 1 | 0.01 | 42 |
| remove | $2+5L$ | 0.02 | 90 |
| nub | $2+3.5N+2.5N^2$ | 0.12 | 938 |
| **File: dyade.raml** | | | |
| multList | $2+4n$ | 0.01 | 23 |
| dyade | $2+4n_1n_2+6n_1$ | 0.06 | 221 |
| **File: eratosthenes.raml** | | | |
| filter | $2+5n$ | 0.01 | 30 |
| eratos | $2+3.5n+2.5n^2$ | 0.04 | 148 |
| **File: bitvectors.raml** | | | |
| bitToInt' | 1 | 0.01 | 22 |
| bitToInt | 2 | 0.01 | 25 |
| sum | 7 | 0.01 | 32 |
| add' | $2+11M$ | 0.03 | 67 |
| add | $3+11n_1$ | 0.03 | 70 |
| diff | 5 | 0.03 | 20 |
| sub' | $4+13M$ | 0.06 | 70 |
| sub | $5+13n_1$ | 0.06 | 73 |
| mult | $2+11n_1n_2+9n_1$ | 0.12 | 409 |
| compare | $1+3M$ | 0.06 | 48 |
| leq | $2+3n_1$ | 0.07 | 53 |
| **File: flatten.raml** | | | |
| flatten | $2+2nl+2n^2l+2n$ | 0.42 | 3353 |
| insert | $6+4n$ | 0.02 | 36 |
| insertionsort | $2+4n+2n^2$ | 0.05 | 152 |
| flattensort | $4+6nl+2n^2l+2n^2l^2+2n$ | 2.43 | 19084 |
| **File: listsort.raml** | | | |
| isortlist | $2+3.5n+2.5n^2$ | 0.11 | 704 |
| **File: longestCommonSubsequence.raml** | | | |
| firstline | $2+5M$ | 0.02 | 51 |
| right | 1 | 0.01 | 12 |
| max | 1 | 0.01 | 5 |
| newline | $2+7M$ | 0.08 | 276 |
| lcstable | $8+5L+7LM+10M$ | 0.14 | 782 |
| lcs | $9+5L+7LM+10M$ | 0.15 | 806 |
| **File: mergesort.raml** | | | |
| msplit | $6+5M$ | 0.03 | 97 |
| merge | $8L+8M$ | 0.05 | 212 |
| mergesortBuggy | – | – | *fail* |
| mergesort | $2-18.67M+23.33M^2$ | 0.14 | 805 |
| **File: minsort.raml** | | | |
| findMin | $2+8M$ | 0.03 | 104 |
| minSort | $4+10M+4M^2$ | 0.04 | 188 |
| **File: queue.raml** | | | |
| empty | 6 | 0.01 | 4 |
| enqueue | 6 | 0.03 | 12 |
| enqueues | $6M$ | 0.1 | 184 |
| copyover | $4+6M$ | 0.08 | 214 |
| dequeue | $10+18M$ | 0.15 | 647 |
| children | 24 | 0.23 | 309 |
| breadth | $12+54BR+54BRT+54LMY+38M+54MY+20R$ | 3.76 | 10447 |
| startBreadth | $15+25M+27M^2$ | 3.99 | 10704 |
| depth | $16+24L+28LM$ | 4.13 | 18650 |
| startDepth | $2+10M+14M^2$ | 4.39 | 18676 |
| **File: quicksort_mutual.raml** | | | |
| part | $8+6L+8LY+4L^2+10M+8MY+4M^2+14Y+4Y^2$ | 0.15 | 843 |
| quicksortMutual | $2+6M+4M^2$ | 0.15 | 843 |
| **File: rationalPotential.raml** | | | |
| zip3 | $2+7L$ | 0.02 | 52 |
| group3 | $2+2.33M$ | 0.02 | 38 |
| **File: sizechange.raml** | | | |
| r1 | $4M$ | 0.05 | 114 |
| rev | $2+4M$ | 0.04 | 117 |
| f | $4LM-1.33M+1.33M^3$ | 0.12 | 545 |
| g | $4L+4LM+1.33M+4MY+1.33M^3$ | 0.12 | 545 |
| f2 | – | – | *fail* |
| last | $2+6M$ | 0.04 | 96 |
| f2' | $2+6L+6M$ | 0.08 | 303 |
| g3 | $4M$ | 0.05 | 114 |
| f3 | $2+8L+4M$ | 0.08 | 267 |
| **File: splitandsort.raml** | | | |
| insert | $14+6M$ | 0.13 | 411 |
| split | $2+11M+3M^2$ | 0.18 | 587 |
| splitqs | $6+8M$ | 0.05 | 161 |
| quicksort | $2+8M+6M^2$ | 0.26 | 1556 |
| sortAll | $2+8LM+6L^2M+8M$ | 0.49 | 3101 |
| splitAndSort | $4+27M+9M^2$ | 0.66 | 3688 |
| **File: subtrees.raml** | | | |
| subtrees | $2+9M+2M^2$ | 0.12 | 651 |
| **File: tuples.raml** | | | |
| attach | $2+6M$ | 0.07 | 85 |
| pairs | $2-3M+5M^2$ | 0.28 | 1395 |
| pairsAux | $-3M+5M^2$ | 0.25 | 1006 |
| pairsSlow | $2+0.33M+M^2+0.67M^3$ | 0.27 | 1395 |
| triples | $2+9.67M-9M^2+3.33M^3$ | 1.04 | 5736 |
| quadruples | $2-4.83M+14.75M^2-7.17M^3+1.25M^4$ | 2.95 | 15510 |
| **File: array_dijkstra.raml** | | | |
| makeGraph | $4+2N+N^2$ | 0.03 | 437 |
| dijkstra | $6+8.5M+9.5M^2$ | 0.1 | 2607 |
| **File: CompCert** | | | |
| **File: String0.ml** | | | |

| |
|---|
| string_dec |
| prefix |
| **File: Tuples.ml** |
| uncurry |
| **File: Specif.ml** |
| projT1 |
| projT2 |
| value |
| **File: EquivDec.ml** |
| equiv_dec |
| **File: Datatypes.ml** |
| implb |
| xorb |
| negb |
| fst |
| snd |
| length |
| app |
| coq_CompOpp |
| coq_CompareSpec2Type |
| coq_CompSpec2Type |
| **File: Bool.ml** |
| bool_dec |
| eqb |
| iff_reflect |
| **File: Ring.ml** |
| bool_eq |
| **File: Peano.ml** |
| plus |
| max |
| min |
| nat_iter |
| **File: List0.ml** |
| hd |
| tl |
| in_dec |
| nth_error |
| remove |
| rev |
| rev_append |
| rev' |
| list_eq_dec |
| map |
| fold_left |
| fold_right |
| existsb |
| forallb |
| filter |
| **File: EqNat.ml** |
| beq_nat |
| **File: Compare_dec.ml** |
| le_lt_dec |
| le_gt_dec |
| nat_compare |
| **File: BinPosDef.ml** |
| succ |
| add |
| add_carry |
| pred_double |
| pred |
| pred_N |
| mask_rect |
| mask_rec |
| succ_double_mask |
| double_mask |
| double_pred_mask |
| pred_mask |
| sub_mask |
| sub_mask_carry |
| sub |
| mul |
| iter |
| pow |
| square |
| div2 |
| div2_up |
| size_nat |
| size |
| compare_cont |
| compare |
| min |
| max |
| eqb |
| leb |
| ltb |
| sqrtrem_step |
| sqrtrem |
| sqrt |
| gcdn |
| gcd |
| ggcdn |
| ggcd |
| coq_Nsucc_double |
| coq_Ndouble |
| coq_lor |
| coq_land |
| ldiff |
| coq_lxor |
| shiftl_nat |
| shiftr_nat |
| shiftl |
| shiftr |
| testbit_nat |
| testbit |
| iter_op |
| to_nat |
| of_nat |
| of_succ_nat |
| digits2_pos |
| coq_Zdigits2 |
| **File: BinNat.ml** |
| succ_double_n |
| double_n |
| succ_n |
| pred_n |
| succ_pos_n |
| add_n |
| sub_n |
| mul_n |
| compare_n |
| eqb_n |
| leb_n |
| ltb_n |
| min_n |
| max_n |
| div2_n |
| even_n |
| odd_n |
| pow_n |
| square_n |
| log2_n |
| size_n |
| size_nat_n |
| pos_div_eucl |
| div_eucl |
| div |
| modulo |
| gcd_n |
| ggcd_n |
| coq_lor_n |
| coq_land_n |
| ldiff_n |
| coq_lxor_n |
| shiftl_nat |
| shiftr_nat |
| shiftl_n |
| shiftr_n |
| testbit_nat_n |
| testbit_n |
| to_nat_n |
| of_nat_n |
| iter_n |
| discr |
| binary_rect |
| binary_rec |
| leb_spec0 |
| ltb_spec0 |
| log2_up |
| lcm |
| eqb_spec |
| b2n |
| setbit |
| clearbit |
| ones |
| lnot |
| max_case_strong |
| max_case |
| max_dec |
| min_case_strong |
| min_case |
| min_dec |
| max_case_strong_pd |
| max_case |
| max_dec_pd |
| min_case_strong_pd |
| min_case |
| min_dec_pd |

## A.4 Tick Bounds

| Name | Tick Bound | Analysis Time | #Constraints |
|---|---|---|---|
| **File: WorkingWithLists.raml (99 Problems in OCaml)** | | | |
| last | $M$ | 0.01 | 30 |
| lastTwo | $M$ | 0.01 | 42 |
| at | $M$ | 0.01 | 29 |
| natAt | $K$ | 0.01 | 35 |
| length | $M$ | 0.01 | 20 |
| rev | $M$ | 0.01 | 26 |
| eqList | $L$ | 0.02 | 40 |
| isPalindrome | $2M$ | 0.02 | 71 |
| flatten | – | – | $fail$ |
| compress | $M$ | 0.03 | 39 |
| pack | $2M$ | 0.05 | 92 |
| encode | $2M$ | 0.07 | 82 |
| decode | – | – | $fail$ |
| duplicate | $M$ | 0.06 | 20 |
| replicate | – | – | $fail$ |
| drop | $M$ | 0.08 | 33 |
| split | $2M$ | 0.12 | 91 |
| slice | $1+M$ | 0.13 | 61 |
| concat | $M$ | 0.12 | 24 |
| rotate | $3M$ | 0.2 | 156 |
| removeAt | $M$ | 0.18 | 27 |
| insertAt | $M$ | 0.25 | 76 |
| constructList | – | – | – |
| random | 1 | 0.24 | 3 |
| min | 1 | 0.25 | 6 |
| randSelect | $1+2M+M^2$ | 0.52 | 484 |
| lottoSelect | – | – | – |
| snd | 1 | 0.48 | 2 |
| fst | 1 | 0.45 | 2 |
| map | $M$ | 0.43 | 60 |
| insert | $M$ | 0.45 | 85 |
| sort | $0.5M+0.5M^2$ | 0.48 | 160 |
| compare | 1 | 0.48 | 10 |
| lengthSort | $LM+2M+2M^2$ | 0.69 | 859 |
| **File: LogicAndCodes.raml** | | | |
| eval2 | $1+2K+2L+M$ | 0.29 | 980 |
| table2 | $4+8K+8L+4M$ | 1.53 | 5594 |
| assoc | $M$ | 0.05 | 50 |
| eval | $1+L+M+MX+MY+2X+2Y$ | 0.32 | 1809 |
| tableMake | – | – | $fail$ |
| **File: echelon_form.raml** | | | |
| size | $n$ | 0.01 | 18 |
| getElem | $n$ | 0.01 | 23 |
| get2Elems | $L$ | 0.01 | 36 |
| subtract_row_helper | $L$ | 0.02 | 35 |
| subtract_row | $L+M$ | 0.03 | 84 |
| subtract_helper | $M+2MY$ | 0.16 | 771 |
| concat | $n_1$ | 0.03 | 24 |
| tail | $n$ | 0.03 | 26 |
| hd_helper | $M$ | 0.03 | 31 |
| reverse_helper | $M$ | 0.03 | 24 |
| reverse | $n$ | 0.03 | 26 |
| head | $2n$ | 0.04 | 59 |
| split_helper | $L+2M$ | 0.06 | 60 |
| split | $2n$ | 0.06 | 62 |
| subtract | $2LM+M$ | 0.3 | 1453 |
| echelon_helper | $2LMY+4MY$ | 1.82 | 8369 |
| echelon_form | $2m^2n+4m^2$ | 1.81 | 8639 |
| **File: matrix.raml** | | | |
| check_lists | $LM+M$ | 0.04 | 133 |
| check_mat | $LM+2M$ | 0.05 | 210 |
| check_matrix | $LM+2M$ | 0.06 | 212 |
| construct_matrix | $LM+2M$ | 0.07 | 287 |
| getElemMatrix | $mn+m$ | 0.05 | 135 |
| op | 1 | 0.03 | 10 |
| rec_list | $2M$ | 0.05 | 42 |
| rec_mat | $2LM+Y$ | 0.14 | 471 |
| check_sanity | $LM+2M+RY+2Y$ | 0.15 | 534 |
| plus | $4m_1n_1+2m_1+m_2n_2+5m_2$ | 0.42 | 1632 |
| minus | $4m_1n_1+2m_1+m_2n_2+5m_2$ | 0.51 | 1632 |
| append | $n$ | 0.25 | 20 |
| append_col | $mn+l$ | 0.33 | 357 |
| transpose_helper | $0.5LM+0.5LM^2+M+MRY$ | 0.63 | 1764 |
| transpose | $0.5mn+0.5m^2n+m$ | 0.58 | 1766 |
| prod | $M$ | 0.32 | 32 |
| prod_mat | $L+LY$ | 0.37 | 347 |
| mult_slow | $m_1+2.5m_1m_2n_2+0.5m_1m_2^2n_2+m_1m_2+2.5m_2n_2+0.5m_2^2n_2+m_2$ | 6.57 | 42169 |
| lineMult | $M$ | 0.41 | 44 |
| computeLine | $LY+M$ | 0.48 | 426 |
| mat_mult_jan | $M+MRY+MY$ | 1.09 | 3836 |
| check_mult_sanity | $LM+2M+RY+2Y$ | 0.69 | 526 |
| mult | $m_1n_1+5m_1+m_1m_2n_2+m_1m_2+m_2n_2+2m_2$ | 1.93 | 7249 |
| delete | $n$ | 1.76 | 27 |
| submat | $mn+m$ | 1.86 | 357 |
| **File: power_radio.raml** | | | |
| sendmsg msg | $200+32n$ | 0.01 | 21 |
| main1_events | $200L'+32L'Y$ | 0.1 | 302 |
| main2_events | $32L'Y+200Z'$ | 0.21 | 1177 |
| main3_events | $16L'Y+16L'MY+200R'+200Z'$ | 0.31 | 7254 |
| main4_events | $32L'Y+200Z'$ | 0.43 | 2317 |
| main5_events | $40L'+32L'Y$ | 0.41 | 2207 |
| **File: avanzini.raml** | | | |
| partition | | | |
| quicksort | | | |
| rev_sort | | | |
| **File: append_all.raml** | | | |
| append_all | | | |
| append_all2 | | | |
| append_all3 | | | |
| **File: bfs.raml** | | | |
| dfs | | | |
| bfs | | | |
| **File: rev_pairs.raml** | | | |
| pairs | $-0.50M+0.5*M^2$ | 0.02 | 284 |
| **File: binary_counter.raml** | | | |
| add_one | | | |
| add_many | | | |
| add_list | | | |
| **File: array_fun.raml** | | | |
| nat_iterate | | | |
| nat_fold | | | |
| apply_all | | | |
| **File: calculator.raml** | | | |
| add | $1+M$ | 0.04 | 109 |
| sub | $1+M$ | 0.03 | 145 |
| mult | $0.5M + 0.5M^2$ | 0.06 | 409 |
| eval_simpl | $KLN + 0.5KN + L + X$ | 0.23 | 9498 |
| eval | 0 | 0.2 | 10678 |
| **File: mergesort.raml** | | | |
| split | 0 | 0.05 | 133 |
| merge | $L + M$ | 0.05 | 378 |
| mergesort | $-0.5K + 0.5K^2$ | 0.16 | 4567 |
| mergesort_list | $-0.5K + 0.5K^2$ | 0.42 | 14946 |
| **File: quicksort.raml** | | | |
| partition | $M$ | 0.05 | 170 |
| quicksort | $-0.5M + 0.5M^2$ | 0.07 | 1687 |
| quicksort_pairs | $-0.5M + 0.5M^2$ | 0.09 | 2046 |
| quicksort_list | $-0.5M + 0.5M^2$ | 0.26 | 8628 |
| **File: square_mult.raml** | | | |
| square_mult | $K' + 2L'$ | 0.07 | 199 |
| **File: subsequence.raml** | | | |
| lcs | | | |
| **File: running.raml** | | | |
| abmap | | | |
| asort | | | |
| asort' | | | |
| btick | $2.5L$ | 0.05 | 877 |
| abfoldr | | | |
| cons_all | | | |
| **File: ocaml_sort.raml** | | | |
| merge | | | |
| list | | | |
| **File: ocaml_list.raml** | | | |
| length | | | |
| cons | | | |
| hd | | | |
| tl | | | |
| nth | | | |
| append | | | |
| rev_append | | | |
| rev | | | |
| flatten | | | |
| concat | | | |
| map | | | |
| mapi | | | |
| rev_map | | | |
| iter | | | |
| iteri | | | |
| fold_left | | | |
| fold_right | | | |
| map2 | | | |
| rev_map2 | | | |
| iter2 | | | |
| fold_left2 | | | |
| fold_right2 | | | |
| for_all | | | |
| exists | | | |
| for_all2 | | | |
| exists2 | | | |
| mem | | | |
| memq | | | |
| assoc | | | |
| assq | | | |
| mem_assoc | | | |
| mem_assq | | | |
| remove_assoc | | | |
| remove_assq | | | |
| find | | | |
| find_all | | | |
| filter | | | |
| partition | | | |
| split | | | |
| combine | | | |
| merge | | | |
| chop | | | |
| stable_sort | | | |
| sort | | | |
| fast_sort | | | |
| **File: aws.raml** | | | |
| average_grade | $M$ | 0.04 | 136 |
| greater_eq | $2M$ | 0.05 | 610 |
| sort_students | $LM + LM^2$ | 0.31 | 8733 |
| make_table | $LM$ | 0.07 | 900 |
| find | 0 | 0.06 | 75 |
| lookup | 0 | 0.05 | 224 |
| average_grade' | 0 | 0.09 | 1000 |
| greater_eq' | 0 | 0.13 | 2365 |
| sort_students_efficient | $LM$ | 0.72 | 19030 |
| **File: PROTOTYPE** | | | |
| **File: appendAll.raml** | | | |
| appendAll | $n_1 n_2 + n_1$ | 0.04 | 184 |
| appendAll2 | $n_1 n_2 + 2n_1 n_2 n_3 + n_1$ | 0.2 | 1525 |
| appendAll3 | $n_1 n_2 + 3n_1 n_2 n_3 n_4 + n_1 n_2 n_3 + n_1$ | 1.57 | 12724 |
| **File: duplicates.raml** | | | |
| eq | $n_2$ | 0.01 | 40 |
| remove | $L + LY$ | 0.1 | 583 |
| nub | $-0.5Nn + 0.5N^2 n + 0.5N + 0.5N^2$ | 0.52 | 3864 |
| **File: dyade.raml** | | | |
| multList | $n$ | 0.01 | 21 |
| dyade | $n_1 n_2 + n_1$ | 0.05 | 217 |
| **File: eratosthenes.raml** | | | |
| filter | $n$ | 0.01 | 27 |
| eratos | $0.5n + 0.5n^2$ | 0.03 | 143 |
| **File: bitvectors.raml** | | | |
| bitToInt' | $M$ | 0.02 | 22 |
| bitToInt | $n$ | 0.01 | 24 |
| sum | 1 | 0.01 | 18 |
| add' | $2M$ | 0.03 | 50 |
| add | $2n_1$ | 0.03 | 52 |
| diff | 1 | 0.03 | 16 |
| sub' | $2M$ | 0.06 | 55 |
| sub | $1 + 2n_1$ | 0.06 | 58 |
| mult | $2n_1 n_2 + n_1$ | 0.11 | 387 |
| compare | $M$ | 0.06 | 42 |
| leq | $1 + n_1$ | 0.07 | 47 |
| **File: flatten.raml** | | | |
| flatten | $0.5nl + 0.5n^2 l + n$ | 0.41 | 3351 |
| insert | $n$ | 0.01 | 24 |
| insertionsort | $0.5n + 0.5n^2$ | 0.04 | 144 |
| flattensort | $nl + 0.5n^2 l + 0.5n^2 l^2 + n$ | 2.32 | 19074 |
| **File: listsort.raml** | | | |
| isortlist | $-0.5nl + 0.5n^2 l + 0.5n + 0.5n^2$ | 0.39 | 2621 |
| **File: longestCommonSubsequence.raml** | | | |
| firstline | $M$ | 0.02 | 48 |
| right | 1 | 0.02 | 12 |
| max | $1$ | 0.01 | 5 |
| newline | $3M$ | 0.08 | 272 |
| lcstable | $L+3LM+M$ | 0.15 | 768 |
| lcs | $1+L+3LM+M$ | 0.15 | 791 |
| **File: mergesort.raml** | | | |
| msplit | $0.5M$ | 0.04 | 85 |
| merge | $L+M$ | 0.05 | 203 |
| mergesortBuggy | – | – | $fail$ |
| mergesort | $-1.5M+1.5M^2$ | 0.14 | 778 |
| **File: minsort.raml** | | | |
| findMin | $M$ | 0.04 | 93 |
| minSort | $1.5M+0.5M^2$ | 0.04 | 175 |
| **File: queue.raml** | | | |
| empty | $1$ | 0.01 | 1 |
| enqueue | $1$ | 0.03 | 10 |
| enqueues | $2M$ | 0.09 | 183 |
| copyover | $M$ | 0.07 | 210 |
| dequeue | $1+2M$ | 0.16 | 631 |
| children | $1$ | 0.22 | 294 |
| breadth | $7BR+7BRT+7LMY+4M+7MY+2R$ | 3.54 | 10413 |
| startBreadth | $2.5M+3.5M^2$ | 3.83 | 10663 |
| depth | $2L+3LM$ | 4.02 | 18631 |
| startDepth | $1-1.5M+1.5M^2$ | 4.33 | 18656 |
| **File: quicksort_mutual.raml** | | | |
| part | $1+L+2LY+L^2+2M+2MY+M^2+3Y+Y^2$ | 0.14 | 836 |
| quicksortMutual | $M+M^2$ | 0.14 | 836 |
| **File: rationalPotential.raml** | | | |
| zip3 | $L$ | 0.02 | 47 |
| group3 | $0.33M$ | 0.01 | 33 |
| **File: sizechange.raml** | | | |
| r1 | $M$ | 0.05 | 113 |
| rev | $M$ | 0.04 | 115 |
| f | $LM+1.67M+0.33M^3$ | 0.12 | 546 |
| g | $1+L+LM+1.67M+MY+0.33M^3$ | 0.12 | 546 |
| f2 | – | – | $fail$ |
| last | $M$ | 0.04 | 93 |
| f2’ | $1+L+M$ | 0.09 | 302 |
| g3 | $M$ | 0.04 | 113 |
| f3 | $1+2L+M$ | 0.08 | 265 |
| **File: splitandsort.raml** | | | |
| insert | $M$ | 0.13 | 397 |
| split | $0.5M+0.5M^2$ | 0.17 | 573 |
| splitqs | $M$ | 0.05 | 152 |
| quicksort | $M^2$ | 0.27 | 1543 |
| sortAll | $L^2M+M$ | 0.49 | 3085 |
| splitAndSort | $1.5M+1.5M^2$ | 0.61 | 3658 |
| **File: subtrees.raml** | | | |
| subtrees | $0.5M+0.5M^2$ | 0.12 | 646 |
| **File: tuples.raml** | | | |
| attach | $M$ | 0.06 | 82 |
| pairs | $M^2$ | 0.24 | 1391 |
| pairsAux | $M^2$ | 0.25 | 1003 |
| pairsSlow | $0.83M+0.17M^3$ | 0.3 | 1391 |
| triples | $1.83M-1.5M^2+0.67M^3$ | 1.05 | 5728 |
| quadruples | $-0.67M+2.75M^2-1.33M^3+0.25M^4$ | 2.79 | 15498 |
| **File: array_dijkstra.raml** | | | |
| makeGraph | | | |
| dijkstra | | | |
| **File: CompCert** | | | |

| |
|---|
| **File: String0.ml** |
| string_dec |
| prefix |
| **File: Tuples.ml** |
| uncurry |
| **File: Specif.ml** |
| projT1 |
| projT2 |
| value |
| **File: EquivDec.ml** |
| equiv_dec |
| **File: Datatypes.ml** |
| implb |
| xorb |
| negb |
| fst |
| snd |
| length |
| app |
| coq_CompOpp |
| coq_CompareSpec2Type |
| coq_CompSpec2Type |
| **File: Bool.ml** |
| bool_dec |
| eqb |
| iff_reflect |
| **File: Ring.ml** |
| bool_eq |
| **File: Peano.ml** |
| plus |
| max |
| min |
| nat_iter |
| **File: List0.ml** |
| hd |
| tl |
| in_dec |
| nth_error |
| remove |
| rev |
| rev_append |
| rev' |
| list_eq_dec |
| map |
| fold_left |
| fold_right |
| existsb |
| forallb |
| filter |
| **File: EqNat.ml** |
| beq_nat |
| **File: Compare_dec.ml** |
| le_lt_dec |
| le_gt_dec |
| nat_compare |
| **File: BinPosDef.ml** |
| succ |
| add |
| add_carry |
| pred_double |
| pred |
| pred_N |
| mask_rect |
| mask_rec |
| succ_double_mask |
| double_mask |
| double_pred_mask |
| pred_mask |
| sub_mask |
| sub_mask_carry |
| sub |
| mul |
| iter |
| pow |
| square |
| div2 |
| div2_up |
| size_nat |
| size |
| compare_cont |
| compare |
| min |
| max |
| eqb |
| leb |
| ltb |
| sqrtrem_step |
| sqrtrem |
| sqrt |
| gcdn |
| gcd |
| ggcdn |
| ggcd |
| coq_Nsucc_double |
| coq_Ndouble |
| coq_lor |
| coq_land |
| ldiff |
| coq_lxor |
| shiftl_nat |
| shiftr_nat |
| shiftl |
| shiftr |
| testbit_nat |
| testbit |
| iter_op |
| to_nat |
| of_nat |
| of_succ_nat |
| digits2_pos |
| coq_Zdigits2 |
| **File: BinNat.ml** |
| succ_double_n |
| double_n |
| succ_n |
| pred_n |
| succ_pos_n |
| add_n |
| sub_n |
| mul_n |
| compare_n |
| eqb_n |
| leb_n |
| ltb_n |
| min_n |
| max_n |
| div2_n |
| even_n |
| odd_n |
| pow_n |
| square_n |
| log2_n |
| size_n |
| size_nat_n |
| pos_div_eucl |
| div_eucl |
| div |
| modulo |
| gcd_n |
| ggcd_n |
| coq_lor_n |
| coq_land_n |
| ldiff_n |
| coq_lxor_n |
| shiftl_nat |
| shiftr_nat |
| shiftl_n |
| shiftr_n |
| testbit_nat_n |
| testbit_n |
| to_nat_n |
| of_nat_n |
| iter_n |
| discr |
| binary_rect |
| binary_rec |
| leb_spec0 |
| ltb_spec0 |
| log2_up |
| lcm |
| eqb_spec |
| b2n |
| setbit |
| clearbit |
| ones |
| lnot |
| max_case_strong |
| max_case |
| max_dec |
| min_case_strong |
| min_case |
| min_dec |
| max_case_strong_pd |
| max_case |
| max_dec_pd |
| min_case_strong_pd |
| min_case |
| min_dec_pd |